\documentstyle[aps,preprint]{revtex}
\input epsf
\global\firstfigfalse  % kludge to bypass foolishness of revtex to place
\tighten

\def\NN{{\cal N}}
\def\nzero{$\NN=0$}
\def\none{$\NN=1$}
\def\noneplus{$\NN=1^*$}
\def\ntwo{$\NN=2$}
\def\nfour{$\NN=4$}

\def\ZZ{{\bf Z}}
\def\slz{$SL(2,\ZZ)$}
\def\tr{{\rm tr}}
\def\half{{1\over2}}

\def\bel{\begin{equation}\label}
\def\ee{\end{equation}}
\newcommand\eref[1]{(\ref{#1})}
\newcommand\Eref[1]{Eq.~\eref{#1}}

\def\mod{\ {\rm mod}\ }

\def\RR{{\bf R}}

\renewcommand{\*}{ &=& }

\newcommand{\xto}{ &\to& }
\newcommand{\OL}[1]{ \hspace{1pt}\overline{\hspace{-1pt}#1
   \hspace{-1pt}}\hspace{1pt} }

\begin{document}
\pagestyle{empty}

\preprint{
\begin{minipage}[t]{3in}
\begin{flushright} IAS--TH--00/18,  NSF-ITP-00-16
\\
hep-th/0003136
\\
March 2000
\end{flushright}
\end{minipage}
}

\title{
The String Dual of a
Confining Four-Dimensional Gauge Theory}

\author{Joseph Polchinski
\thanks{Institute for Theoretical Physics, University of California, Santa
Barbara CA 93106-4030} and
Matthew J. Strassler
\thanks{School of Natural Sciences, Institute for Advanced Study, Olden
Lane, Princeton NJ 08540}
}

\maketitle

\begin{abstract}

We study \none\ gauge theories obtained by adding finite mass terms to
\nfour\ Yang-Mills theory.  The Maldacena dual is nonsingular: in each
of the many vacua, there is an extended brane source, arising from
Myers' dielectric effect.  The source consists of one or more $(p,q)$
5-branes.  In particular, the confining vacuum contains an NS5-brane;
the confining flux tube is a fundamental string bound to the 5-brane.
The system admits a simple quantitative description as a perturbation
of a state on the \nfour\ Coulomb branch.  Various nonperturbative
phenomena, including flux tubes, baryon vertices, domain walls,
condensates and instantons, have new, quantitatively precise, dual
descriptions.  We also briefly consider two QCD-like theories.  Our
method extends to the nonsupersymmetric case.  As expected, the
\nfour\ matter cannot be decoupled within the supergravity regime.
\end{abstract}

\newpage
\pagestyle{plain}
\narrowtext
\baselineskip=18pt

\setcounter{footnote}{0}

\section{Introduction}

The proposal of 't Hooft \cite{tHooftlargeN}, that large-$N$ non-abelian
gauge theory can be recast as a string theory, has taken an interesting
turn with the work of Maldacena \cite{maldacon}.
The principal Maldacena duality applies not to confining theories but to
conformal
\nfour\ gauge theories, which are dual to IIB string theory on $AdS_5
\times S^5$.  Starting with this duality
one can perturb by the addition of mass terms preserving a smaller
supersymmetry, or none at all, and in this way obtain a confining gauge
theory.   The problem is that the perturbation of the dual string theory
appears to produce a spacetime with a naked singularity \cite{gppz}.  As
a consequence, even basic quantities such a condensates are incalculable.

In this paper we show that the situation is actually much better.
There is no naked singularity, but rather an expanded brane source,
and all physical quantities are calculable.  We believe that this is
the first example of a dual supergravity description of a
four-dimensional confining gauge theory.  It is also gives new insight
into the resolution of naked singularities in string theory.

We focus on perturbations that preserve \none\ supersymmetry, though
in fact our solutions are stable under the addition of small
additional masses that break the supersymmetry completely. The \nfour\
vector multiplet contains an \none\ vector multiplet and three \none\
chiral multiplets.  We will add finite \none\ supersymmetry-preserving
masses to the three chiral multiplets.  For brevity we will refer to
this theory as `\noneplus'.  This theory has been studied by many
authors \cite{cvew,rdew,mjsnotes,doreya,doreykumar}.  It is known to
have a rich phase structure \cite{cvew,rdew,lattice}, which includes
confining phases that are in the same universality class as those of
pure \none\ Yang-Mills theory.  We will show that the rich structure
of this theory is reflected in supergravity in remarkable ways.

To study pure \nzero\ or \none\ Yang-Mills theories would require
working at small 't Hooft coupling and taking the masses of the extra
multiplets to infinity.  This is not tractable without an understanding of
classical string theory in Ramond-Ramond backgrounds at large
curvature. At large 't Hooft coupling, where supergravity is valid, the
masses of the extra multiplets must be kept finite.  However, we
emphasize these multiplets are four-dimensional and the ultraviolet
theory is conformal. An alternative approach to obtaining a string dual
of confining theories is via high-temperature five-dimensional
supersymmetric field theories \cite{highTa,highTb,highTc,grossooguri}, whose
low-energy limit is four-dimensional strongly-coupled
non-supersymmetric Yang-Mills theory.  The dual spacetime is
non-singular, and the infrared cutoff provided by the temperature does
indeed lead to confinement of electric flux tubes.  In this case,
however, there is a full set of massive five-dimensional states that
do not decouple.

Our work was motivated by the observation of Myers \cite{myers}, that
D-branes in a transverse Ramond-Ramond (RR) potential can develop a
multipole moment under fields that normally couple to a
higher-dimensional brane.  This `dielectric' property is analogous to
the induced dipole moment of a neutral atom in an electric field.  For
example, a collection of $N$ D0-branes in an electric RR 4-form flux
develops a dipole moment under the corresponding 3-form potential.
One can think of them as blowing up into a spherical D2-brane, and in
a strong field the latter is the effective description.  This happens
because the D0-brane coordinates become noncommutative.  The original
D0-brane charge $N$, which of course is conserved in this process,
shows up as a nonzero world-volume field strength on the D2-branes.
Even earlier, Kabat and Taylor \cite{ktsphere} had observed that $N$
D0-branes with noncommuting position matrices could be used to build a
spherical D2-brane in matrix theory, generalizing the flat membranes
of matrix theory~\cite{bfss}.  For finite $N$ the sphere is `fuzzy';
or better, perhaps, it is somewhat granular.  The equations describing
this sphere bear a marked similarity to those which appear in the
\noneplus\ theory, which were first analyzed in \cite{cvew}.

It is then natural to guess that Myers' mechanism is at work in this
theory.  The mass perturbation corresponds to a magnetic RR 3-form
flux, which is dual to an electric RR 7-form flux.  The
latter couples to the D3-brane in the same fashion as the electric
4-form flux does to the D0-brane, and so the D3-branes polarize
into D5-branes with world-volume $\RR^4\times S^2$.  One difference is
that Myers considers D-branes in flat spacetime ($gN$ small), whereas
for the gauge/gravity duality the background is $AdS_5 \times S^5$.
In Myers' case a small field produced a small D2-sphere, but in the
conformal field theory there is no invariant notion of a small mass
perturbation, and on the supergravity side there is no such thing as a
small transverse two-sphere.  Rather, the D5-spheres, which are
dynamically (though not topologically) stable, wrap an equator of
the $S^5$.  We will show that there exist supergravity solutions in
which the only `singularity' is that due to the D5-brane source on the
$S^5$.

However, this is far from the whole story.  First, the classical
\noneplus\ theory has many isolated vacua \cite{cvew}.  For each
partition of $N$ into integers $n_i$, there must be a separate
solution involving multiple D5-branes with D3-branes charges $n_i$,
each wrapped on an equator of $S^5$ but at different $AdS$ radii $r_i$
proportional to $n_i$.  We will study these vacua, and their
properties, in our discussion below.  Second, the quantum theory has
even more vacua, which are permuted under the \slz\ duality the field
theory inherits from \nfour \cite{rdew}.  In particular, the
transformation $\tau \to -{1\over\tau}$, which takes the maximally
Higgsed vacuum into the confining vacuum, will replace the D5-brane
sphere with an NS5-brane sphere: {\it this is the effective string
description of the confining vacuum.}  The confining flux tubes are
bound states of a fundamental string to the NS5-brane, or
equivalently, instantons of the 5-brane world-volume noncommutative
gauge theory.  Meanwhile, the leading nonperturbative condensate
corresponds to the three-form field generated by the NS5-brane's
magnetic dipole moment.

Our removal of the singularity resembles phenomena that occur on the
Coulomb branch~\cite{coul1,coul2,coul3,coul4} and with the repulson
singularity that arises in \ntwo\ supergravity duals~\cite{jpp}.  There are
certainly connections which need to be developed further, but the
detailed mechanism is different.  In particular, the appearance of
NS-branes is new.  Our result also gives insight into perturbations of
the Randall-Sundrum compactification \cite{RS}, and into recent
proposals for the solution to the cosmological constant problem
\cite{stanford}.

We begin in section II with a review of the classical and quantum
field theory vacua, and a discussion of the corresponding brane
configurations.  In fact, there are more brane configurations than
vacua, but later we will argue that only one configuration is
applicable for any given value of the parameters.  In section III we
review perturbations of the $AdS$/CFT duality, with attention to the
issue of the naked singularity.  We show that there is a small
parameter: the system can be regarded as a perturbation of one that
has only D3-brane charges.  This enables us to obtain a quantitative
description even for the rather asymmetric and nonlinear supergravity
configuration that results from the expansion of the branes. In
section IV we study a simplified calculation, in which $n \ll N$ probe
D3-branes are introduced into a fixed background.  We find that their
potential has minima where they form a D5-brane or NS5-brane, or more
generally one or more $(c,d)$ 5-branes, wrapped on an equator of the
$S^5$. In section V we consider the case that all $N$ D3-branes expand
into 5-branes.  Although this substantially deforms the geometry,
serendipitous cancellations allow us to find the effective potential
in a simple form: it is the same as in the probe case.  We discuss the
stability of the solution, arguing that it survives even when
supersymmetry is broken completely.  In section VI we use the dual
description to discuss the physics of the gauge theory, including flux
tubes and confinement, baryons, domain walls, condensates, instantons,
and glueballs. In section VII we briefly discuss extensions, including
the \nzero\ case and orbifolds, and in section VIII we discuss
implications and future directions.

\section{\noneplus\ Ground States}

\subsection{Field Theory Background}

In the language of four-dimensional \none\ supersymmetry, the \nfour\
theory consists of a vector multiplet $V$ and three chiral multiplets
$\Phi_i$, $i=1,2,3$, all in the adjoint representation of the gauge
group.  In addition to the usual gauge-invariant kinetic terms for
these fields, the theory has additional interactions summarized in the
superpotential\footnote
{The K\"ahler potential is normalized $(2/ g_{\rm YM}^{2}) \tr\, \OL\Phi_i
\Phi_i$.}

\begin{equation}
W = \frac{2\sqrt 2}{g_{\rm YM}^2} \tr ([\Phi_1,\Phi_2]\Phi_3) \ .
\end{equation}
The theory has an $SO(6)$ $R$-symmetry which is partially
hidden by the \none\ notation; only the $U(1)$ $R$-symmetry of
the \none\ supersymmetry and the $SU(3)$ that rotates the $\Phi_i$
are visible.  However, if we write the lowest component of $\Phi_i$ as
\begin{equation}
\phi_i = \frac{A_{i+3} + i
A_{i+6}}{\sqrt{2}}
\end{equation}
(the reason for this notation will become evident
later), then the potential energy
for the scalar fields $A_m$, $m=4,\dots,9$,
is explicitly $SO(6)$ invariant:
\begin{equation}
V(A_m) \propto
\sum_{m,n=4}^9\tr\left([A_m,A_n][A_m,A_n]\right)\ .
\label{supmass}
\end{equation}
The theory is
conformally invariant, and consists of a continuous set of
theories indexed by a marginal coupling $\tau = {\theta\over 2\pi} + i
{4\pi\over g_{\rm YM}^2}$, where $\theta $ and $g_{\rm YM}$ are the theta
angle and gauge coupling of the theory.

We can partially break the supersymmetry by adding arbitrary
terms to the superpotential.  Consider the addition of mass terms
\begin{equation}\label{Wmass}
\Delta W = \frac{1}{g_{\rm YM}^2} (m_1\, \tr\,\Phi_1^2 + m_2\, \tr\,
\Phi_2^2 + m_3
\, \tr\, \Phi_3^2)\ .
\end{equation}
If $m_1=m_2$ and $m_3=0$ the theory has \ntwo\ supersymmetry;
otherwise it has \none.  If $m_1=m_2=0$ and $m_3\neq 0$ then the
theory flows to a conformal fixed point with a smooth moduli space and
\slz\ duality \cite{twoadj,karch,coul3,ALIS}.  With two nonzero masses, the
theory has a moduli space containing special subspaces where charged
particles are massless and the K\"ahler metric is singular.  However,
in \noneplus, where all three masses are non-zero,
there is no moduli space; the theory has a number of isolated vacua.
In the limit
\bel{nonelimit}
\tau\to i\infty\ ,\quad m_i\to\infty\ , \quad
\Lambda^3 = m_1m_2m_3 e^{2\pi i\tau/N} \ {\rm fixed}
\ee
the theory becomes pure \none\ Yang-Mills theory.
For gauge group $SU(N)$ the pure
\none\ theory has
$N$ vacua related by a spontaneously broken discrete $R$-symmetry.
Note that this $R$-symmetry is not present in \noneplus; it
is an accidental symmetry present only in the limit \Eref{nonelimit}.

The classical vacua were described by Vafa and Witten \cite{cvew}.
Assuming all masses are nonzero,
we may rescale the fields $\Phi_i$ so as to make all the masses
equal; having computed the vacua in this case one may
undo this rescaling.  In this case the $F$-term equations
for a supersymmetric vacuum read
\begin{equation}
[\Phi_i,\Phi_j] = -\frac{m}{\sqrt 2}\epsilon_{ijk} \Phi_k \ .
\label{fterm}
\end{equation}
Consider the case of $SU(N)$.  Recalling that the $\Phi_i$ are
$N\times N$ traceless matrices, it is evident that the solutions to
these equations are given by $N$-dimensional, generally reducible,
representations of the Lie algebra $SU(2)$.  The irreducible spin
$(N-1)/2$ representation is one solution;
$N$ copies
of the trivial representation give another ($\Phi_i=0$).  Since for every
positive integer $d$ there is one irreducible $SU(2)$ representation of
dimension $d$, each vacuum corresponds to a partition of $N$ into
positive integers:
\begin{equation}\label{partN}
\{k_d \in \ZZ\geq 0\}\ {\rm such\ that\ } \sum_{d=1}^N\ dk_d = N\ ,
\end{equation}
where $k_d$ is the number of times the dimension $d$ representation
appears.  The number of classical vacua of the
theory is given by the number of such partitions.

Generally, for a given partition, the unbroken gauge group is
$\left[\otimes_d U(k_d)\right]/ U(1)$.  For example, if $k_d=1$ and
$k_{N-d}=1$, then the $\Phi_i$ are block diagonal with blocks of
dimension $d$ and ${N-d}$; the diagonal traceless matrix which is
${\bf 1}$ in each block generates an unbroken $U(1)$ gauge symmetry.
Clearly we obtain $U(1)^{k-1}$ if there are $k$ such blocks.  However,
if $k_d=2$, then the two blocks of size $d$ can be rotated into each
other by additional generators, giving altogether an $SU(2)$ instead
of a $U(1)$.  More generally we obtain $SU(k_d)$.  Among these vacua
there is a unique one which we will call the `Higgs' vacuum, in
which the $SU(N)$ gauge group is completely broken.  This is the only
`massive vacuum' (meaning that it has a mass gap) at the classical
level. For each divisor $d<N$ of $N$ we may take $k_d = N/d$ with all
others zero, giving a vacuum with a simple unbroken gauge group
$SU(N/d)$.  All other vacua have one or more $U(1)$ factors; these are
`Coulomb vacua.'

Quantum mechanically, the story is even richer.  Donagi and Witten
\cite{rdew} found an integrable system which permitted them to
write the holomorphic curve and Seiberg-Witten form describing the
quantum mechanical moduli space of the \ntwo\ theory with $m_1=m_2$
and $m_3=0$.\footnote{It would be very interesting to find this
integrable system in the supergravity dual description of this
theory.}  They considered the effect of breaking the supersymmetry to
\none\ through nonzero $m_3\ll m_1,m_2$, and showed that the theory
has a number of remarkable properties.  Each classical vacuum
which has unbroken gauge symmetry $SU(k)$
splits into $k$ vacua, all of which have a mass
gap.  (Coulomb vacua with non-abelian group factors split as well,
although a complete accounting of these vacua was not given in
\cite{rdew}; since the photons remain massless, such vacua do not
have mass gaps.) The vacuum with $SU(N)$ unbroken
($k_1=N$, $\Phi_i=0$) splits into $N$ massive vacua, exactly the
number which would be needed in the \none\ Yang-Mills theory obtained
in the limit \Eref{nonelimit}. The massive quantum vacua are those
without
$U(1)$ factors, and as noted above are associated with the divisors of
$N$.  Their total number is obviously given by
the sum of the divisors of
$N$; it therefore depends in an interesting way, one which does not have
a large-$N$ limit, on the prime factors of $N$.  The number of
Coulomb vacua is exponential in $\sqrt{N}$.

Donagi and Witten showed the massive vacua were in a beautiful
one-to-one correspondence with the phases of gauge theories classified
by 't Hooft.  Let us review this classification \cite{tHclass}.
$SU(N)$ gauge theories with only adjoint matter can be probed by
sources which carry electric charges in the $\ZZ_N$ center of $SU(N)$
and magnetic charges in the $\ZZ_N = \pi_1[SU(N)/\ZZ_N]$ which
characterizes possible Dirac strings.  We may think of these charges
as lying in an $N\times N$ lattice, a $\ZZ_N\times \ZZ_N$ group $L$. 't
Hooft showed that the possible massive phases of $SU(N)$ gauge
theories are associated to the dimension-$N$ subgroups $P$ of $L$.  In
each phase, the charges corresponding to the $N$ elements of $P$ are
screened, and all others are confined; the flux tubes which do the
confining are represented by the elements of $L/P$.  For example, if
the ordinary Higgs mechanism creates a mass gap, all sources with
magnetic charge are confined; the only unconfined elements of $L$ are
the $(m,0)$, $m=0,\dots,N-1$.  Thus $P$ is generated by the single
element $(1,0)$.  Every magnetic flux tube carries a $\ZZ_N$ charge
$n=0,\dots, N-1$ and confines the sources with charge $(m,n)$ for any
$m$.  In an ordinary confining vacuum, the roles of $m$ and $n$ are
reversed, but otherwise the story is the same.  Vacua with oblique
confinement are given by groups $P$ generated by $(m,1)$, where
$m=0,\dots,N-1$.

More generally, however, the vacua are more complex.
As mentioned earlier, each classical vacuum with unbroken
$SU(k)$ symmetry splits into $k$ vacua.  These vacua correspond to
subgroups $P$ generated by $(k,0)$ and
$(s,d)$, where $dk=N$ and $s=0,1,\dots, k-1$.  This map of vacua to
subgroups is one-to-one and onto.  Note the Higgs vacuum is the case
$d=N$, while the $N$ vacua which survive in the pure \none\ Yang-Mills
theory are the cases $d=1$ for $ s=0,1,\dots ,N-1$, with $s=0$ being the
confining vacuum.

The action of \slz\ on the massive vacua is then straightforward
\cite{rdew}.  The $T$ transformation $\tau\to\tau + 1$ shifts each
element $(m,n)$ of the group $L$ to $(m+n\mod N,n)$; all electric
charges shift by their magnetic charge, through the Witten effect
\cite{ewtheta}.  The $S$ transformation $\tau\to -{1\over \tau}$
reverses electric and magnetic charges \cite{om}: $(m,n)\to (-n\mod
N,m)$.  Thus $S$ and $T$ map $L$ to itself, but act nontrivially on
its subgroups $P$.  This action then corresponds to a permutation of
the massive vacua.  In particular, note that the Higgs and confining
vacua are exchanged by $S$, while $T$ rotates the confining and
oblique confining vacua into each other while leaving the Higgs vacuum
unchanged. $S$ and $T$ then generate the entire \slz\ group and its
action on the vacua.  The Coulomb vacua have not been fully
classified, and the action of \slz\ on them has not yet been
understood.\footnote{In section~VI.C we will show that some of
the Coulomb vacua are transformed in a simple way by certain elements
of \slz.  However, we will not obtain the full story.}

We close the discussion of field theory by noting that this theory is
very different from \none\ Yang-Mills theory in certain respects.
(Recently, many of these qualitative points were emphasized in
\cite{doreykumar}.)  Although it is a four-dimensional theory, it
still has massive degrees of freedom (three Weyl fermions and six real
scalars in the adjoint representation) with masses of order $m$.
These massive states ensure that far above the scale $m$ (actually, as
we will see, above $mg_{{\rm YM}}^2N$ in the confining phase) the theory
becomes conformal, with gauge coupling $\tau$.  The important
$\ZZ_{2N}$ non-anomalous $R$-symmetry of the pure \none\ Yang-Mills
theory, a $\ZZ_2$ of which is unbroken and a $\ZZ_N$ of which permutes
the $N$ vacua of the theory, is broken explicitly by the presence of the
massive fields.  Consequently the confining and oblique confining vacua,
although still permuted by $\tau\to\tau + n$ with $n$ an integer, are
not related by a discrete $R$-symmetry and are not isomorphic.  In
particular their superpotentials have different magnitudes and the
domain walls between them have a variety of tensions
\cite{doreykumar}.  In the limit of \Eref{nonelimit}, for fixed $N$,
the strong coupling scale and the corresponding gluino condensate,
domain wall tension, and string tension are all much below the scale
$m$ of the masses, and so the strong dynamics is not affected by the
massive fields.  However, we want to study the gravity dual of this
theory, which requires large $g_{\rm YM}^2 N$.  In this limit $\Lambda
= m \exp({-8\pi^2/ g_{\rm YM}^2 N})$ is of order $m$, and so all of
the physics of the theory takes place near the scale $m$.  We will not
find the exponentially large hierarchy expected from dimensional
transmutation; this can only be seen at small $g_{\rm YM}^2 N$,
outside the supergravity regime.

\subsection{Brane Representations}

Consider the Higgs phase, in which
\begin{equation}
A_7 = - m L_1\ ,\quad
A_8 = - m L_2\ ,\quad
A_9 = - m L_3\ ,
\end{equation}
where $L_i$ is the $N$-dimensional irreducible representation of
$SU(2)$.  The scalars $A_m$ are the collective coordinates of the
D3-branes, normalized $x^m = 2\pi\alpha'A_m$ \cite{dbranes}.  These are
therefore noncommutative, but lie on a sphere of radius $r = \pi \alpha' m N$
\begin{equation}
x^m x^m = (2\pi \alpha' m)^2 L_i L_i \approx \pi^2 \alpha'^2 m^2 N^2\ .
\end{equation}
The nonzero commutator of the
collective coordinates corresponds to higher-dimensional brane charge,
a fact familiar from matrix theory.  Specifically \cite{ktsphere} the
D3-branes can be equivalently represented as a single D5-brane of
topology $\RR^4\times S^2$, the two-sphere having radius $r$, with $N$
units of world-volume magnetic field on the two-sphere.  The Higgs
vacuum of the four-dimensional theory is represented by this
D5-brane.

Similarly, a vacuum corresponding to the reducible representation $\{
k_d \}$, defined as in \Eref{partN}, corresponds to concentric
D5-branes, where $k_d$ have radius $\pi\alpha' md$ for each $d$.
Consider the case of two spheres, with $k_d = k_{N-d}=1$. If $d \neq
N-d$ then the spheres have different radii; the gauge group of the
field theory is $U(1)$.  However, if $d = N-d$, the two spheres
coincide and the field theory has gauge group $SU(2)$.  For $N-2d$
small, the $SU(2)$ is broken at a low scale and its W-bosons have mass
proportional to $N-2d$.  More generally, $k$ coincident D5-branes
correspond to a classical vacuum with $SU(k)$ symmetry.  Just as in
the case of flat branes with sixteen supercharges, the curved
D5-branes with four supercharges and only four-dimensional Lorentz
invariance show enhanced gauge symmetry when they coincide, and when
separated have W-bosons with masses of order the separation
distance.\footnote{The absence of an overall center-of-mass $U(1)$ in
the brane configuration, in parallel with the absence of a $U(1)$ in
the gauge theory, is not completely understood, although we will
comment on it in section~VI.H.}  Each classical vacuum of the theory
is given by a set of D5 branes of radius $n_i$, with $\sum_i n_i=N$.

Quantum mechanically the situation is much more complicated.  The $S$
transformation $\tau\to -{1\over \tau}$ should exchange the Higgs and
confining vacua; therefore by Type IIB duality the confining vacuum is
a single NS5-brane.  A $T^n$ transformation ($\tau\to\tau+n$) leaves
D5-branes unchanged and shifts an NS5-brane to a $(1,n)$ 5-brane.  It
follows that the $n^{\rm th}$ oblique confining vacuum is given by a
$(1,n)$ 5-brane.  The $S$-duality implies also that there should be
vacua with multiple NS5-branes, or generally $(1,n)$ 5-branes,
possibly coincident.  In fact we may expect there to be vacua in which
different types of 5-brane coexist.  For example, suppose we partition
$N$ using $k_s=1$ and $k_1= N-s$, so that the lower $(N-s)\times(
N-s)$ block of the fields $\Phi_i$ is zero, leaving $SU(N-s)$
unbroken.  In this case we would expect a D5-brane of radius $s$
representing the broken part of the gauge group, and an NS5-brane (or
a $(1,q)$ 5-brane) representing an (oblique) confining phase of the
unbroken $SU(N-s)$ subgroup.

We will show that all of these brane configurations do indeed appear
in the dual of the \noneplus\ theory.  This is a puzzle, however: the
number of brane configurations is much larger than the number of
phases.  For $N=pq$, for example, the vacuum with $k_p = q$ is
described by $q$ D5-branes of radii $\pi \alpha' m p$.  In
supergravity, this is clearly $S$-dual to the vacuum with $k_q = p$,
which is therefore described by $q$ NS5-branes.  However, from
investigation of the field theory \cite{rdew}, this vacuum also has
a description in terms of $p$ D5-branes.  We will see, in this and
other examples, that our solutions exist only in limited ranges of
parameter space, such that only one of the descriptions is valid at a
time.  Ideally, however, a more complete understanding
of how the theory resolves this puzzle would be desirable.

\section{Perturbations on $A{\lowercase{ d}}S_5 \times S^5$}

In this section we first review deformations of the
$AdS$/CFT duality with attention to the issue of singularities,
introduce the small parameter that makes the problem tractable, and
discuss the field theory perturbation and its supergravity dual. We
then give the IIB field equations, develop the necessary tensor spherical
harmonics, and solve the field equations to first order in
an expansion around $AdS_5 \times S^5$.

\subsection{$AdS$/CFT and its Deformations}

The $d=4$, \nfour\ Yang-Mills theory is dual to IIB string theory on
$AdS_5 \times S^5$ \cite{maldacon}.  The Yang-Mills coupling is related
to the string coupling by $g_{\rm YM}^2 = 4\pi g$, and the common radius
of the two factors of spacetime is $R = (4\pi gN \alpha'^2)^{1/4}$.
To each local operator ${\cal O}_i$ of dimension $\Delta_i$ in the CFT
corresponds two solutions of the linearized field equations
\cite{witads,GPK}, a nonnormalizable solution which scales as
$r^{\Delta_i - 4}$ with $r$ the $AdS$ radius, and a normalizable
solution which scales as $r^{-\Delta_i}$.  A supergravity solution which
behaves at large $r$ as
\begin{equation}
a_i r^{\Delta - 4} + b_i r^{-\Delta}   \label{linear}
\end{equation}
is dual to a field theory with Hamiltonian
\begin{equation}
H = H_{\rm CFT} + a_i {\cal O}_i\ ,
\end{equation}
and where the vacuum expectation value (vev) is \cite{bala}
\begin{equation}
\langle 0 | {\cal O}_i | 0 \rangle = b_i\ .
\end{equation}

We will be interested in relevant perturbations, those with $\Delta <
4$.  In the field theory these are unimportant in the UV, while in the
IR they become large and take the theory to a new fixed point or
produce a mass gap.  Correspondingly the perturbation \eref{linear}
is small at large $r$, but at small $r$ it becomes large and nonlinear
effects become important.

For a theory with a unique (or at least isolated) vacuum, the dynamics
should determine the vev once the Hamiltonian is specified.  This is
in accord with the general experience with second order differential
equations, where some condition of nonsingularity at small $r$ would
give one relation for each pair $a_i$ and $b_i$.

Now let us summarize what is known, with attention first to two special
cases that make sense:

{\bf 1.}  In the \nfour\ theory, $a_i = 0$, it is actually
possible to vary the particular $b$ that corresponds to $\cal O$ being
a scalar bilinear.  The point
is that the \nfour\ theory does not have an isolated vacuum, and varying
$b$ gives a state on the Coulomb branch.  It is important to note
that the supergravity solution is still singular, but that the
singularity is physically acceptable, corresponding to an extended
D3-brane source \cite{coul1,coul2,coul3,coul4}.

{\bf 2.}  Certain perturbations give a nontrivial fixed point in the
IR.  These correspond to supergravity solutions with $AdS$ behavior at
large and small $r$, with a domain wall interpolating
\cite{ir1,ir2,twoadj,karch,ir3}. The vacua do have moduli, but most or all
analyses have imposed symmetries which determine a unique vacuum and restrict
to a single pair $(a,b)$. In these cases the differential equation does
indeed determine $b$.  The condition of $AdS$ behavior in the IR gives a
boundary condition, which takes the form of an initial condition for damped
potential motion.

{\bf 3.}  More generally, for perturbations that produce a mass gap
and destroy the moduli space, the known solutions are singular for all
values of
$b_i$ \cite{gppz1,gppz}; for a recent discussion see \cite{gubs}.  It does not
make sense, however, that such singularities can all be understood as
physically acceptable brane or other sources, because that would mean that
the vevs are undetermined even though the vacua are isolated.  This is
another example of the important observation made by Horowitz and Myers in
the context of negative mass Schwarzschild \cite{garyrob}: string theory does
not repair all singularities; many singular spacetimes do not correspond to
any state in string theory.

We will show that the perturbations corresponding to the masses
\eref{Wmass} actually produce spacetimes with
extended brane sources.  The spacetime geometry is singular, but in a
way that is fixed by the source, and so in particular the values of
$b_i$ are determined.

This resembles the case {\bf 1} in that there are extended branes, and
could in principle be analyzed by supergravity means as in that case: for
some subset of the supergravity solutions the singularity will have an
acceptable physical interpretation as a brane source.  There has in
fact been a search for just such solutions \cite{priv1,gubs}; it has
thus far been unsuccessful, but some features of our solution have been
anticipated.  This approach is extremely difficult, and has
generally been restricted to special solutions with constant dilaton.  In
fact, the  branes in our solution couple to the dilaton, which is
therefore position-dependent.

We are able to treat these rather asymmetric geometries without facing
the full nonlinearity of supergravity because of the existence of a
small parameter.  Consider the case of a single D5-brane with D3-brane
charge $N$, wrapped on an equator of the $S^5$.  The area of the
two-sphere is of order $R^2$, so the density of D3-branes is
\begin{equation}
\frac{N}{R^2} \sim \frac{N^{1/2}}{g^{1/2} \alpha'}\ . \label{densest}
\end{equation}
Under the rather weak condition $N/g \gg 1$, this is large in string
units and the effect of the D3-brane charge dominates that of the
D5-charge charge.\footnote{This estimate~\eref{densest} ignores the
warping of the geometry by the expanded brane, but should be correct
in order of magnitude almost everywhere.  In fact, very close to the
surface of the two-sphere the effect of the D5-brane dominates.
However, in this regime we can match onto the exact solution for a
flat D5-brane with D3-brane charge, as we develop further in
section~V.D.}  The system is therefore well approximated by a Coulomb
branch configuration of the parent \nfour\ theory, where the general
solution is given by linear superposition in the harmonic function.
Thus we can work by treating the D5-brane charge, and the 3-form field
strengths that are generated by it, as perturbations.  It is less
obvious, but will be seen in section~IV.A, that the full 3-form field
strength is effectively proportional to the same small parameter.

For the NS5 solution the corresponding condition is given by $g \to
1/g$ and so $Ng \gg 1$.  This is precisely the condition for the gauge
theory to be strongly coupled.  We then recognize the earlier condition
$N/g \gg 1$ as the condition for the dual gauge theory to be strongly
coupled.  When both of these conditions are satisfied the supergravity
description is valid, so the D5 and NS5 solutions are both valid in the
entire supergravity regime.

We will begin with a simpler problem, where we place a probe
D5-brane of D3-brane charge $n$ into the linearized perturbation of the
$AdS_5\times S^5$ background.  In this case the condition for the
D5-brane solution to be valid is similarly
\begin{equation}
\frac{n^2}{gN} \gg 1\ .\label{condit}
\end{equation}
We will use this condition at several points.
In section V.B we will infer that
this condition is not just a convenience but in fact a
necessity in order for the solution to exist.

\subsection{Field Equations and Background}

The IIB field equations can be derived from the Einstein frame action
\cite{iibact}
\begin{eqnarray}
&&\frac{1}{2\kappa^2} \int d^{10}x (-G)^{1/2} R - \frac{1}{4\kappa^2}
\int \biggl( d\Phi \wedge *d\Phi + e^{2\Phi} dC \wedge *dC +
\nonumber\\
&&\qquad g e^{-\Phi} H_{\it 3} \wedge * H_{\it 3} +
g e^{\Phi} \tilde F_{\it 3} \wedge * \tilde F_{\it 3}
+ \frac{g^2}{2} \tilde F_{\it 5} \wedge * \tilde F_{\it 5}
+ g^2 C_{\it 4} \wedge H_{\it 3} \wedge F_{\it 3}\biggr) \ ,\qquad
\end{eqnarray}
supplemented by the self-duality condition
\begin{equation}
* \tilde F_{\it 5} = \tilde F_{\it 5}\ .
\end{equation}
Here
\begin{eqnarray}
\tilde F_{\it 3}\* F_{\it 3} - C H_{\it 3}\ ,\quad F_{\it 3}
= d C_{\it 2}\ ,
\nonumber\\
\tilde F_{\it 5}\* F_{\it 5} - C_{\it 2} \wedge H_{\it 3}\ ,\quad F_{\it 5}
= d C_{\it 4}\ .
\label{fstrens}
\end{eqnarray}
We define the Einstein metric
by $(G_{\mu\nu})_{\rm Einstein} = g^{1/2}e^{-\Phi/2}
(G_{\mu\nu})_{\rm string}$, so that it is equal to the string metric in this
constant background.  As a result $g$ appears in the action, explicitly
and also through $2\kappa^2 = (2\pi)^7 \alpha'^4 g^2$.

The field equations are \cite{schwarz}
\begin{eqnarray}
\nabla^2 \Phi \* e^{2\Phi} \partial_M C \partial^M C
-\frac{ge^{-\Phi}}{12} H_{MNP} H^{MNP} +\frac{ge^{\Phi} }{12}
 \tilde F_{MNP} \tilde F^{MNP} \ ,
\nonumber\\
\nabla^M( e^{2\Phi} \partial_M C) \* - \frac{ge^{\Phi} }{6}
H_{MNP} \tilde F^{MNP} \ ,
\nonumber\\
d{*}(e^{\Phi} \tilde F_{\it 3}) \* gF_{\it 5} \wedge H_{\it 3} \ ,
\nonumber\\
d{*}(e^{-\Phi} H_{\it 3} - C e^{\Phi} \tilde F_{\it 3}) \*
- g F_{\it 5}
\wedge F_{\it 3} \ ,
\nonumber\\
d{*}\tilde F_{\it 5} \* - F_{\it 3} \wedge H_{\it 3}\ ,
\nonumber\\
 R_{MN} \* \frac{1}{2} \partial_M \Phi \partial_N \Phi +
\frac{e^{2\Phi}}{2} \partial_M C \partial_N C + \frac{g^2}{96}
\tilde F_{MPQRS} \tilde F_N{}^{PQRS}
\nonumber\\
&&\qquad +\frac{g}{4}(e^{-\Phi} H_{MPQ} H_N{}^{PQ} +
e^{\Phi} \tilde F_{MPQ} \tilde F_N{}^{PQ}) \nonumber\\
&&\qquad
- \frac{g}{48} G_{MN} (e^{-\Phi} H_{PQR} H^{PQR} +
 e^{\Phi} \tilde F_{PQR} \tilde F^{PQR})\ . \label{feldeq}
\end{eqnarray}
We use indices $M,N,\ldots$ in ten dimensions.
The Bianchi identities are
\begin{eqnarray}
d\tilde F_{\it 3}\* - dC \wedge H_{\it 3}
\nonumber\\
d\tilde F_{\it 5}\* - F_{\it 3} \wedge H_{\it 3}\ .
\end{eqnarray}

One class of solutions is
\begin{eqnarray}
ds^2 \* ds^2_{\rm string} =
Z^{-1/2} \eta_{\mu\nu} dx^\mu dx^\nu + Z^{1/2} dx^m
dx^m\ ,\nonumber\\
\tilde F_{\it 5} \* d\chi_{\it 4} + * d\chi_{\it 4}\ ,\quad
\chi_{\it 4} = \frac{1}{gZ}  dx^0 \wedge dx^1
\wedge dx^2
\wedge  dx^3\ , \nonumber\\
e^\Phi \* g \ ,\quad C = \frac{\theta}{2\pi}\ ,\label{backg}
\end{eqnarray}
with $g$ and $\theta$ constant and other fields vanishing.
Here $\mu,\nu = 0,1,2,3$, and $m,n = 4,\ldots,9$.  Also, $Z$ is any
harmonic function of the $x^m$, $\partial_m \partial_m Z = 0$.  For $AdS_5
\times S^5$,
\begin{equation}
Z = \frac{R^4}{r^4}\ ,\quad R^4 = 4\pi g N\alpha'^2\ .
\end{equation}
This fails to be harmonic at the origin, but this is a horizon,
dual to a D3-brane source at the origin.  More generally a nonharmonic $Z$
corresponds to a distributed D3-brane source.

We will need to expand the field equations around this solution.  The
equations for linearized $F_{\it 3}$ and $H_{\it 3}$ perturbations are
conveniently written in terms of
\begin{equation}
G_{\it 3} = F_{\it 3} - \hat\tau H_{\it 3}\ .
\end{equation}
Here
\begin{equation}
\tau = C + i e^{-\Phi}\ ,
\end{equation}
and a\ $\hat{}$\ denotes unperturbed fields, so that
\begin{equation}
\hat\tau = \frac{\theta}{2\pi} + \frac{i}{g}\ .
\end{equation}
The linearized field and Bianchi equations in a general background are
\begin{eqnarray}
d \hat{*} G_{\it 3} + i G_{\it 3} \wedge \hat{\tilde F}_{\it 5} \* 0\ ,
\nonumber\\
d G_{\it 3} \* 0\ .
\end{eqnarray}
We will only be interested in the transverse
($mnp$) components of $G_{\it 3}$.
For the background~(\ref{backg}) and a transverse 3-form field,
\begin{equation}
\hat{*} G_{\it 3} =  Z^{-1} {*_6} G_{\it 3} \wedge
dx^0 \wedge dx^1 \wedge dx^2 \wedge  dx^3 \ .
\end{equation}
where the dual $*_6$ acts in the six-dimensional transverse space
with respect
to the flat metric $\delta_{mn}$.
Then, in the solution~(\ref{backg}) with general $Z$, the
field equation for a transverse 3-form field can be written simply as
\begin{equation}
d [ Z^{-1} ( *_6 G_{\it 3} - i  G_{\it 3})] = 0\ . \label{field}
\end{equation}

The duality
of the field strengths implies that the 7-form field strength is
\begin{equation}
- {*} \tilde F_{\it 3}
= dC_{\it 6} - H_{\it 3} \wedge C_{\it 4}\ . \label{f7}
\end{equation}
This is parallel in form to the other field strengths~\eref{fstrens}.
The relative sign of the two terms on the right can be deduced by
noting that the D5-brane action, which we will write in section~IV.A,
and the field strength are both invariant under $\delta C_{\it 4} =
d\chi_{\it 3}$ provided that $\delta C_{\it 6} = - H_{\it 3} \wedge
\chi_{\it 3}$.  The relative sign of the two sides is obtained by
acting with $d$ and comparing with the field equation~\eref{feldeq}.

For the 6-form we write
\begin{equation}
d(B_{\it 6} - \hat\tau C_{\it 6}) = \frac{i}{g} {*}G_{\it 3} + C_{\it 4}
\wedge G_{\it 3}\ .\label{a6a}
\end{equation}
The imaginary part of this equation is just~\Eref{f7}, while the real part
defines $B_{\it 6}$; the meaning of $B_{\it 6}$ will become clear in the
section~IV.B.  For the background~\eref{backg} this becomes
\begin{equation}
d(B_{\it 6} - \hat\tau C_{\it 6}) = \frac{i}{gZ} ({*_6}G_{\it 3} - i
G_{\it 3}) \wedge dx^0 \wedge dx^1 \wedge dx^2 \wedge dx^3
\ .\label{a6b}
\end{equation}

\subsection{Fermion Masses and Tensor Spherical Harmonics}

The \nfour\ theory has Weyl fermions $\lambda_\alpha$ transforming as a {\bf
4} of the
$SO(6)$ $R$-symmetry.  We will add a mass term
\begin{equation}
m^{\alpha\beta} \lambda_\alpha \lambda_\beta + {\rm h.c.}
\end{equation}
(spinor indices suppressed), which we can assume to be diagonal,
$m^{\alpha\beta} = m_\alpha \delta^{\alpha\beta}$.
When one of the masses, say $m_4$, vanishes, the Hamiltonian has an
\none\ supersymmetric completion, as given by the
superpotential~\eref{supmass}. The fermion $\lambda_4$ is then the
gluino.\footnote{Even when all four masses are nonvanishing, this operator
is still chiral and has a supersymmetric completion to linear order in
$m$.  However, the Hamiltonian at order $m^2$ is nonsupersymmetric.  This
case will be discussed in section~VII.B.}

The fermion bilinear transforms as the $({\bf 4} \times {\bf 4})_{\rm
sym} = {\bf 10}$ of
$SO(6)$, and the mass matrix as the $\OL{\bf 10}$.
The $\bf 10$ and $\OL{\bf 10}$ are imaginary-self-dual
antisymmetric 3-tensors,
\begin{equation}
*_6 T_{mnp} \equiv \frac{1}{3!} \epsilon_{mnp}{}^{qrs} T_{qrs} = \pm i
T_{mnp}
\ ,\label{dual}
\end{equation}
with $+$ for the ${\bf 10}$ and $-$ for the $\OL{\bf  10}$.
The indices again run from 4 to 9.

To relate the fermion mass
to a tensor,
it is convenient to adopt complex coordinates $z^i$:
\begin{equation}
z^1 = \frac{x^4 + i x^7}{\sqrt 2}\ ,\quad z^2 = \frac{x^5 + i x^8}{\sqrt 2}\
,\quad z^3 = \frac{x^6 + i x^9}{\sqrt 2}\ .
\end{equation}
Under a rotation $z^i \to e^{i\phi_i} z^i$
the spinors in the {\bf 4} transform
\begin{eqnarray}
\lambda_1 \xto e^{i(\phi_1 - \phi_2 - \phi_3)/2} \lambda_1\ , \nonumber\\
\lambda_2 \xto e^{i(-\phi_1 + \phi_2 - \phi_3)/2} \lambda_2\ , \nonumber\\
\lambda_3 \xto e^{i(-\phi_1 - \phi_2 + \phi_3)/2} \lambda_3\ , \nonumber\\
\lambda_4 \xto e^{i(\phi_1 + \phi_2 + \phi_3)/2} \lambda_4\ .
\end{eqnarray}
~From this it follows that a diagonal mass term transforms in the same way
as the form
\begin{equation}
T_{\it 3} = m_{1} dz^1 \wedge d\bar z^2 \wedge d\bar z^3 +
m_{2} d\bar z^1 \wedge dz^2 \wedge d\bar z^3 +
m_{3} d\bar z^1 \wedge d\bar z^2 \wedge dz^3 +
m_{4} dz^1 \wedge dz^2 \wedge dz^3 \ . \label{tenso}
\end{equation}
In \none\ language, $m_{4}$ is a gluino mass and the other $m_{\alpha}$ are
chiral superfield masses.
In the supersymmetric case the nonzero components are
\begin{equation}
T_{1\bar2\bar3} = m_1\ ,\quad
T_{\bar1 2\bar3} = m_2\ ,\quad
T_{\bar1\bar2 3} = m_3 \label{susyt}
\end{equation}
and permutations, and in the equal-mass case
\begin{equation}
T_{\bar \imath\bar \jmath k} =T_{i\bar \jmath \bar k} =
T_{\bar \imath j\bar k} =  m
\epsilon_{ijk} \ . \label{tensor}
\end{equation}
These satisfy $*_6 T = -iT$.

One might guess, correctly, that the fermion mass is associated with the
lowest spherical harmonic of the field $G_{\it 3}$ \cite{witads,KRv,gunmar}.
To make a 3-tensor field transforming in the same way as any given tensor $T$,
we can use the constant $T$ itself, or combine it with the radius vector to
form
\begin{equation}
V_{mnp} = \frac{x^q}{r^2} ( x^m T_{qnp} + x^n T_{mqp} + x^pT_{mnq})
\end{equation}
where $r^2 = x^m x^m$.
Define the forms
\begin{eqnarray}
T_{\it 3} \* \frac{1}{3!}  T_{mnp} dx^m\wedge dx^n\wedge dx^p \ ,\quad
V_{\it 3} = \frac{1}{3!}  V_{mnp} dx^m\wedge dx^n\wedge dx^p \ ,
\nonumber\\
S_{\it 2} \* \frac{1}{2} T_{mnp} x^m dx^n\wedge dx^p\ .
\end{eqnarray}
One then finds
\begin{eqnarray}
dS_{\it 2} \* 3T_{\it 3}\ ,\quad d(\ln r)\wedge S_{\it 2} = V_{\it 3}\ ,
\quad
d(r^p S_{\it 2}) = r^p(3 T_{\it 3} + p V_{\it 3})\ ,\nonumber\\
dT_{\it 3} \* 0\ ,\quad dV_{\it 3} = -3 d(\ln r) \wedge T_{\it 3}\ ,
\end{eqnarray}
and
\begin{equation}
{*_6} T_{\it 3} = \pm i T_{\it 3}\ ,\quad
{*_6} V_{\it 3} = \pm i (T_{\it 3}-V_{\it 3})\ . \label{uvdual}
\end{equation}

\subsection{Linearized Solutions}

We specialize to perturbations on the $AdS_5 \times S^5$ case $Z = R^4
/r^4$, which is invariant under the transverse $SO(6)$.  This will be
applicable to the probe calculation of the next section.
A general form for the perturbation is
\begin{equation}
G_{\it 3} = r^p (\alpha T_{\it 3} +
\beta V_{\it 3} )\ ,
\end{equation}
where for now we take $T$ to be an arbitrary constant tensor in the {\bf 10}
or $\OL{\bf 10}$.
The Bianchi identity gives
\begin{equation}
0 = dG_{\it 3} = (p\alpha - 3\beta) d(\ln r) \wedge S_{\it 2}
\ \Rightarrow\ \beta = p\alpha/3\ ,
\end{equation}
corresponding to
\begin{equation}
G_{\it 3} =  (\alpha/3) d ( r^p S_{\it 2})\ .
\end{equation}
Using the duality properties~\eref{uvdual} we then have
\begin{equation}
{*_6} G_{\it 3} - i G_{\it 3}=
-ir^p (\alpha/3)
[ (3 \mp p \mp 3) T_{\it 3} + (p\pm p) V_{\it 3} ]\ ,
\end{equation}
and so the equation of motion~\eref{field} gives
\begin{equation}
p^2 -10 p + (12 \mp 12) = 0\ .
\end{equation}

For the lower sign, the $\OL{\bf 10}$, there are two solutions:
\begin{eqnarray}
p \* -4\ ,\quad G_{\it 3} = \alpha r^{-4} (T_{\it 3} - 4V_{\it 3}/3)
\ ,\nonumber\\
p \* -6\ ,\quad G_{\it 3} = \alpha r^{-6} (T_{\it 3} - 2V_{\it 3})\ .
\label{perts}
\end{eqnarray}
In interpreting these, note that a factor $Z^{-3/4} = (r/R)^3$ must be
included to translate the tensors to an inertial frame.  These solutions then
have the falloffs appropriate to the nonnormalizable and normalizable solutions
for a operator of $\Delta=3$.  The former thus corresponds to the perturbation
of $m$, and the latter to the vev of $\bar\lambda\bar\lambda$.
The mass perturbation therefore corresponds at first order to
\begin{equation}
G_{\it 3} = \frac{\zeta}{g}
\biggl( \frac{R}{r} \biggr)^4 (T_{\it 3} - 4V_{\it 3}/3)
= d\Biggl[ \frac{\zeta}{3g}
\biggl( \frac{R}{r} \biggr)^4 S_{\it 2} \Biggr]
\label{first}
\end{equation}
with $T_{\it 3}$ given in \Eref{tenso}.
 The factors of $R$ are necessary for
the dimensions, and the factor of $g^{-1}$ arises from the overall
$g_{\rm YM}^{-2}$ in the superpotential.  The numerical coefficient $\zeta$
appearing in the relation between the fermion bilinear and the
supergravity field will eventually be determined to take the value $\zeta
= -3\sqrt 2$.  Note also that as a consequence of the equation of
motion~\eref{field},
\begin{equation}
Z^{-1} ({*_6} G_{\it 3} - i G_{\it 3}) = \frac{2i\zeta}{3g} T_{\it 3}
= \frac{2i\zeta}{9g} d S_{\it 2}
\label{dufirst}
\end{equation}
is exact.

For fields in the
${\bf 10}$, the upper sign, there are again two solutions:
\begin{eqnarray}
p \* 0\ ,\quad G_{\it 3} = \alpha  T_{\it 3}
\ ,\nonumber\\
p \* -10\ ,\quad G_{\it 3} = \alpha
r^{-10} (T_{\it 3}- 10 V_{\it 3}/3)\ .
\end{eqnarray}
The first of these corresponds to the coefficient of  $\bar\lambda\bar \lambda
F^2$, and the second to the vev of $\lambda \lambda F^2$.

\section{Five-brane Probes}

In this section we consider probes in the background given by $AdS_5
\times S^5$ plus the linear $G_{\it 3}$ perturbation.  The probes are
5-branes with world-volume $\RR^4 \times S^2$ and D3-brane charge $n
\ll N$, with $n\gg \sqrt{gN}$.  We consider first D5-brane probes, and
then use \slz\ duality to extend to a general $(c,d)$ 5-brane.  For
all such probes we find that there is a supersymmetric minimum at
nonzero $AdS$ radius $r$.

\subsection{The D5 Probe Action}

The relevant terms in the action for a D5-brane are
\cite{lido,dbranes,bergd}
\begin{equation}
S = -\frac{\mu_5}{g}
\int d^6\xi\, \Bigl[-\det (G_\parallel)
\det(g^{-1/2}e^{\Phi/2}G_\perp + 2\pi\alpha'{\cal F})\Bigr]^{1/2}
+ \mu_5\int ( C_{\it 6} + 2\pi\alpha'{\cal F}_{\it 2} \wedge
C_{\it 4} )
\ , \label{d5act}
\end{equation}
where
\begin{equation}
2\pi\alpha'{\cal F}_{\it 2} = 2\pi\alpha'F_{\it 2} - B_{\it
2}\ .
\end{equation}
Here $G_\parallel$ is the metric in the $\RR^4$
directions of the world-volume and $G_\perp$ is the metric
in the
$S^2$ directions, pulled back from spacetime.  It is
convenient to note that $\det G_\parallel=Z^{-2}$ and that
$\det{\cal F} = \half {\cal
F}_{ab}{\cal F}^{ab}\det G_\perp$.

The D3-brane charge of the probe is $n$, so that
\begin{equation}
\int_{S^2} F_{\it 2} = 2\pi n\ . \label{quant}
\end{equation}
This is assumed in this section to be small compared to $N$ so that
the effect of the probe on the background can be ignored.  If the
internal directions are a sphere, rotational symmetry and the
quantization~(\ref{quant}) give $F_{\theta\phi} = \frac{1}{2} n
\sin\theta$, or $F_{ab} F^{ab} = n^2/2Zr^4$.

Let us first consider the action in the absence of the $G_{\it 3}$ 
background so in particular ${\cal F}_{ab} = F_{ab}$.  The first term in the
Born-Infeld action is dominated by the second, since for $Z=R^4/r^4$
\begin{equation}
 4 \pi^2 \alpha'^2 F_{ab} F^{ab} =
 2 \pi^2 \alpha'^2 n^2/2R^4
\sim n^2/gN \gg 1
\end{equation}
That the field strength dominates reflects the physical input that the
D3-brane charge dominates.  It is then useful to write
\begin{eqnarray}
\sqrt{\det(G_\perp + 2\pi\alpha'{ F})}
\*
 2\pi\alpha'\sqrt{\det{ F}}
\left[1 + {1\over(2\pi\alpha')^2 { F}_{ab}{ F}^{ab}}\right]
\nonumber\\
\* 2\pi\alpha'\sqrt{\det{ F}} + 
{\det G_\perp \over4\pi\alpha'\sqrt{\det { F}}}\ .
\end{eqnarray}
If the D5-brane
is a sphere in the $x_\perp$ directions, then in spherical coordinates
$\det G_\perp = Zr^4\sin^2 \theta$.
Since a D3-brane probe feels no
force from D3-branes, there is a large cancellation between the Born-Infeld
and Chern-Simons terms.  The leading nonvanishing term in
the D5-action gives a potential density of
the form
\begin{equation}
\frac{\mu_5}{g} \int_{S^2} d^2 \xi \,
{\sqrt{\det G_\parallel} \det G_\perp \over
 4\pi \alpha' \sqrt{\det { F}}}
 = \frac{\mu_5}{g} \int_{S^2} d\cos\theta\, d\phi\,{r^4\over
2\pi \alpha' n} =\frac{\mu_5}{g} \frac{2r^4}{n\alpha'}\ ,
\label{quarticnew}
\end{equation}
where in the last two equations we have assumed the 5-brane is
a two-sphere in the $x_\perp$ directions.  
Notice the $Z$ factors cancel explicitly; if the metric takes the form
in \Eref{backg}, the energy density of the 5-brane goes as $r^4$.  This is
consistent with the fact that the D3-branes see this energy as coming
from the square of a commutator term, $([\Phi,\Phi^\dagger])^2$.

Now let us add the perturbation back in. 
For the linear perturbation, \Eref{first} immediately gives the
potentials (up to an irrelevant gauge choice) as
\begin{equation}
C_{\it 2} - \hat\tau B_{\it 2} = \frac{\zeta}{3g} \biggl( \frac{R}{r}
\biggr)^4 S_{\it 2} \ . \label{potent}
\end{equation}
For the 6-form,
Eqs.~\eref{dufirst} and \eref{a6b} then give
\begin{equation}
C_{\it 6} = \frac{2\zeta}{9g}
 dx^0 \wedge dx^1 \wedge dx^2 \wedge  dx^3 \wedge {\rm Im}( S_{\it 2})
\label{dupotent}
\ ,
\end{equation}
up to gauge choice.

The effect of $B_2$ in the D5-brane action is subleading and can be
ignored. Using the flux~\eref{quant} and the potential~\eref{potent},
one finds the ratio of the two terms in ${\cal F}_{ab}$ is
\begin{equation}
{B_{ab}}/{2\pi\alpha' F_{ab}} \sim \frac{mR^4}{r^3} \biggl/
\frac{\alpha' n}{r^2} \sim \frac{mgN\alpha'}{nr}\ . \label{terms}
\end{equation}
Looking ahead, the minimum of interest is located at
\begin{equation}
r \sim m n \alpha' \ ,\label{rapp}
\end{equation}
and so the ratio~\eref{terms} becomes $gN / n^2$ which is just the
small parameter.  Thus, at the $AdS$ radii~\eref{rapp} or greater, the
field strength term in ${\cal F}_{ab}$ dominates: ${\cal F}_{ab}
\approx { F}_{ab}$.  The cancellation between the Born-Infeld
and Chern-Simons terms is unaffected; $B$ need merely be inserted in
\Eref{quarticnew}, where it is negligible. 

Inserting the perturbed $C_{\it 6}$ from \Eref{dupotent} into the D5-brane
action gives an additional potential density
\begin{equation}
-\frac{\Delta S}{V} = -\frac{\mu_5}{g} 
\int_{S^2} \frac{2\zeta}{9}
{\rm Im}( S_{\it 2})
\ .\label{cubicnew}
\end{equation}
which is cubic in $r$, linear in $m$, and independent of $Z$.

The two terms in Eqs.~\eref{quarticnew} and \eref{cubicnew} can be
identified with the quartic $\phi^4$ and cubic $m\phi^3$ terms in the
\none\ supersymmetric potential, as we will see in more detail in
section~IV.C.  For consistency we must also keep the term of order
$m^2\phi^2$.  This arises from the second-order perturbations of the
dilaton, metric, and four-form potential.  In fact, supersymmetry
makes it possible to write the second-order term in the potential
directly:
\begin{eqnarray}
-\frac{S}{V} \* \frac{\mu_5}{g}\Biggl\{ \int_{S^2} d^2 \xi \,
{\sqrt{\det G_\parallel} \det G_\perp \over
 4\pi \alpha' \sqrt{\det F}}
- \frac{\zeta}{9}\int_{S^2}
{\rm Im}( T_{mnp} x^m dx^n\wedge dx^p)
\nonumber\\
&&\qquad\qquad\qquad\qquad\qquad\qquad
+ \frac{\pi\alpha'\zeta^2}{18}
T_{i\bar \jmath \bar k}\OL{T}_{\bar l j k} \int_{S^2} F_{\it 2} z^i \bar
z^{\bar l}
\Biggr\}
\ . \label{pot2}
\end{eqnarray}
The form of this term is readily understood.  The integral $\int_{S^2} F_{\it
2}$ essentially sums over D3-branes, while the tensor structure gives the
\none\ scalar mass $\sum_i |m_i|^2 |\phi_i|^2$.  The coefficient will be
deduced in section~IV.C.

Before we go on, let us address two puzzles.  The first
is the expansion around $AdS_5 \times S^5$, and why we need to keep
terms precisely through second order.
A measure of the square of the size of the perturbation is the ratio of the
energy density in the perturbation $|F_{\it 3}|^2$ with that in the unperturbed
$|F_{\it 5}|^2$:
\begin{equation}
{|F_{\it 3}|^2}/{|F_{\it 5}|^2} \sim
\frac{m^2 R^2}{g^2 r^2} \biggl/\frac{1}{ g^2 Z^{1/2} r^2}
\sim \frac{m^2 gN\alpha'^2}{r^2} \sim \frac{gN}{n^2}\ , \label{enrat}
\end{equation}
which is the controlling small parameter, basically the effective
ratio of brane charge densities $\sigma_5^2/\sigma_3^2$.  The three terms in
the potential~\eref{pot2} are respectively of zeroth, first, and
second order in the perturbation.  The zeroth order term is the remainder
after cancellation between the Born-Infeld and Chern-Simons terms,
and, since the D5 and D3 tensions add in quadratures, is of order
\begin{equation}
\sqrt{\sigma_5^2 + \sigma_3^2} - \sigma_3 \sim \frac{\sigma_5^2}{\sigma_3}\
.
\label{magnit}
\end{equation}
The linear perturbation is of order $\sigma_5/\sigma_3$ and
couples to $\sigma_5$, so the first order term is again of
magnitude~\eref{magnit}.  The second order perturbation is felt by the
D3-branes and so this term is of order $\sigma_3
(\sigma_5/\sigma_3)^2$, again the same.  Note that this analysis does not
use supersymmetry, and so will apply to the \nzero\ case as well.

The second puzzle is that the second order term in the
potential~\eref{pot2} makes reference to complex coordinates in
spacetime, and these are not intrinsic.  In particular, when all four
fermion masses are nonvanishing (\nzero) there is no special complex
structure.  The point\footnote
{See also section~5 of ref.~\cite{ir3}.}
is that the supergravity equations have {\it
homogeneous} second order solutions, corresponding to the traceless
scalar bilinear $A_m A_n - \frac{1}{6} \delta_{mn} A_pA_p$.  The
coefficients of these solutions are determined by boundary conditions,
so the inhomogeneous solution with $(G_{\it 3})^2$ as source determines
only the trace part $A_m A_m$.  Thus, the general form for the second order
term, not imposing \none\ supersymmetry, is given by replacing
\begin{equation}
T_{i\bar \jmath \bar k}\OL{T}_{\bar l j k} z^i z^{\bar l} \to
T_{mnp} \OL{T}_{mnp} \frac{r^2}{18} + \mu_{mn} x^m x^n
\label{ambig}
\end{equation}
with arbitrary traceless $\mu_{mn}$.  Note that both $T_{mnp}$ and $\mu_{mn}$
are intrinsic (determined by the boundary conditions).

\subsection{The $(c,d)$ Probe Action}

A given background can also be given in an $S$-dual description,
\begin{equation}
\tau' = \frac{a \tau + b}{c\tau + d}\ .
\end{equation}
Specifically,
\begin{eqnarray}
g' \* g |M|^2\ ,\quad G_{MN}' = G^{\vphantom{\prime}}_{MN} |M|\ ,
\quad C_{\it 4}' = C^{\vphantom{\prime}}_{\it 4}\ ,\nonumber\\
G_{\it 3}' \* G^{\vphantom{\prime}}_{\it 3} M^{-1}\ ,
\quad B_{\it 6}' -\hat\tau' C_{\it 6}' = (B_{\it 6} -\hat\tau C_{\it 6})
M^{-1}\ ,
\label{doof}
\end{eqnarray}
where $M = c\tau + d$.
A D5-brane in the primed description has the action
\begin{equation}
-{S} = \mu_5\int d^4x \Biggl\{ \frac{1}{g'}\int_{S^2} d^2 \xi \,
{\sqrt{\det G'_\parallel} \det G'_\perp \over
 4\pi \alpha' \sqrt{\det F}} 
- \int_{S^2}  C'_{\it 6} + O(T^2)  
\Biggr\}
\ .
\end{equation}
%\begin{equation}
%-\frac{S}{V'} = \frac{\mu_5}{g'}\Biggl\{  \int_{S^2} d^2 \xi \,
%{\sqrt{\det G'_\parallel} \det G'_\perp \over
% 4\pi \alpha' \sqrt{\det F}}
%- \frac{\zeta}{9}\int_{S^2} {\rm Im}( T'_{mnp} x^m dx^n\wedge dx^p)
%+ O(T^2)
%\Biggr\}
%\ .
%\end{equation}
Under the duality~(\ref{doof}), this translates into
\begin{eqnarray}
-\frac{S}{V} \* \frac{\mu_5}{g}\Biggl\{  |M|^2 \int_{S^2} d^2 \xi \,
{\sqrt{\det G_\parallel} \det G_\perp \over
 4\pi \alpha' \sqrt{\det F}}
- \frac{\zeta}{9} \int_{S^2} {\rm Im}(\OL{M} T_{mnp} x^m dx^n\wedge dx^p)
 \nonumber\\ &&\qquad\qquad\qquad\qquad
+  \frac{\pi\alpha'\zeta^2}{18}
T_{i\bar \jmath \bar k}\OL{T}_{\bar l j k} \int F_{\it 2} z^i \bar
z^{\bar l}
\Biggr\}
\ .\label{cdact}
\end{eqnarray}

The probe couples to
\begin{equation}
C'_{\it 6} = -g' {\rm Im}(B'_{\it 6} - \hat\tau C'_{\it 6})
= -g{\rm Im}(\OL{M} [B_{\it 6} - \hat\tau C_{\it 6}])
= B_{\it 6} c + C_{\it 6} d\ . \label{bccd}
\end{equation}
This is the coupling of a $(c,d)$ 5-brane, a bound state of $c$
NS5-branes and $d$ D5-branes.  In the first term of the potential, the
factor $|c\tau + d|^2$ is the tension-squared of the $(c,d)$ 5-brane,
squared from the addition in quadratures in the Born-Infeld term.  The
second is the coupling to the background~\eref{bccd}.  The final term
has again been added by hand in the form required by supersymmetry,
which is in fact independent of $(c,d)$.  This is because it is the
interaction of the D3-brane charge with the second-order background,
and so does not depend on the 5-brane quantum numbers. The duality
transformation only gives relatively prime $(c,d)$, but the result
holds generally, by superposition.

\subsection{The Probe Potential and Minima}

We now focus on the $SO(3)$-invariant \none\ equal-mass case.  The
general $SO(3)$-invariant brane configuration is
\begin{equation}
z^i = z e^i\ ,\quad e^i = \OL{e^{i}}\ ,\quad e^i e^i = 1\ .
\label{config}
\end{equation}
This is a sphere of coordinate radius $|z|/\sqrt{2}$, obtained from the
sphere
$(x^4)^2 +  (x^5)^2 + (x^6)^2 = \frac{1}{2}|z|^2$ by a simultaneous phase
rotation of the
$z^i$. Rotational symmetry and the quantization~(\ref{quant}) give
$F_{\theta\phi} = \frac{1}{2} n \sin\theta$, or $F_{ab} F^{ab} =
n^2/2Zr^4$. Inserting
this configuration into the action~(\ref{cdact}) gives
\begin{eqnarray}
-\frac{S}{V} \*  \frac{\mu_5}{g} \Biggl[
 \frac{8}{\alpha' n} |M|^2 |z|^4 + \frac{8\pi\zeta}{3} {\rm Im}(\OL M\bar
z^2 mz) + \frac{2\pi^2 n \alpha' \zeta^2}{9} |m|^2|z|^2
\Biggr] \nonumber\\
\* \frac{4}{\pi g n} |M \phi^2 + i \zeta m n \phi /12 |^2\ .
\label{finpot}
\end{eqnarray}
Here $\phi = z/2\pi\alpha'$ is the normalization of the gauge theory scalar
relative to the D3-brane collective coordinate.  This is of the form required
by \none\ supersymmetry; the second order term was normalized to give this
result.

For $M = 1$, the D5-brane, we can compare to the classical \none\ potential.
We can use the Ansatz
\begin{equation}
\Phi_i = \frac{2}{n} \Phi L_i\ ,
\end{equation}
where $\Phi$ is a scalar (not a matrix) complex superfield,
so that $\sum_i \Phi_i \Phi_i = \Phi^2{\bf 1}$.  The K\"ahler potential and
superpotential are then
\begin{equation}
K = \frac{n}{2\pi g} \OL \Phi\Phi \ , \quad
W = \frac{mn}{4\pi g} \Phi^2 + \frac{i \sqrt{2} }{3\pi g} \Phi^3\ .
\label{KW}
\end{equation}
The potential then agrees with that found in the brane calculation
provided $\zeta = -3 \sqrt 2$.  This could be checked by various
independent means, such as the fermionic terms in the D3-brane action in a
$G_{\it 3}$ background.

Returning to general $M$, there is supersymmetric minimum at
\begin{equation}
z = \frac{ \pi \alpha' i m n}{\sqrt 2M}\ .
\end{equation}
For a D5-brane, $(c,d) = (0,1)$ and $z= {i \pi \alpha' m n/\sqrt 2}$.
For illustration let $m$ be real.
The $i$ reflects the fact that the two-sphere lies in the 789-directions,
where $\tilde F_{\it 3}$ is maximized.   For an
NS5-brane, taking $C = 0$ for convenience,
$z= {\pi \alpha' m g n/\sqrt 2}$.  This is smaller by $g$, and lies
in the 456-directions where
$H_{\it 3}$ is maximized.  Note that the potential in each case has another
minimum at $z = 0$, where the probe has dissolved into the source
branes; our
approximation is not valid at $z=0$, but it is valid far enough to show that
the potential becomes attractive at small $z$.

We can also introduce several probes of arbitrary types, and each
will independently sit at the minimum of its own potential.  Note that in the
$AdS$ geometry we should not think of these as concentric, but rather arranged
along the $AdS$ coordinate $r$ while wrapped at various angles on equators of
the $S^5$.

An $S^2$ on $S^5$ can be contracted to a point, but it is
energetically unfavorable to do so.  The first term in the potential
vanishes in this limit (since $\det G_\perp$ goes to zero), and the
second does as well, leaving only the positive third term. This is
because the pointlike D5-brane retains only its D3 charge, which feels
a positive potential.

\section{The Full Problem}

We now consider the fields and self-energy of the full set of $N$
D3-branes, when these are in the configuration $\RR^4 \times S^2$ (or a sum
of several two-spheres) with 5-brane charges.  As an intermediate step we
consider a probe moving in such a background.
One might expect these calculations to be much harder that the previous
probe problem, as the symmetry is greatly reduced.  Remarkably, however,
all of the work has already been done.  The expanded brane configuration is
reflected in a less symmetric warp factor $Z$, but we will see that this
drops out of all terms in the potential.

In this section we also work out the first-order correction to the background.
In addition we show that our approximation breaks down close to the 5-brane
shell, and give the corrected form.

\subsection{The Warped Geometry}

Consider $N$ D3-branes spread on a two-sphere of $AdS$ radius $r_0$ in some
3-plane in the six transverse dimensions.  This Coulomb
branch background is again of the form~(\ref{backg}), with the
$Z$-factor given by harmonic superposition.  The $Z$-factor at any point
can depend only on its radii $w$ in the 3-plane and $y$ in the orthogonal
3-plane:
\begin{eqnarray}
Z \* \frac{1}{2} \int_{-1}^1 d\cos\theta\, \frac{R^4}{(w^2 + y^2 + r_0^2 -
2r_0 w \cos\theta)^2} \nonumber\\
\* \frac{R^4}{(y^2 + [w + r_0]^2)(y^2 + [w - r_0]^2)}\ .
\label{zwarp}
\end{eqnarray}
This is normalized to agree with the $AdS$ $Z$-factor at large $w,y$.  When
the D3-brane charge is divided among several two-spheres, then $Z$ is a sum
of such terms, with total coefficient $R^4$.  At $r=0$ this $Z$ goes to a
constant, so  for
$w,y\ll r_0$ we find flat ten-dimensional spacetime, with no
nontrivial topology.

To next order we consider linearized $G_{\it 3}$ fields in this background.  The
field equation is again
\begin{equation}
d [ Z^{-1} ( *_6 G_{\it 3} - i  G_{\it 3})] = 0\ , \label{field2}
\end{equation}
and the Bianchi identity is $dG_{\it 3} = 0$.  The origin is now a smooth
point and the perturbation will be nonsingular there.  It has a specified
nonnormalizable behavior at infinity, corresponding to the perturbation of
the gauge theory Hamiltonian, and a specified source at the 5-branes.
Note that this is a magnetic source, appearing in the Bianchi identity but not
the field equation.  Note also that
\begin{equation}
d {*_6} [ Z^{-1} ( *_6 G_{\it 3} - i  G_{\it 3})] =
 d [-i Z^{-1} ( *_6 G_{\it 3} - i  G_{\it 3})] = 0\ .
\end{equation}
Thus, the combination $Z^{-1} ( *_6 G_{\it 3} - i  G_{\it 3})$ is annihilated
by both $d$ and $d {*}_6$.  Further, at infinity it approaches the constant
value~\eref{dufirst} which is just governed by the boundary condition on the
nonnormalizable solution:
\begin{equation}
Z^{-1} ( *_6 G_{\it 3} - i  G_{\it 3}) \to - i\frac{2\sqrt{2}}{g} T_{\it 3} \ .
\end{equation}
It follows that it takes this
constant value everywhere, independent of the warp factor $Z$ and of the
configuration of the brane.

The field $G_{\it 3}$ itself does depend on the brane configuration,
and we will determine it in section~V.C, but it is not relevant
here. The brane dominantly couples only to the integral of the potential
$B_{\it 6} - \hat \tau C_{\it 6}$ which is already determined
by~\Eref{a6b} to be independent of $Z$.  Thus it too is independent of
the brane configuration.

\subsection{The Potential and Solutions}

Let us consider again a probe, but now moving in the warped geometry
just described.  The potential felt by the D3-brane charge of the
probe is again zero, for the usual supersymmetric reasons, so the
Born-Infeld and Chern-Simons terms again nearly cancel, leaving behind
the first term in the potential~\eref{pot2}.  As we noted, this term
is independent of $Z$.  The second term in the potential comes from
the coupling to $B_{\it 6} - \hat \tau C_{\it 6}$, and we have found that this 
too
is independent of $Z$.  The third term, given by supersymmetry, must
then also be $Z$-independent.  Thus, {\it a probe feels exactly the
same potential in the warped geometry formed by $\RR^4 \times S^2$
sources, as when all the sources are at the origin.}

Now consider the potential felt by the full set of $N$ D3-branes with 5-brane
charges.  As is familiar from electrostatics, we cannot simply take the
coupling of the branes to their self-field.  Rather, we must think of dividing
them into infinitesimal fractions and assembling the configuration by
bringing these together one at a time; in electrostatics this produces
the familiar factor of $\frac{1}{2}$.  In the present case, however, there is no
`charging up' effect because as just shown the potential felt by each
fractional `probe' is unaffected by the distribution of the earlier fractions.
Thus the potential is the same as in the probe case.  If the brane configuration
consists of two-spheres of respective D3-charges $n_I$ (with $\sum_I n_I = N$),
5-brane charges $(c_I,d_I)$, and radii and orientations $z_I = 2\pi\alpha'
\phi_I$, the potential is
\begin{equation}
-\frac{S}{V} = \sum_I \frac{4}{\pi g n_I} |M_I \phi_I^2 - i m n_I \phi_I /2\sqrt
2 |^2\ .
\label{fullpot}
\end{equation}
Thus, for every collection of 5-branes of total D3-brane
charge $N$ there is a solution with nonzero radii,
\begin{equation}
z_I = \frac{\pi\alpha' i m n_I  }{M_I \sqrt 2} \ .
\label{finrad}
\end{equation}
For a D5 sphere this is $AdS$ coordinate radius $r = \pi\alpha' m n_I$.
For an NS5-sphere it is $r = \pi\alpha' m gn_I$, smaller by a factor $g$
(when $C=0$).

It is important to check the validity of these solutions.  We have
already argued that for all $N$ D3-branes in a single D5 or NS5
two-sphere the solution is valid in the entire supergravity regime.
Now let us consider the problematic case discussed in section~II.B,
namely $p$ D5-branes each of charge $q$, which is supposed to
represent the same state as $q$ NS5-branes each of charge $p$.  For
the former solution, each D5-brane has charge $N/p = q$ and so the
central condition~\eref{condit} becomes
\begin{equation}
\frac{q^2}{gN} = \frac{q}{gp} \gg1\ .  \label{dcon}
\end{equation}
For the NS5-brane solution we can simply interchange $g \leftrightarrow 1/g$
and $p \leftrightarrow q$ via $S$-duality to obtain
\begin{equation}
\frac{gp}{q} \gg1\ .  \label{nscon}
\end{equation}
The conditions~\eref{dcon} and~\eref{nscon} are beautifully complementary, so
that only one solution is valid at a time.  At weak coupling the state is
described by a D5-brane and at strong coupling by an NS5-brane.

This example also provides the evidence that the condition~\eref{condit} is a
necessity, not a convenience: if the solutions persisted beyond this range
there would be too many, as compared to the known vacua of the gauge
theory.  Thus, we require that for each sphere
\begin{equation}
\frac{n_I}{g |M_I|^2} \gg 1\ .
\end{equation}

It would be extremely interesting to understand the crossover
between the D5 and NS5 representations of the above phase.  At a minimum
this will require the full nonlinear supergravity solutions, but it may
involve nonperturbative brane dynamics beyond this.  Note that at the
crossover coupling  the D5 and NS5 two-spheres have the same $AdS$ radii but
different and nonoverlapping orientations.

There should be a similar story for the minima of the potential at $\phi = 0$.
These are outside the range of validity of the approximation, and should not
correspond to true solutions because these would again have no duals in the
gauge theory.  Rather, a 5-brane at small $\phi$ should transmute into a
different kind of 5-brane.

As another example consider the oblique solutions $(c,d) = (1,s)$.  The
condition that the 5-brane energy density, added in quadratures, be much less
than the D3-brane energy density, is [see \Eref{enrat}]
\begin{equation}
\frac{1}{\alpha'} \biggl( \frac{1}{g^4} + \frac{s^2}{g^2} \biggr)^{1/2}
\ll \frac{N^{1/2}}{g^{3/2}\alpha'}\quad \Rightarrow\quad 1+g^2 s^2 \ll gN\ .
\label{pqcon}
\end{equation}
For small $s$ this is valid in most of the supergravity regime, but for
$s \sim
N$ it is valid nowhere.  This resolves the overcounting, that $(1,s)$ and
$(1,s+N)$ represent the same state.  Note that for $s \gg 1$ there is a range
of $g$ where supergravity is valid but the $(1,s)$ brane solution is not;
the
\slz\ duality (which acts on these vacua in an intricate way) gives other
candidate brane configurations.

There is one final issue connected with the stability of the brane solutions.
Let us focus on the D5-brane.  At opposite points on the two-sphere, the D5
world-volumes are antiparallel.  Intuition from flat space D5-branes
\cite{dbranes} would suggest that this configuration is not supersymmetric,
but this must be wrong. The supersymmetry transformation related to the D5
charge must be offset by the effect of the background on the much larger D3
charge.

We leave the analysis of supersymmetry for the future, but do address
a related point: the self-force of the D5-brane.  Again, intuition
suggests that there should be an attractive force between opposite
sides of the two-sphere, rendering the state unstable, but if the
configuration is supersymmetric then this must vanish.  Let us see how
this works. In the D5-brane action~\eref{d5act}, the strongest
couplings to bulk fields are those of the D3-brane charge to
$G_\parallel$ and to $C_{\it 4}$.  The self-force from these cancels
as usual due to the supersymmetry of D3-branes. The next strongest
coupling is of the D5-brane charge to $G_{\it 3}$.  It is this that
might give an attractive force, but in fact it does not:~\Eref{a6b}
shows that the field sourced by the D5-brane does not act back on the
D5-brane.  The $C_{\it 4}$ background induces mixing between $F_{\it
3}$ and $H_{\it 3}$ in such a way that the self-force cancels for any
orientation!\footnote{This might seem to
contradict claims that there is a large-$N$ limit of $AdS$ space which gives
flat-spacetime physics
\cite{smat}, since nonparallel D5-branes do attract in flat spacetime.
The point is that this large-$N$ limit includes going to small $AdS$
distances.  This would bring us into the `near-shell' region of the
D5-brane (to be discussed in section V.D), where the above no-force
analysis does not apply.} Finally, the dilaton and metric couple to the
quadrature term; this is second order in $\sigma_5/\sigma_3$, and so the
exchange force would be fourth order.  In the supersymmetric case this
should actually vanish, but because it is in any event small we will not
show this. Moreover, even for a nonsupersymmetric perturbation the
arguments for the vanishing of the forces from $G_\parallel$, $C_{\it 4}$,
and
$G_{\it 3}$ continue to hold, so only the small residue from the
dilaton and $G_\perp$ remains.  This is too small to destabilize the
solution, as the potential~\eref{fullpot} is a second order effect.

\subsection{First Order $G_{\it 3}$ Background}

Here we work out the first order correction to the
background, which appears only in the field $G_{\it 3}$.
In addition to the earlier result~\eref{a6b},
\begin{equation}
*_6 G_{\it 3} - i  G_{\it 3} = -i\frac{2\sqrt{2}}{g} Z T_{\it
3}\ , \label{a62}
\end{equation}
we have the Bianchi identity with magnetic source,
\begin{equation}
dG_{\it 3} = J_{\it 4}\ . \label{biaj}
\end{equation}
Let us adopt a coordinate system in which the brane is a sphere of radius
$r_0$ in the $w^{1,2,3}$ directions and at the origin in the $y^{1,2,3}$
directions.  Then
\begin{equation}
J_{\it 4} = 4\pi^2\alpha' M \delta^3(y) \delta(w-r_0) dw \wedge d^3y\ ,
\end{equation}
where $w$ is the radius in the $w$-plane, $d^3 y = dy^1 \wedge dy^2 \wedge
dy^3$, and the factor $4\pi^2\alpha'$ arises as $2 \kappa^2 \mu_5 / g^2$.
Note that the quantum numbers $M$ appear in a simple way.
In place of~\Eref{a62}, we can use its exterior derivative,
\begin{equation}
d{*_6} G_{\it 3} = i J_{\it 4}  -i\frac{2\sqrt{2}}{g} dZ \wedge T_{\it
3}\ . \label{deq}
\end{equation}
This and the Bianchi identity determine $G_{\it 3}$; they can be
solved in terms of potentials.

Write
\begin{equation}
G_{\it 3} = {*_6} d\omega_{\it 2} + i d\omega_{\it 2} + d \eta_{\it 2}
\label{gsol}
\end{equation}
with the gauge choice
\begin{equation}
d {*_6} \omega_{\it 2} = d {*_6} \eta_{\it 2} = 0\ .
\end{equation}
Then
\begin{eqnarray}
\partial_m \partial_m \omega_{\it 2} \* {*_6} J_{\it 4}
= \frac{2\pi^2\alpha' M}{r_0} \delta^3(y) \delta(w-r_0)
\epsilon_{ijk} w^i dw^j \wedge dw^k\ , \nonumber\\
\partial_m \partial_m \eta_{\it 2} \* - \frac{2i\sqrt{2}}{g} {*_6} (dZ \wedge
T_{\it 3})
= -\frac{\sqrt{2}}{g} T_{mnp} \partial_m Z dx^n \wedge dx^p\ .
\end{eqnarray}
The solutions are
\begin{eqnarray}
\omega_{\it 2} \*  -\frac{\alpha' M}{4w^3}
\epsilon_{ijk} w^i dw^j \wedge dw^k
\partial_t \biggl( \frac{1}{t} \ln \frac{y^2 + w^2 + r_0^2 + 2 r_0 wt}{y^2
+ w^2 +
r_0^2 -  2 r_0 wt}
\biggr)\biggr|_{t = 1} \ ,
\nonumber\\
\eta_{\it 2} \* \frac{R^4}{8 g r_0 \sqrt{2}} T_{mnp} dx^n \wedge dx^p
\partial_m \biggl( \frac{1}{w} \ln \frac{y^2 + [w+r_0]^2}{y^2 + [w-r_0]^2}
\biggr)\ .
\label{onsol}
\end{eqnarray}
These do not seem very enlightening, but we can obtain their forms at large
$r^2$:
\begin{eqnarray}
\omega_{\it 2} &\approx&  -\frac{8\alpha' M r^3_0}{3 r^6}
\epsilon_{ijk} w^i dw^j \wedge dw^k
 \ ,
\nonumber\\
\eta_{\it 2} &\approx& -\frac{R^4}{g \sqrt{2}r^4 } T_{mnp} x^m dx^n \wedge
dx^p
\ . \label{onasy}
\end{eqnarray}
These scale as the normalizable and nonnormalizable solutions respectively.  The
latter,
$\eta_2$, matches the boundary condition~\eref{first}.

\subsection{The Near-Shell Solution}

Our small parameter guarantees that our solution is good over most of
spacetime, but it must break down as we approach the 5-brane shell.
The metric in the directions parallel to the 5-brane and orthogonal to
the D3-branes expands, diluting the D3-brane charge so that close to
the 5-brane it no longer dominates.  One also sees this in the ratio
of energy densities, where the metric has the same effect.  Since this
occurs only close to the 5-brane, we can approximate the solution in
this region by a flat 5-brane+D3-brane solution.  Specializing to $p$
D5-branes,\footnote {See for example Eq.~(6), and for the NS5 brane
Eq.~(35), of Ref.~\cite{AOS-J}.  Note that these equations arise after
taking the limit where a noncommutative gauge theory describes the
5-brane dynamics.  We will return to this issue briefly in our conclusions.}
\begin{eqnarray}
ds^2_{\rm string} \* \frac{\alpha' u}{agp} \Bigl[ \eta_{\mu\nu} dx^\mu dx^\nu
+ h(d\tilde x^4 d\tilde x^4  + d\tilde x^5 d\tilde x^5) \Bigr] +
\frac{\alpha' a gp}{u} (du^2 + u^2 d\Omega_3^2)\ , \nonumber\\
e^{2\Phi} \* g^2 \frac{a^2 u^2}{1 + a^2 u^2}\ ,
\quad ds^2 = g^{1/2}e^{-\Phi/2} ds^2_{\rm string}\ ,
 \label{aosj}
\end{eqnarray}
where
\begin{equation}
h = (1 + a^2 u^2)^{-1}
\ .
\end{equation}
Let us compare with the near-shell metric based on the harmonic
function~\eref{zwarp}, near the point $(w_1, w_2, w_3) =
(0,0,r_0)$:
\begin{equation}
ds^2_{\rm string} =  \frac{2r_0 \rho}{R'^2} \eta_{\mu\nu} dx^\mu dx^\nu
+ \frac{R'^2}{2r_0 \rho} (dw \cdot dw + dy \cdot dy)\ , \label{mid}
\end{equation}
where
\begin{equation}
\rho^2 = (w_3 - r_0)^2 + y^2\  .
\end{equation}
We have also defined $R'^4 = 4\pi g n \alpha'^2$ to include the case that
the shell does not carry the full D3 charge $N$; we do not assume that $n$ is
small. The metrics agree away from the shell, $au \gg 1$, provided that
\begin{equation}
u = \frac{\rho}{\alpha'}\ ,\quad
a = \frac{R'^2}{2 g p r_0}\ ,\quad
\tilde x^{4,5} = \frac{R'^8}{16 g^3 p^2 r_0^4 \alpha'^2} w^{1,2}\ .
\end{equation}

With these identifications, the solution~\eref{aosj} gives the continuation
to $au < 1$.
As a check, the crossover distance $au = 1$ is
\begin{equation}
\rho_{\rm c} = \alpha' a^{-1} = \frac{2 g p r_0 \alpha'}{R'^2}
\sim \frac{p g^{1/2}}{n^{1/2}} r_0 \sim p m (gn\alpha')^{1/2}\ .
\end{equation}
Thus the shell is indeed thin: $\rho_{\rm c}$ is smaller than the radius
$r_0$ by $p(g/n)^{1/2}$, which is precisely our controlling
parameter~\eref{dcon} for the D5 solution.   As a reminder,
$r_0 = m\pi \alpha'n/p$ for this shell.  In summary, the components of the
metric tangent to the two-sphere, and the
dilaton, are multiplied by a factor
$
{\rho^2}/(\rho^2 + \rho_c^2),
$
\begin{eqnarray}
ds_{\rm string}^2 \*  \frac{2r_0 \rho}{R'^2} \eta_{\mu\nu} dx^\mu dx^\nu
+ \frac{R'^2\rho}{2r_0(\rho^2 + \rho_c^2)}
(dw^1 dw^1 + dw^2 dw^2)
+ \frac{R'^2}{2r_0 \rho} (dw^3 dw^3 + dy \cdot dy)\ , \nonumber\\
e^{2\Phi} \* g^2 \frac{\rho^2}{\rho^2 + \rho_c^2}\ ,
\quad ds^2 = g^{1/2}e^{-\Phi/2} ds^2_{\rm string}\ . \label{dshell}
\end{eqnarray}
This interpolates between the D3- and D5-brane metrics.

Similarly for $q$ NS5-branes, the solution interpolates between the D3- and
NS5- solutions.  The crossover radius is now
\begin{equation}
\rho'_{\rm c} =  \frac{2 r_0 q \alpha'}{R'^2}
\sim \frac{q}{(gn)^{1/2}} r_0 \sim q m (gn\alpha')^{1/2}\ ,
\end{equation}
the $AdS$ radius is
$r_0 = m\pi \alpha'gn/q$ for this shell,
and the solution is
\begin{eqnarray}
ds^2_{\rm string} \* \frac{2r_0 (\rho^2 + \rho'^2_{\rm c})^{1/2}}{R'^2}
\eta_{\mu\nu} dx^\mu dx^\nu + \frac{R'^2}{2r_0 (\rho^2 + \rho'^2_{\rm
c})^{1/2}} (dw^1 dw^1 + dw^2 dw^2) \nonumber\\
&&\qquad\qquad
+\frac{R'^2 (\rho^2 + \rho'^2_{\rm c})^{1/2}}{2r_0 \rho^2} (dw^3 dw^3 + dy
\cdot dy)\ , \nonumber\\
e^{2\Phi} \* g^2 \frac{\rho^2 + \rho'^2_c}{\rho^2}\ ,
\quad ds^2 = g^{1/2}e^{-\Phi/2} ds^2_{\rm string}\ . \label{nshell}
\end{eqnarray}
For $\rho<\rho_c'$ the metric develops the usual throat
for $q$ NS5-branes \cite{CHS2}.  The string coupling becomes strong at
$\rho/\rho_c'\sim g$, a proper distance $\ln 1/g$ from the
crossover region.

It is important to see where the supergravity solution is valid.  A
crude but simple measure is that the radius of a transverse sphere (fixed
$\rho$) must be large in string units.  (We assume $g \leq 1$ so that the
F-string scale is the relevant one.)  At the crossover point, the D5 and NS5
radii-squared are respectively
\begin{equation}
gp\alpha'\ ,\quad q\alpha'\ .
\end{equation}
The NS5 solution is valid for $q \gg 1$ and marginal for $q = 1$
(these properties continue to hold down the throat, until the dilaton
diverges). The D5 solution has a limited range of validity for $p \gg 1$ but
none for
$p=1$ (not even $g \gg1$, because the dual string theory is strongly
curved).  Thus the low energy physics of the Higgs phase is given by the
dual field theory description.

\subsection{The Complete Metric and Dilaton}

The pieces of our solution are scattered through this paper.
The zeroth order solution is the D3-brane
background~\eref{backg} with harmonic function~\eref{zwarp}, with the brane
locations and orientations~\eref{finrad}.  The first order correction is given
by Eqs.~\eref{gsol} and~\eref{onsol}.  The correction near the brane is
given in
Eqs.~\eref{dshell} and~\eref{nshell}.  For convenience we give here the full
solution for the metric and dilaton in a form that interpolates between the
zeroth order solution and the near-shell solution.  We emphasize that these have
overlapping ranges of validity,
$\rho > \rho_{\rm c}, \rho'_{\rm c}$ versus $\rho < r_0$.

We focus on a single
shell of D5 or NS5 type, but the generalization is straightforward.
The solution is be conveniently written using coordinates
$x^\mu$ for spacetime, $w^i$ for the three coordinates in which the
brane is embedded, and $y^i$ for the other three.  Write $w, \Omega_w$
as spherical coordinates for the $w^i$, and similarly for the $y^i$.
Both the Higgs and confining metrics, in string frame, can be conveniently
written
\begin{equation}
Z_{x}^{-1/2} \eta_{\mu\nu}dx^\mu dx^\nu
+
Z_{y}^{1/2} (dy^2 + y^2 d\Omega_y^2 + dw^2)
+
Z_\Omega^{1/2} w^2 d\Omega_w^2\ .
\end{equation}
For the Higgs (D5) vacuum, the $w^i$ are $x^{7,8,9}$ and the $y^i$ are
$x^{4,5,6}$; for the confining (NS5) vacuum at $\theta = 0$ this is
reversed.

For the D5 brane we have
\begin{equation}\label{hmetric}
Z_x = Z_y = Z_0 \equiv { R^{4}\over\rho_+^2\rho_-^2}
 \ , \quad
Z_\Omega = Z_0 \left[{\rho_-^2\over \rho_-^2 + \rho_c^2}\right]^2 \ ,
\end{equation}
where
\begin{equation}
R^4 = 4\pi g N\ ,\quad\rho_\pm = (y^2 + [w \pm r_0]^2) \ ,\quad \rho_c =
{2gr_0\alpha'\over R^2}
\ ,\quad r_0 = \pi \alpha' mN\ .
\end{equation}
The dilaton is
\begin{equation}\label{hdilaton}
e^{2\Phi} = g^2 {\rho_-^2\over \rho_-^2 + \rho_c^2} \ .
\end{equation}

For the NS5-brane, we have
\begin{equation}\label{cmetric}
Z_x=Z_\Omega = Z_0 {\rho_-^2\over \rho_-^2 + \rho^2_c}
 \ ,\quad
Z_y = Z_0 {\rho_-^2 + \rho^2_c\over \rho_-^2} \ ,
\end{equation}
where
\begin{equation}
\rho_c = {2r_0\alpha'\over R^2}
\ , \quad r_0 = \pi \alpha' mgN\ .
\end{equation}
Meanwhile the dilaton is
\begin{equation}\label{cdilaton}
e^{2\Phi} = g^2 {\rho_-^2 + \rho_c^2\over \rho_-^2}\ .
\end{equation}
Note $\rho_c = m R^2 / 2 \sim m\sqrt{gN}\alpha'$ for both branes.

\section{Gauge Theory Physics}

In this section, we consider some of the non-perturbative objects in
the field theory --- strings, baryon vertices, domain walls,
condensates, instantons and glueballs, --- and discuss their
appearance in the supergravity representation. Although objects of
this type have appeared in a number of previous incarnations
\cite{highTa,highTb,highTc,ewbaryons,grossooguri,gktII,dibaryons},
they arise here in novel forms.  We will
also consider a vacuum with massive fundamental matter and mention
some of its amusing properties.

\subsection{Flux Tubes: A First Pass}

Many of the vacua of the \noneplus\ field theory have stable flux
tubes.  At weak coupling, the Higgs vacuum, where the $SU(N)/\ZZ_N$
gauge group is completely broken, has semiclassical vortex solitons in
which certain components of the adjoint scalars wind at infinity.  The
topological charge associated with this winding takes values in
$\pi_1(SU(N)/\ZZ_N) = \ZZ_N$; it measures the magnetic flux carried by
the vortex.  The confining vacuum has electric flux tubes carrying
flux in the $\ZZ_N$ center of $SU(N)$.  These become semiclassical
solitons in the $S$-dual description of the theory as $\tau\to0$.
Similar statements apply for the oblique confining vacua.  In the
other massive vacua \cite{rdew} there are both electric and magnetic
flux tubes, and in the Coulomb vacua there may or may not be any
stable flux tubes.  We will return to these cases in a later section.
For the moment we focus our attention on the strings of the Higgs and
confining vacua.

One of the surprising features of Maldacena's duality is that it
relates string theory to a conformal rather than a confining gauge
theory.  Unconfined electric flux lines between two charged sources in
the conformal \nfour\ field theory are represented by a string in the
gravity dual \cite{reyyee,maldastring}.  The string in question droops
into the $AdS_5$ space, rather than lying at a fixed $AdS$ radius $r$.
Since small $r$ corresponds to large distances in the field theory,
the drooping string represents flux lines which spread out in the
region between the sources, as expected in a nonconfining theory.  The
symmetries of $AdS$ space suffice to show that the energy of the
string scales as a constant plus a term inversely proportional to the
separation of the sources.

In the realization of confining gauge theories via high-temperature
five-dimensional field theories \cite{highTa,highTb,highTc}, the
temperature provides an IR cutoff on $r$.  The flux between two
charged sources in the field theory now is represented by a string
which droops only part way into the $AdS_5$ space, becoming stuck at a
radius of order the temperature $R^2 T$; consequently the
string represents flux lines trapped in a physical string-like object,
of definite tension and width.  In this way the confinement of this
theory, which is hoped to be in the same universality class as
asymptotically free Yang-Mills theory, was established.  The same
happens in our dual description of \noneplus\ gauge theory.

Before treating the supergravity picture carefully, we begin with an
intuitive argument.  Let us assume, as we will shortly show, that a
$(p,q)$ string, with its world-sheet oriented in the $x^\mu$
directions, can bind to a $(p,q)$ 5-brane with D3-brane charge, in a
state of finite width and nonzero tension.  We claim that this object
is a confining flux tube of the gauge theory; since its $AdS$ radius
is by construction constant, it certainly has a definite tension. Let
us consider $p=0$, $q=1$, the Higgs vacuum.  The potential between
charged electric sources, given by suspending a fundamental string
from two points on the $AdS$ boundary, is highly suppressed: the
string can split into two strings joining the D5-brane to the
boundary, meaning there is little energy cost to moving the endpoints
of the string apart.  By contrast, a D1-brane cannot end on the
D5-brane.  However, it can link up with our putative D1-D5/D3 bound
state.  This makes the potential between two magnetic sources linear
in the distance between them, with a coefficient set by the tension of
the bound state.  Note also that any $(p,q)$ string with $q\neq0$ is
similarly confined --- its $p$ F1 charges ending on the D5, its $q$
charges connected to $q$ flux tubes (or a bound state of such tubes)
on the D5 brane. It follows that monopoles and dyons, represented by
strings with D-charge, are confined in the Higgs vacuum, while
electric charges are screened.  This is as expected on general grounds
from the field theory.

By $S$-duality, the confining vacuum sports F1-NS5 bound states.  All
strings except those having only D1-charge will bind to the
D3-NS5-brane.  These bound states are the electric flux tubes of the
gauge theory.  In this vacuum it is fundamental string charge which is
confined and D-charge which is screened, in agreement with
expectations.  Similar conclusions hold in the oblique confining
vacua.

We now turn to the supergravity description of this physics, and
demonstrate that these bound states truly exist.  In our solutions the
function $Z$, given in \Eref{zwarp}, diverges at the branes, so all
strings can lower their tensions by drooping inward toward one of the
branes.  However, we have seen that there is a crossover point near
each brane, where the universal D3-brane behavior ceases to hold and
5-brane behavior takes over.  An F-string, representing electric flux,
couples to the string metric. The string stretches in a noncompact
direction, so the relevant metric component is $G_{\mu\nu}$.  In the
D5-solution~\eref{dshell} this still goes to zero at $\rho = 0$, so
electric flux is unconfined.  In the NS5-solution~\eref{nshell} it
takes the minimum value $ r_0^2 q / \pi g n \alpha' = \pi \alpha' m^2
g n q$, so for the confining phase, where $n=N$ and $q =1$, the
tension is
\begin{equation}
\tau_{\rm e} = \pi \alpha' m^2 g N\frac{1}{2\pi\alpha'}
= \frac{m^2 gN}{2}\ .
\end{equation}
This satisfies 't Hooft scaling, as expected in a confining vacuum.
The F-string lowers its tension, but only by a finite amount, by binding to
the NS5-brane.

A D-string, representing magnetic flux, couples to $e^{-\Phi}$ times
the string metric.  For the NS5-brane this now vanishes at the
brane,\footnote{This `magnetic screening' is required both by physical
intuition and by $S$-duality, but notice that it requires that the
string coupling diverge in the NS5-brane throat.  For multiple
coincident NS5-branes, this can be seen in supergravity alone.  For
one NS brane, however, the very nature of the throat is in dispute
\cite{throat} and it is not clear whether supergravity, worldsheet
CFT, or semiclassical brane physics gives a good description.  In
any case, the D-string must dissolve in the NS5-brane, one way or
another.} 
  but for the D5-brane there is a
minimum value $r_0^2 p / \pi n \alpha' = \pi \alpha' m^2 n q$.  The
Higgs phase magnetic tension is then
\begin{equation}
\tau_{\rm m} = \pi \alpha' m^2 N \frac{1}{2\pi\alpha'g} =
\frac{m^2 N }{2 g} \ . \label{tmag}
\end{equation}
The $g$ and $N$ scaling appears to be the same as for a classical
Nielsen-Oleson vortex.  The action scales as $N^3/g$, and the change in the
field, which appears squared, is presumably of order $1/N$ for a $\ZZ_N$
vortex.  The D-string is bound to the D5-brane.

Though satisfying, these results are partly outside the range of
validity of the supergravity description.  We have seen in section~V.D
that for D5-branes this description breaks down before the crossover
point, while for NS5-branes it is marginal (we assume $g < 1$; for $g
> 1$ the $S$-dual is true.)

Let us discuss the bound state more carefully in the D5 case, by
considering the limit in which the
D5-brane is flat.  Note
first that D1-branes outside a D5-brane are BPS saturated and not
attracted to the D5-brane. By contrast, D1-branes outside and parallel
to a set of D3-branes are attracted to the D3-branes; upon reaching
the D3-branes they appear as tubes of magnetic flux inside an \nfour\
gauge theory, which, since flux is unconfined, expand to infinite
radius.  Combining the D5 and D3 branes, the D1-brane is attracted to
the D5/D3 object but upon reaching it cannot expand to arbitrary size.
Its behavior within the D3-brane field theory is determined by the
semiclassical calculation of the vortex soliton which confines
magnetic flux.  Alternatively, it should be a semiclassical
instanton of the noncommutative field theory on the 5-brane \cite{ncinst}.

Another intuitive way to see the bound state is to use $T$-duality.
Begin with a D1-brane extended in the 01-directions, a 012345
D5-brane, and 0123 D3-branes which are distributed in the
45-directions with density $\sigma$.  A $T$-duality in the 5-direction
converts the D1-brane into a 015 D2-brane and the D5/D3 system into a
D4 brane, which fills 0123 plus a line in the 45-plane.  The line
makes an angle $\theta$ with the 5-axis, where
\begin{equation}
\tan \theta = (4\pi^2\alpha'\sigma)^{-1}\ .
\end{equation}
If $\theta=\pi/2$ ($\sigma = 0$), the D2 and D4 are perpendicular and
BPS, so there is no force on the D2.  If $\theta=0$ ($\sigma \to
\infty$), then the D2 can be absorbed by the D4.  But if
$0<\theta<\pi/2$ the D2 brane is attracted to the D4 but is misaligned
with it, and so cannot be completely absorbed.  Instead, only a part
of it is absorbed, leading to a D2-D4 bound state of finite energy and
size.  As shown in figure~1, the tension is reduced by a factor
$\sin\theta$.
\begin{figure}
\begin{center}
\leavevmode
\epsfbox{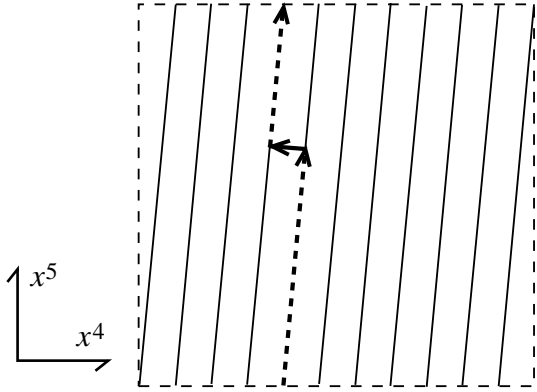}
\caption[]{D4/D2 system projected on the 45-plane.  The D4-brane, which is
also extended in 0123, wraps multiple times ($\tan\theta$).  The D2-brane,
which is oriented in the 05 directions, can partly dissolve, leaving a
piece connecting two leaves of the D4-brane.}
\end{center}
\end{figure}

In fact,  this bound state is
supersymmetric: as one sees from figure~1, after the flux dissolves to
the maximal extent, the remaining state is a D2-brane ending on a
D4-brane, a familiar supersymmetric configuration.  The BPS bound
\cite{dbranes} is
\begin{equation}
\tau \geq \Bigl[(V \tau_{\rm D4} + \tau_{\rm D2})^2 + V \tau_{\rm
D4}^2 \cot^2\theta  \Bigr]^{1/2} - \Bigl[V \tau_{\rm D4}^2 + V \tau_{\rm
D4}^2 \cot^2\theta \Bigr]^{1/2} = \frac{\tau_{\rm D_2} }{\sin\theta}\ ,
\end{equation}
where $V$ is a large regulator volume in the 123 directions.

Having established that a BPS state exists in this limit, we should
now verify that a nearly-BPS state of essentially the same mass is
still present when the D5-brane has the shape of a two-sphere. The
D1-D5/D3 bound state is in a rather difficult region of parameter
space because the effect of gravity is large (because the D3 charge is
large) but the gravity description is not valid everywhere (because
the D1 and D5 charges are small).  At large $r$ the supergravity
description is valid, while at small $r$ the effective description is
the field theory on the brane, as in examples in ref.~\cite{IMSY}.  It
appears that a correct treatment requires that we match these two
descriptions, in the spirit of the correspondence
principle~\cite{corr}.  By the logic of section V.D, the crossover
between the two descriptions occurs at a radius
\begin{equation}
\hat\rho = \eta \frac{\alpha' r_0}{R^2}\ .
\end{equation}
We will see that it is interesting to retain an undetermined constant
$\eta$ in the crossover point.  At the crossover, the metrics in the
0123 and the 45 ($w^{1,2}$) directions are
\begin{equation}
\frac{2 \eta\alpha' r_0^2}{R^4} \eta_{\mu\nu} dx^\mu dx^\nu
+ \frac{R^4}{2 \eta\alpha' r_0^2}(dw^1 dw^1 + dw^2 dw^2)\ . \label{cormet}
\end{equation}
The area of the two-sphere is then
\begin{equation}
4\pi r_0^2 \frac{R^4}{2 \eta\alpha' r_0^2} = \frac{8\pi^2 g N \alpha'}{\eta}\ ,
\end{equation}
giving
\begin{equation}
4\pi^2\alpha' \sigma = \frac{\eta}{2 g }\ .
\end{equation}
Combining the D1 tension, the rescaling of $G_{\mu\nu}$, and the
effect shown in figure~1 gives the tension
\begin{equation}
\tau_{\rm m} = \frac{1}{2\pi\alpha' g} \frac{2 \eta\alpha' r_0^2}{R^4}
\frac{2 g }{\eta} = \frac{ m^2 N }{2 g}\ .
\end{equation}

This is the same as the estimate~\eref{tmag} which came
from the purely gravitational picture; it is independent of the
precise crossover $\eta$ (a necessary, though not sufficient,
condition for correspondence arguments to give a correct numerical
value); and, one gets the same result if one ignores the gravitational
effect entirely and takes unit coefficients in the
metric~\eref{cormet}.

It has been suggested that the $\ZZ_N$ strings of supersymmetric QCD
might be nearly BPS saturated in the limit of infinite $N$.  In
\noneplus\ this hope is realized, although we see that large $gN$ is
necessary for this to be the case.  But we have not yet explained why
the strings carry charges which are conserved only mod $N$. To do so,
we turn to the construction of the baryon vertex.

\subsection{Baryon Vertex}

To put $N$ sources in the fundamental representation into a gauge
invariant configuration requires a baryon vertex.  In the $AdS_5\times
S^5$ supergravity dual of \nfour\ Yang-Mills this vertex is given by a
D5-brane which wraps the entire $S^5$ \cite{ewbaryons,grossooguri}.
We will see that in \noneplus, by contrast, the baryon vertex is a
D3-brane with the topology of a ball $B^3$, whose boundary is the
two-sphere of the 5-branes which form the vacuum.  The link between
these two pictures is the Hanany-Witten brane-creation mechanism
\cite{ahew}.

One way to derive the nature of the baryon vertex in the \noneplus\
theory is to begin in the ultraviolet. The ultraviolet theory is \nfour\
supersymmetric and the spacetime is approximately $AdS_5\times S^5$.
Let us consider the confining vacuum, represented by an NS5-brane
two-sphere.  A baryon vertex joining charged sources in a small
spatial region corresponds to a D5-brane wrapping the $S^5$
near the $AdS_5$ boundary, with $N$ fundamental strings joining the
boundary and the D5-brane. Now let the region containing the sources
grow comparable to the IR cutoff distance $m^{-1}$ (we will give a more
precise estimate in section~VI.H).  The
$AdS_5$ radius of the D5-brane decreases until it eventually crosses the
NS5-brane.  Since drawing the sources further apart than
$m^{-1}$ should lead to a large energy cost, something dramatic
must happen at this scale.  And indeed, it does: the crossing of the D5-
and NS5-brane produces a D3-brane which connects the two.

To see this, consider the configuration more carefully.  The
brane-creation process is local, so let us consider a nearly-flat
portion of the NS5-brane, which extends in the
$12345$-directions.  The D5-brane locally extends in the
$45678$-directions.  The distance vector between the two branes lies in the
$6$-direction.  In this arrangement, the crossing of the branes leads to
the creation of a D3-brane which fills the dimensions $456$. The
transition is shown in figure 2.
\begin{figure}
\begin{center}
\leavevmode
\epsfbox{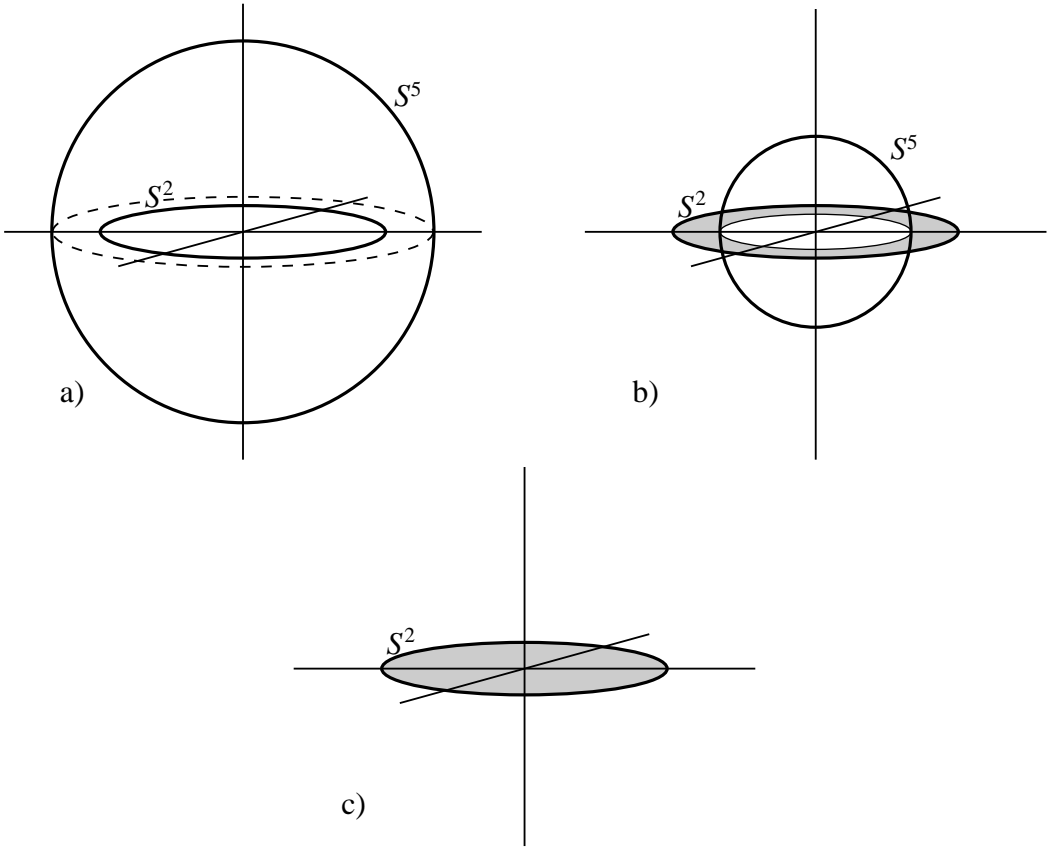}
\caption[]{a) A small baryon vertex, at $r> r_0$: the $S^5$ baryon vertex is
outside the $S^2$ of the vacuum brane (the vacuum brane is also extended in
123, while the baryon is not). b) The baryon $S^5$ contracted to
$r < r_0$: a 3-brane (shaded) has been created.  c) The $S^5$ has contracted
to nothing, leaving a 3-brane baryon filling the $S^2$.}
\end{center}
\end{figure}

Looking globally at the two-sphere, we see that the D3-brane fills the
part of the NS5-brane two-sphere which lies outside the D5-brane.
But the space inside the NS5-brane is topologically flat; the
radius of the five-sphere shrinks to zero inside.  The D5-brane
therefore is topologically unstable and can be shrunk to zero radius,
leaving a D3-brane which fills the entire two-sphere inside the
NS5-brane.  Like the D5-brane which created it, this D3-brane
is a particle in the four-dimensional spacetime; more precisely, it
is a localized object whose size is of order $m^{-1}$.
If the charged sources are taken to lie further apart than this,
then they will connect to the D3-brane not directly but through F1-NS5
flux tubes; thus the baryon vertex behaves dynamically as we would
expect in a confining theory.

The D5-brane baryon vertex in the \nfour\
theory has no preferred size or energy.  Here, the D3-brane actually
represents a physical excitation of definite size and mass.  The mass is
\begin{eqnarray}
\frac{\mu_3}{g} \int_{B^3} d^3x\, e^{-\Phi} {G^{1/2}_{\rm string}}
\* \frac{\mu_3 R^2}{g} \int_0^{r_0} dw\, \frac{4\pi w^2}{(r_0^2 +
\rho'^2_{\rm c} - w^2)}
\nonumber\\
&\approx& N\frac{m \sqrt{gN}}{2 \pi^{3/2}} \ln (gN)
\ .
\end{eqnarray}
Note that this diverges, due to a net factor $Z^{1/2}$ in the integrand,
until the near-shell form is taken into account.  The result is a factor
of $N$ times 't Hooft scaling, as would be expected.

To see directly that the D3-ball is a baryon vertex, note that the
NS5-brane world-volume action includes a Chern-Simons term
\begin{equation}
\int F_{\it 2} \wedge F_{\it 2} \wedge B_{\it 2} \ .
\end{equation}
This is the $S$-dual of the term
\begin{equation}
\int F_{\it 2} \wedge F_{\it 2} \wedge C_{\it 2}
\end{equation}
in the D5-brane action, which is familiar as it implies that a world-volume
gauge instanton is a dissolved D1-brane.  The D3-brane ending on the
NS5-brane is a magnetic monopole source for $F_{\it 2}$ (the $S$-dual of a
familiar fact for D3- and D5-branes), while the dissolved D3-branes become
$N$ units of $F_{\it 2}$.  Under $\delta B_{\it 2} = d\chi_{\it 1}$,
\begin{equation}
\delta \int F_{\it 2}(\mbox{monopole}) \wedge F_{\it 2}
(\mbox{dissolved})  \wedge B_{\it 2}
= - \int dF_{\it 2}(\mbox{monopole}) \wedge F_{\it 2}
(\mbox{dissolved})  \wedge
\chi_{\it 1} \ .
\end{equation}
This violation, proportional to the number of dissolved D3-branes, must be
offset by $N$ fundamental strings ending at the D3-NS5 junction.

Note that this now also explains the $\ZZ_N$ quantum numbers of the
flux tubes.  If we place $N$ flux tubes close and parallel to each
other, pair-creation of these D3-branes can occur.  This allows
the flux tubes to annihilate in groups of $N$.

Application of $S$-duality allows us to form the same construction for
other similar vacua.  Notice that the baryon vertex is always a
D3-brane.  However, the 5-brane on which it ends determines its
properties.  For example, if we are in the Higgs vacuum, the magnetic
flux and its magnetic baryon form the $S$-dual of what we just
considered.  By contrast, the electric baryon is completely shielded:
the $N$ fundamental string sources, which in the confining vacuum were
forced to end on the D3-brane, are no longer forced to do so, since
they may end anywhere on the D5-brane.  This is of course consistent
with field theory expectations.

Now let us consider some other vacua.  Suppose that we take a vacuum
where the classical unbroken gauge group was $SU(N/k)$, with $k\ll\sqrt N$
a small divisor of $N$.  Since only the $SU(N/k)$ confines, and since
a fundamental representation of the $SU(N)$ parent breaks up into $k$
copies of the fundamental representation of $SU(N/k)$, we should
expect that $N$ sources would now be joined by not one but $k$
different baryon vertices.  To see this in the supergravity is
straightforward.  The relevant vacuum is given by $k$ coincident
NS5-branes, so when the D5-brane baryon vertex of \nfour\ crosses the
NS5-branes, $k$ D3-branes are created.  Each of these carries $N/k$
units of string charge (since each NS5-brane has $N/k$ units of
D3-brane charge) and so $N/k$ strings must end on each of them.  On
the other hand, the $k$ D3-branes are not bound together and may be
separated spatially from one another.  Each one represents a separate,
dynamical, massive baryon vertex of $SU(N/k)$.  Note also that pair
creation of these objects ensures that the electric flux tubes in this
vacuum carry only $\ZZ_{N/k}$ quantum numbers.

\subsection{Flux Tubes: A Second Pass}

Here we will look at Coulomb vacua to understand how the baryons and
strings behave, and obtain the correct flux tube quantum numbers.

We have already noted the flux tubes present when the vacuum is
massive --- that is, when the classical vacuum is given by $k$
copies of the $N/k$-dimensional representation of $SU(2)$.
The baryons ensured that the electric flux tubes carry flux in
$\ZZ_{k}$ and the magnetic or dyonic flux tubes have
charge in $\ZZ_{N/k}$.

Let us consider instead a general Coulomb vacuum, given by choosing
$p_i$ copies of the $q_i$-dimensional representation, with $\sum
p_iq_i= N$. The unbroken gauge group is $[U(p_1)\times U(p_2)\times
...\times U(p_k)] / U(1)$, corresponding to $p_i$ D5-branes of $AdS$
radius proportional to $q_i$. Let $r$ to be the greatest common
divisor of the $p_i$, and $s=\gcd(q_i)$.  Simple field theoretic
arguments then determine the properties of possible flux tubes.  The
topology of the breaking pattern of the gauge group permits magnetic
flux tubes to carry a $\ZZ_s$ charge, while the massive $W$-bosons of
the theory will break all electric flux tubes down to those carrying a
$\ZZ_r$ charge.

How do we see these flux-charges in the supergravity?  For the
magnetic flux, it is straightforward.  Consider a collection of $k$
magnetic flux tubes.  A magnetic flux tube can be moved with impunity
from one D5-brane to another, since two flux tubes on different
D5-branes can be connected by a D1-string in the radial direction,
corresponding to a magnetic gauge boson.  A magnetic baryon vertex
connecting to a D5-brane of radius $q_1$ can remove or add $q_1$ flux
tubes from the $k$ that we started with.  Since we may move all the
flux tubes from the first group of D5-branes to the second group, we
may also remove any multiple of $q_2$ flux tubes from our collection.
Removing $q_i$ flux tubes in any combination, we are left with a
number $\hat k$ with $0\leq \hat k<\gcd(\{q_i\})$.  This confirms what
we set out to prove.

To see the charges of the electric flux tubes requires $S$-duality,
which in not understood for the general Coulomb vacuum.  However, we
conjecture that the $\tau\to -1/\tau$ transformation acts in a simple
way in an important subclass of the vacua.  In particular, consider
those classes of vacua {\it where all $\{p_i\}$ are distinct integers
and all $\{q_i\}$ are distinct integers.}  In this case we claim that
the $S$-dual of this vacuum is that with $q_i$ NS5-branes of radius
$p_i$.  This is of course consistent with the known transformation of the
massive vacua \cite{rdew}, for which $p_1q_1=N$.  The $S$-dual of
the argument in the previous paragraph then shows that the electric
flux tubes for these vacua is indeed $\ZZ_r$.  Indeed, this is our main
evidence for the conjecture.

If the $p_i$ or the $q_i$ are not distinct integers, then
the $S$-duality transformation we have suggested is ambiguous.
We do not know what happens in this case, either in
field theory or in supergravity.

\subsection{Domain Walls}

Since the theory has many isolated vacua, it also has a large number
of domain walls which can separate two spatial regions in different
vacua.  If the walls are spatially uniform then they may be BPS
saturated \cite{dvalishif,ewMQCD}.

Between the oblique confining vacuum represented by a $(1,1)$ 5-brane
sphere and the confining vacuum represented by an NS5-brane, there
must be a BPS domain wall which carries off one unit of D5-brane
charge.  We may therefore conjecture that a
BPS junction of three 5-branes --- the NS5-brane sphere for $x^1>0$, the
$(1,1)$ 5-brane sphere for $x^1<0$, and a D5-brane at $x^1=0$ which fills the
two-sphere --- describes this domain wall.  That is, the world-volume of
the D5-brane is the $023$-plane of the domain wall times the three-ball
spanning the two-sphere.
At small $g$ the NS5-brane and
the $(1,1)$ brane are nearly coincident (their $AdS$ radius and orientation
on the $S^5$ differ only at order $g$) and the effect of
the D5-brane on the much denser NS5-brane is very small.

We can see that this reproduces some known properties of the domain
wall.  First \cite{dvalishif,ewMQCD}, the flux tubes of the theory (F1
strings) obviously can end on the domain wall (a D5-brane).
Furthermore, consider dragging a $(1,1)$ dyonic string, representing a
dyonic source in the gauge theory, across the wall. For $x^1<0$, the
dyon is screened; it can end happily on the $(1,1)$ 5-brane.  For
$x^1>0$, the dyon is confined; its monopole charge ends on the
NS5-brane, but its electric charge must join onto a flux tube --- an
F1-NS5 bound state --- which in turn ends on the D5-brane domain wall.
Finally, note that we may dissolve an $N$-string vertex (a D3-brane)
into this domain wall (a D5-brane), leaving $N$ strings which end on
the wall and are free to move around on it.  If we then permit this
domain wall to annihilate with an antidomain wall with no strings
attached, then the annihilation will leave a D3-brane behind on which
the strings may end, as in the well-known process described in
\cite{Sen}.

More generally, if for $x^1<0$ the system is in the phase corresponding to
a $(c,d)$ 5-brane, and for $x^1 > 0$ it is in the phase corresponding to a
$(c',d')$ 5-brane, then a $(c-c',d-d')$ 5-brane must fill the 2-sphere
where they meet.
In general the branes on the right and left have
different orientations and radii, and so must bend as they meet as depicted
in figure~3.
\begin{figure}
\begin{center}
\leavevmode
\epsfbox{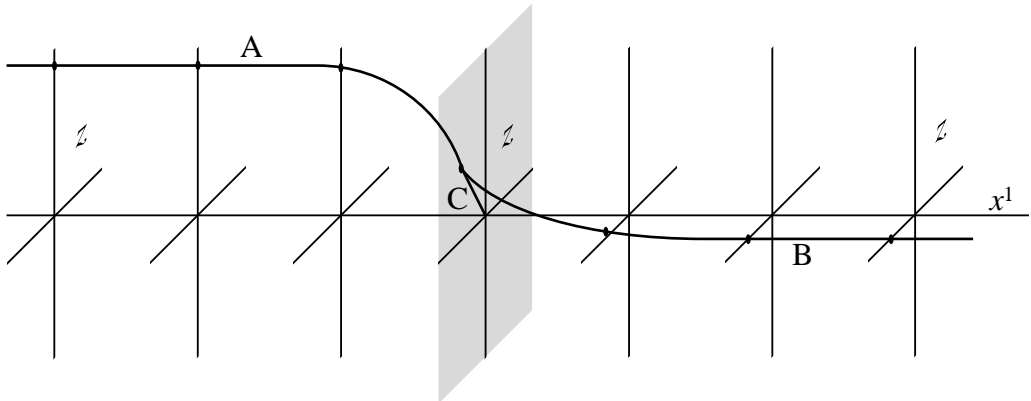}
\caption[]{Triple 5-brane junction, corresponding to a domain wall.
Depicted is $z(x^1)$; the full geometry is obtained by translating in
$x^{2,3}$ and rotating in the transverse $SO(3)$ symmetry.  At $x^1 < 0$ the
system is in the vacuum corresponding to 5-brane A; at $x^1 > 0$ it is in
the vacuum corresponding to 5-brane B, with different radius and orientation.
The domain wall lies in the shaded plane $x^1 = 0$ and is a 5-brane of type
C.  The bending of branes A and B is described by the BPS
equation~\eref{dwbps}.}
\end{center}
\end{figure}
When the left and right phases involve multiple spheres, there
will be a more complicated domain wall, constructed from multiple triple
5-brane junctions.

To discuss the domain wall tensions quantitatively  we need the kinetic
term for the collective coordinate $z = 2\pi\alpha' \phi$.  This arises in
the Born-Infeld action, from
\begin{equation}
G_{\mu\nu}(\mbox{induced}) = G_{\mu\nu} + G_{mn} \partial_{\mu} x^m
\partial_{\nu} x^n\ .
\end{equation}
Then
\begin{equation}
\frac{S}{V} = -\frac{\mu_5}{2g} 2\pi\alpha' \int d^2\xi\, G_{\perp}^{1/2}
(F_{ab} F^{ab})^{1/2}
\eta^{\mu\nu} \partial_\mu x^m \partial_\nu x^n
= -\frac{n}{2\pi g} \eta^{\mu\nu}\partial_\mu \bar\phi \partial_\nu \phi\ .
\end{equation}
This gives the K\"ahler potential $K = n\bar\Phi \Phi/2\pi g$.  This is the
same as~\Eref{KW} for the classical gauge theory, but by an \slz\
transformation one can show that it holds for all $(c,d)$.  This makes
sense, as the main kinetic effect comes from the D3-branes, which are
self-dual.  With this normalization the potential~\eref{fullpot} implies the
superpotential
\begin{equation}
W = \frac{1}{4\pi g} (i\frac{4\sqrt{2}}{3M} \Phi^3 + mn\Phi^2)\ .
\end{equation}
as in \Eref{KW}. 
At the nonzero stationary point this takes the value
\begin{equation}
W \to -\frac{m^3 n^3}{96 \pi g M^2}\ .
\label{oursup}
\end{equation}
For a multi-brane configuration it is summed over $I$.  

We cannot rule out an additional additive contribution to this
superpotential, although from our semiclassical reasoning we know that
any such contribution must be subleading in the Higgs vacuum and
others containing only large D5-branes.\footnote{We thank O.~Aharony
for pointing out an error in our orginal approach.}  Field theory
also suffers from the same ambiguity \cite{doreykumar}.  Up to these
additive contributions, the field theory and supergravity agree.
Consider the massive vacuum corresponding to $p$ D5 branes of radius
$q$; for this vacuum $M = p$.  Using \cite{doreya}, it is easy to
obtain a slight generalization of \cite{doreykumar} (adjusted to match
our conventions, and with the function $A(\tau,N)$ in Eq.~(5) of
\cite{doreykumar} set to zero)
\begin{equation}
W = {m^3N^2\over 24 g_{{\rm YM}}^2}\left[E_2(\tau) -
{q\over p}E_2\left({q\over p}\tau\right)\right]
\to {m^3N^2E_2(\tau)\over 96 \pi g}-{m^3N^3\over 96 \pi gp^2} \ .
\end{equation}
Here $E_2$ is the second Eisenstein series, and
we have used \Eref{dcon} and $E_2(i\infty) = 1$.  Note the
first term is $p$ independent and is subleading for $p\ll\sqrt{N}$.
For the massive vacuum given by $q$ $(1,k)$ 5-branes of radius $p$,
$k\ll p$, in which $M = q(\tau +k)$, the formula is
\begin{equation}
W = {m^3N^2\over 24 g_{{\rm YM}}^2}\left[E_2(\tau) -
{q\over p}E_2\left({q\over p}(\tau+k)\right)\right]\ ,
\end{equation}
but since Eqs.~\eref{nscon} applies and
$x E_2(x) = x^{-1}E_2(-x^{-1}) + (6i/\pi)$,
\begin{equation}
W \to {m^3N^2E_2(\tau)\over 96 \pi g}
- {m^3N^3\over 96\pi g q^2(\tau+k)^2} \  .
\end{equation}
The first term in this expression is the same as in the D5-brane
vacua, so field theory and supergravity agree up to a
classically-subleading $M$-independent function.

For a supersymmetric domain wall between two phases, the tension is
\begin{equation}
\tau_{\rm DW} = 2|\Delta W|\ =
\frac{ |m^3|}{48\pi g} \left| \sum_{I\,\rm left}\frac{n_I^3}{M_I^2} -
\sum_{J\,
\rm right}
\frac{n_J'^3}{M_J'^2}
\right| \ . \label{dwtens}
\end{equation}
Let us consider two examples, to see that the 5-brane junction construction
reproduces this tension.  For both examples we take $g \ll 1$.

The first is the domain wall described above, between the confining and first
oblique confining phase.  The general result~\eref{dwtens} becomes
\begin{equation}
\tau_{\rm DW}  =
\frac{ |m^3| N^3}{48 \pi g} \Bigl|  (ig^{-1})^{-2} - (ig^{-1} + 1)^{-2}
\Bigr|
\approx \frac{ |m^3| g^2 N^3}{24 \pi}
\ .
\end{equation}
In the brane picture, the tension comes from the spanning D5-brane.
This has three transverse and three longitudinal dimensions and so feels no
warp factor, giving simply
\begin{equation}
\tau_{\rm DW}  =
\frac{4\pi r_0^3}{3} \cdot \frac{\mu_5}{g} = \frac{ |m^3| g^2 N^3}{24
\pi}
\ .
\end{equation}
The agreement is quite beautiful, given the very different physics that has
gone into the two calculations.

The second example is the domain wall between the confining and Higgs
phases: a junction between an NS5-brane and a D5-brane, spanned by a
(1,1) 5-brane.  The NS5-brane is at much smaller radius than the D5-brane (by
a factor $g$) and has much greater tension, so the predominant effect is that
the D5-brane bends down to join the NS5-brane.  The bending is described by
the BPS equation
\begin{equation}
\partial_1 \phi = \Omega \OL{\frac{\partial W}{\partial \phi}} \label{dwbps}
\end{equation}
with $\Omega$ any phase.  For $m$ real, $\Omega = -1$ gives a solution that
passes through the origin and the nonzero stationary point, approximating
to order $g$ the solution needed.  The tension comes primarily from this
bending,
\begin{equation}
\tau_{\rm DW} = \int_0^\infty dx^1\,
\biggl(\frac{N}{2\pi g} |\partial_1 \phi|^2 + \frac{2\pi g}{N} |W_\phi|^2
\biggr)\ .
\end{equation}
In this case the general result~\eref{dwtens} follows by construction
\begin{equation}
\tau_{\rm DW} =
\frac{ |m^3| N^3}{48\pi g} \ ,
\end{equation}
where the superpotential in the confining phase is smaller by $O(g^2)$.
It will be an interesting exercise to show that the brane construction
reproduces~\Eref{dwtens} in the general case, without assuming small $g$.

Note that a domain wall is the same as a baryon vertex extended in two
additional directions.
By analogy, we
might expect the D5-brane three-ball which acts as a
domain wall to be
associated with passing a D7-brane through the NS5-brane.
This suggests that D7-branes should be reexamined in the
original $AdS_5\times S^5$ context.

\subsection{QCD-like vacua}

It is amusing that \noneplus\ is rich enough to permit us to study a
theory similar to QCD with heavy quarks.  Suppose we consider a vacuum
with a D5-brane of radius $n$ and one NS5-brane of radius $N-n$.  Here
we assume the usual condition on $n$, $n^2\gg gN$, but take $n\sim
gN$, so the D5-branes have comparable $AdS$ radius to the NS5-brane.
In the field theory this corresponds to a vacuum with a broken $SU(n)$
sector, a $U(1)$ vector multiplet, and a confining $SU(N-n)$ sector.
Among the massive vector multiplets are spin-1 bosons, along with
fermions and scalars, charged as $(\bf{\bar n}, \bf{N-n})$ under
$SU(n)\times SU(N-n)$.  These are strings connecting the D5-brane to
the NS5-brane.  We will refer to these as `quarks'.  Clearly these
theories have no free quarks: the D5-NS5 strings cannot exist in
isolation, since they cannot actually end on the NS5-brane, and
instead must be connected to a flux tube.  Note that the $SU(n)$ acts
as a sort of flavor group for the quarks (analogous to broken
weak isospin), and we will refer to its massive adjoint representation as
`flavor' gauge multiplets.

In order that the supergravity solution be valid, we must have $n \gg
(gN)^{1/2}$.  When $n=gN$, so that the D5 and NS5
sit at equal $AdS$ radii, the quark has mass of order
$$
(\alpha')^{-1}\int \sqrt{G_{00}G_{yy}r_0^2} d\zeta = r_0/\alpha' = mn
$$
which agrees with field theory.  However, it is easy to see that if
$n$ is smaller than $gN$, the quark retains mass $\sim mgN$; indeed, as an
extreme, note that if a D3-brane sits exactly at $r=0$, corresponding
to $(gN)^{1/2} \ll n\ll gN$, the quark mass is just proportional to the
coordinate length of the string, $mgN$.  We therefore see signs that the
physical `constituent' masses of the quarks can be much larger than their
current values.  Of course the quarks are never light
compared to $m$.

We can see easily that QCD-like theories have no stable flux tubes due
to quark pair production.  Recall that we measure the potential $V(L)$
between two electric sources by hanging a probe string by its ends
from the $AdS$ boundary, with the ends a distance $L$ apart.  Take
$L\gg\ m^{-1}(gN)^{-1/2}$; then in the absence of the D5-brane, the
probe would bind to the NS5-brane forming a confining flux tube
between the sources.  In the presence of the D5-brane, however, pair
production of the D5-NS5 strings can occur.  This breaks the flux
tube, which shrinks away allowing the quark to screen the source.  The
probe string ends up as two strings a distance $L$ apart, each
attached to the D5-brane.   Note however that if we take $n\gg
gN$, the quarks become very heavy, and the time scale for their pair
production becomes very long.  In this limit the confining flux tubes
are metastable.

Low-lying quark-antiquark mesons are not stable in this theory.  Highly
excited mesons are represented by two D5-NS5 strings joined by a long
F1-NS5 flux tube.  However, as the mesons deexcite by emission of
glueballs (either supergravity or string states), it eventually
becomes energetically preferable for them to decay to a D5-D5 string,
bypassing the NS5-brane altogether.  In short, the lowest lying mesons
between two quarks always mix with and decay to a massive, but
lighter, flavor particle, in a vector multiplet of the broken $SU(n)$
group.  (Indeed this almost happens in nature; charged pions decay
through isospin gauge multiplets, although not because those gauge
bosons are light but because they couple to light leptonic states ---
which could also be represented here, if there were a need.)

Baryons, on the other hand, carry a conserved charge and are both
stable and interesting.  $N-n$ D5-NS5 strings can end on a D3-brane
filling the NS5-brane two-sphere, forming an object whose mass can be computed.
If we arrange for a more complicated spectrum of quarks by choosing to
use multiple D5-branes of various radii, then there are processes by
which baryons can be built from quarks of different masses, and can
decay by emission of flavored mesons (or the corresponding
gauge multiplets of the `flavor' group.)
Scattering of baryons, or of baryons and antibaryons, could also
be studied.  In addition, it is possible that these baryons
have residual attractive short range interactions (different from the
physical case in that they are dominated by the `flavor' gauge
multiplets) which can cause them to form nuclei.  It would be amusing
to look for such baryon-baryon bound states.  Furthermore, these
baryons and nuclei carry $U(1)$ gauge charges, and in some vacua there are
lepton-like objects which  presumably can combine with them to form
atoms.

We cannot resist mentioning one more possibility, although it
admittedly may not be realized.  Namely, our baryons act as D0 branes in
spacetime, and our domain walls as D2-branes (their structure in the
extra dimensions is identical, so we suppress it.)  While we are not
used to thinking of baryons as places where strings can end, this is
quite natural if there are no light quarks; if all quarks are heavy
then short flux tubes are stable, and physical baryons can be linked
by them.  Turning on condensates of these flux tubes makes the
positions of the baryons noncommuting (note these branes have no
massless world-volume gauge fields, but still have massless
scalars), and through the Kabat and Taylor mechanism
\cite{ktsphere}, an assembly of baryons can be arranged into a
spherical domain wall!  Thus the properties of the gauge theory
recapitulate the method we have used to solve it.  In practise, one
should try to implement this process physically, through Myers'
mechanism \cite{myers}. Here we have a difficulty, as the required
three-form potential is a massive state, a glueball which couples to
domain walls \cite{dvakakgab}, so we cannot create a long-range field
to induce a dipole charge.  However, there may be ways to circumvent
this problem, and create this effect as a thought experiment or even
in a lattice simulation, where hints of domain walls have been observed
\cite{montvay}.
% [The Metric.]  [Spectrum? Baryon vertex?]

\subsection{Condensates}

With the naked singularity banished, the coefficients of the
normalizable terms in the supergravity fields, and so the dual
condensates, become calculable.  We have already determined the
superpotential in our discussion of domain walls, and in principle the
condensates can be determined directly from this function.  The full
field theoretic superpotential, and corresponding condensates, are
also known \cite{doreykumar}.  However there are subtleties
\cite{gubs,ADK}, and our understanding is only partial.

The condensates of the operators $\lambda\lambda$, $\tr
[\Phi_1,\Phi_2]\Phi_3$ and $m_i\tr \Phi_i\Phi_i$ are all related by
the chiral anomaly and operator mixing.  A linear
combination of these must couple, by the $AdS$/CFT correspondence, to
the mode of $G_{\it 3}$ which falls off as $1/r^3$ in invariant units,
which we have identified in \Eref{perts}.  More generally, higher
modes of $G_{\it 3}$ should give the expectation values of all of the
chiral operators $\lambda\lambda\phi^k + \cdots$.  For the lowest mode
of $G_{\it 3}$, \Eref{onasy} for $\omega_2$ gives the $m$ and phase
dependent parts as
\begin{equation}
M r_0^3 \propto \frac{m^3 N^3}{M^2}\ .
\end{equation}
For multiple shells, superposition gives
\begin{equation}
m^3 \sum_I \frac{n_I^3}{M_I^2}\ .
\label{multiplevev}
\end{equation}
Note that there are two $SO(3)$-invariant fermion bilinears, namely
$\sum_{i=1}^3 \lambda_i \lambda_i$ and $\lambda_4 \lambda_4$.  These
correspond to the polarization tensors $\epsilon_{i\bar\jmath\bar k}$ and
$\epsilon_{ijk}$, which have equal overlap with the actual field
$\epsilon_{w^1 w^2 w^3}$.

In the Higgs vacuum, and other vacua with only large D5-branes, all
condensates can be described semiclassically.  The expectation values
for $\tr [\Phi_1,\Phi_2]\Phi_3$ and $m_i\tr \Phi_i\Phi_i$ are known,
and their $m$, $n$ and $M$ scaling agrees with \eref{multiplevev}.  In
the confining cases, however, the situation is more subtle, since
these operators have condensates of different sizes and since the
gluino bilinear is also expected to play a role.  We note the
following facts.  First, careful examination of the field theory
superpotential given in \cite{doreya,doreykumar} reveals no obvious
linear combination of the operators whose expectation values would
have this property --- except the second term in the superpotential
itself, as discussed in section VII.D.  Second, there is every
indication from our study of domain walls that the second term in the
superpotential, proportional to the two-sphere volume times the
5-brane's charge under $*G_{\it 3}$, measures the dipole moment of the
5-branes. Naturally, the lowest normalizable mode of $G_{\it 3}$
couples to the lowest allowable 5-brane multipole moment, which is
indeed a magnetic dipole.  We therefore speculate that perhaps the
$ijk$ components of $G_{\it 3}$ couple to this part of the
superpotential, and to the worldsheet instantons which we will discuss
in section VII.G below.

In any case, the supergravity clearly shows that there are expectation
values for some dimension-three operators which classically would have
had vanishing vevs. It also shows these vevs differ from one confining
vacuum to the next.  These qualitative features certainly agree with
field theory.

The chiral operators $F^2 \phi^k$ are determined by the dilaton background.
The dilaton is nontrivial and is obtained from
\begin{equation}
\nabla^2 \Phi = \frac{2\pi g r_0^4 |M|^2}{NR^2} \delta^3(y)
\delta(w-r_0)
+ \frac{g^2}{12}
{\rm Re}( G_{mnp} {G}^{mnp}) \ .  \label{dilback}
\end{equation}
The first term comes from the coupling of the dilaton to the 5-brane
through the Born-Infeld action.  The second is directly from the
coupling to the bulk fields.  Taking the value of $r_0$ appropriate to
a single shell of quantum numbers $M$, and integrating over a volume
of order $r_0^6$, both terms are of order $m^6 R^2 r_0^4$ and must be
retained.  One can argue that they must cancel in the dilaton
monopole, which gives the expectation value of the derivative of the
Lagrangian with respect to the coupling; this must vanish in a
supersymmetric vacuum. Many of the higher operators $F^2 \phi^k$ are
also highest components of superfields, and for them a similar
argument should apply.  We have not yet shown this cancellation
directly for the background~\eref{dilback}, and it is possible that
there are subtleties.  It is a challenge to include this varying
dilaton in the full nonlinear treatment of supergravity.

\subsection{Instantons}

At weak coupling, many quantities in \nfour\ Yang-Mills theory receive
contributions from instantons.  Holomorphic objects can be written as
an instanton expansion, given by an infinite series in powers of the
parameter $q = e^{2\pi i \tau}$.  This expansion is not useful at
strong coupling, but in that case the same objects can typically be
reexpressed, using a modular transformation, in terms of $\tilde q =
e^{-2\pi i/\tau}$ (or some other $SL(2,\ZZ)$ variant.)  In string
theory, both in perturbation theory for flat D3 branes and in the
$AdS_5\times S^5$ language, D-instantons [D-($-1$) branes], whose action
is $2\pi/g$, play the role of these field theory instantons.  For
large $g$ one uses an expansion in magnetic D-instantons, with action
$2\pi g$, or more generally in dyonic D-instantons.

In \noneplus, the situation is slightly different.  Consider first
weak 't Hooft coupling, such that the strong coupling scale $\Lambda$
is much less than $m$.  In this case the Higgs vacuum has a
superpotential which is an expansion in $q = e^{2\pi i \tau}$, but the
superpotential of the confining vacuum has an expansion in $q^{1/N}$.
The $1/N$ in the exponent is responsible, in the $m\to\infty$ limit,
for $SU(N)$ Yang-Mills having $N$ vacua, related by
$\theta\to\theta+2\pi k$.  It has long been suggested that this
behavior implies the existence of fractional instantons, carrying
$1/N$ units of instanton charge, which, unlike instantons themselves,
remain important in the large $N$ limit.  Some evidence for these
objects has been found in MQCD \cite{brodie} (note that these are
distinct from similar objects which require compactification of
a dimension of spacetime for their existence) and there is even a
claim that they have been seen in lattice nonsupersymmetric Yang-Mills
\cite{narayanan}.

We might hope to find such objects here, but for the same reason that
$\Lambda\sim m$, we cannot do so, for $q^{1/N}$ is of order one, and
the fractional-instanton expansion fails.  We also cannot use the
magnetic instanton expansion, since $g\ll 1$.  But remarkably, as can
be inferred from the results of \cite{doreykumar}, there is yet another
expansion, one which is dual, in the sense of $gN\leftrightarrow
(gN)^{-1}$, to the expansion in fractional instantons.

In particular, Dorey and Kumar show that the superpotential
for the confining vacuum is proportional to
\begin{equation}\label{Etwos}
E_2(\tau) - {1\over N} E_2\left(\tau\over N\right)
\end{equation}
where $E_2$ is the second Eisenstein series.  For large imaginary
arguments, $E_2(z)$ can be written as an expansion in $e^{2\pi iz}$.
The first term in \eref{Etwos} can therefore be interpreted as a sum
over ordinary instantons.  The second term, by contrast, cannot be
expanded in this way, since $|\tau/N|\ll 1$.  However, since $N/\tau$
is large and imaginary, we may make progress as in \cite{doreykumar} by
using the anomalous modular transformation
\begin{equation}\label{Etwoinvert}
E_2(z) = {1\over z^2} E_2\left(-{1\over z}\right)  + {6i\over \pi z} \ .
\end{equation}
from which we learn that the second term in \Eref{Etwos} dominates the
first and that it can be expanded in power of $e^{-2\pi iN/\tau} =
e^{-2\pi gN}$.  This is an expansion in a small quantity, and we need
only provide an interpretation for it.

This is not difficult to obtain, for an NS5-brane of the form $S^2$
times Minkowski space permits the string world-sheet to wrap the $S^2$,
producing instantons.  From the metric \eref{cmetric}
the proper area of the NS5-brane sphere times the
tension of the fundamental string is minimized by
\begin{equation}
(4\pi r_0^2) {R^2\over2 r_0 \rho_c} {1\over 2 \pi \alpha'}=
{R^4\over 2 \alpha'^2} = 2 \pi g N
\end{equation}
which is just as required to explain the expansion in the
superpotential.  As a check, note that if we repeat the calculation
for a vacuum with $q$ coincident spheres, both the area of the spheres
and the exponent in the field theory are reduced by a factor $q$. For
the Higgs vacuum, the metric and dilaton in Eqs.~\eref{hmetric} and
\eref{hdilaton} imply the area of the D5-sphere times the tension of a D1
brane goes at small $\rho$ to
\begin{equation}
\frac{R^4 }{2g^2 \alpha'^2} = 2\pi N/g \ .
\end{equation}
Since the Higgs vacuum superpotential is proportional to \cite{doreykumar}
\begin{equation}\label{Etwosh}
E_2(\tau) - {N} E_2\left(N\tau\right)
\end{equation}
we may again interpret the second term (now much smaller than the
first term, except for its leading $\tau$-independent contribution)
as an expansion in D1-instantons wrapping the D5-brane two-sphere.

It would be interesting to find a string theory interpretation for the
coefficients in the expansion of of the superpotential, and especially
for the anomalous term in the modular transformation of $E_2$.

\subsection{Glueballs and Other Particle States}

The spectrum of states in this theory is complicated, and we
do not yet have a physical understanding of its features.
We will outline its structure and point out some puzzles and
problems which must be solved in future.

First, there is already surprising structure in the \nfour\ theory in
the vacuum with $Z$ given by \Eref{zwarp}, where the D3-branes form a
2-sphere of radius $r_0$.   The typical warp factor
in the region $r \sim r_0 \sim mN(g)\alpha'$ is $Z \sim R^4/r_0^4$.
A supergravity state then has typical
$k_\mu$ given by~\cite{amandajoe}
\begin{equation}
G^{\mu\nu} k_\mu k_\nu \sim G^{mn} k_m k_n
\end{equation}
so that
\begin{equation}
k_\mu \sim Z^{-1/2} k_m \sim
\frac{r_0}{R^2}
\sim m g^{\pm 1/2} N^{1/2} \ . \label{coulombstates}
\end{equation}
where the minus (plus) sign applies for the D3-sphere radius
appropriate to the Higgs (confining) vacuum.  Immediately we have a
puzzle.  The semiclassical field theory of this \nfour\ vacuum would have
led us to expect physics from the $W$-bosons whose
masses lie between the scales $m$ and $Nm$.  There are also
monopoles of masses $m/g$ and $Nm/g$.  These states
are present on the supergravity side as F- and D-strings stretched between
the D3-branes.  But there is no sign
of gauge theory states with masses given in the previous equation.
The situation is not improved by consideration of
excited string states in the bulk, for which one has similarly
\begin{equation}
G^{\mu\nu} k_\mu k_\nu \sim 1/\alpha' \label{kmass}
\end{equation}
and
\begin{equation}
 k_\mu \sim Z^{-1/4} \alpha'^{-1/2} \sim
\frac{r_0}{R \alpha'^{1/2}}
\sim m g^{\pm 1/2} N^{1/2} \times (gN)^{1/4}\ ,\label{coulombstring}
\end{equation}
an odd-looking scale.

Now, what changes when the D5- or NS5-brane charge is added?  For the
D5-brane, essentially nothing happens to these arguments.  This is
despite the fact that a magnetic flux tube has formed, with a tension
whose square root is given by \Eref{coulombstates} with the minus sign.
Presumably the details shifts around slightly, and of course the
massless photons of the D3-branes now develop mass of order $m$, but
apparently the spectrum is otherwise little changed.

For the NS5-brane, the situation is more subtle.  The electric flux
tube has a tension whose square root is given, as in the D5 case by
\Eref{coulombstates}, now with the plus sign.  But in addition, unlike
the D5-brane, the NS5-brane has a throat region.  If we consider a
vacuum with several coincident NS5-branes, then the throat region is
reasonably well understood, since it can be described in conformal
field theory in the region where the string coupling is small
\cite{CHS2}. From this it is known that supergravity states get
string-scale masses from the coupling to the throat geometry. 
Using the metric~\eref{nshell} in the calculation~\eref{kmass} gives
\begin{equation}
k_\mu \sim r_0^{1/2} \rho_{\rm c}'^{1/2} R^{-1} \alpha'^{-1/2} \sim
\frac{r_0}{R^2}
\sim m g^{1/2} N^{1/2}\ .
\end{equation}
This applies to both supergravity and excited string states.  Notice
that this is the same scale as for supergravity states in the bulk,
and for the string tension.

Unfortunately, the confining vacuum has only one NS5-brane, and there
is a long-standing controversy over this object, reflecting the
difficulty of doing any reliable calculations in its presence
\cite{throat}.  There are disputes over whether the throat and its
region of strong coupling even exist.  In any case, neither
supergravity nor conformal field theory is reliable, and we simply do
not know what the spectrum will do in this regime.  We note this may
hint at a profound obstacle to using string theory as a practical
computational tool in QCD.

In both the D5 and NS5 cases, there is one more class of light states,
arising from the massless gauge fields on the 5-brane.  Relative to
the bulk states, the magnetic field on the D5-brane reduces the
velocity of open string states by a factor of the dimensionless field,
$v \sim(N/g)^{-1/2}$: restoring $F_{\mu\nu} F^{\mu \nu}$ and $F_{\mu
a} F^{\mu a}$ to the brane action, one finds that the latter is
multiplied by $v^2$.  The mass gap is reduced by the corresponding
factor, and so is simply
\begin{equation}
k_\mu \sim m \ .
\end{equation}
This agrees with the classical result, since these are the $W$ bosons
and this is the scale of $SU(N)$ breaking.  Meanwhile, the NS5-brane
has a normalizable zero mode \cite{CHS}, which gives rise to the
massless vector required by $S$-duality to the D5-brane.  We assume
that this mode survives when D3-branes are dissolved in the NS5-brane.
Then
\begin{equation}
G^{\mu\nu} k_\mu k_\nu \sim G^{w^1 w^1} k_{w^1} k_{w^1}\ ,
\end{equation}
and
\begin{equation}
k_\mu \sim \frac{r_0 \rho_{\rm c}'}{R^2} k_{w^1} \sim \frac{\rho_{\rm
c}'}{R^2} \sim m  \ .
\end{equation}
Again we find a lighter branch of states localized at the brane.
By analogy to the Higgs vacuum, one might interpret them
as massive magnetic gluons, but in the NS system they are not open strings
but rather closed string states in localized wavefunctions on the throat.
Most of them carry $SO(3)$ quantum numbers and are therefore
not seen in \none\ Yang-Mills theory.  Some or all of
the remainder may mix with bulk states, as we discuss below.

Before doing so, we note that all of the masses we have
obtained in the confining vacuum are consistent with
't Hooft scaling, as they are proportional to $m$ times
a power of $gN$.    This is pleasing, although it means
of course that all of the states merge and mix as $gN$
is taken small, making any quantitative computations in this regime
essentially irrelevant for \none\ Yang-Mills.

Let us note one important interplay between the open and closed string
states. There would appear to be one unbroken $U(1)$ gauge group per
(noncoincident) brane, from the world-sheet gauge field on each brane.  For
an $SU(2)$ representation given by the sum of $k$ distinct irreducible blocks,
$SU(N)$ is broken to $U(1)^{k-1}$.  On the string side there are
$k$ D5-spheres and so apparently a $U(1)^k$.  In fact, one $U(1)$ should be
lifted by the coupling to the bulk states.\footnote{This was suggested by
E. Witten.}  We have not understood all the details, but will indicate the
ingredients.
There is a massless tensor
in
$3+1$ dimensions, with the field $B_{\mu\nu}$ independent of $x^m$.  Its
kinetic term is
\begin{eqnarray}
\int d^{10}x\, \sqrt{G} H_{\mu\nu\lambda} H^{\mu\nu\lambda}
= \int d^{4}x^\mu\,
\eta^{\mu\mu'} \eta^{\nu\nu'} \eta^{\lambda\lambda'}
 H_{\mu\nu\lambda} H_{\mu'\nu'\lambda'} \int d^6 x^m\, Z^2\ .
\end{eqnarray}
The $x^m$ integral converges both at the branes and at infinity, so this is
a discrete state.  By itself, the coupling of this field to $F_{\mu\nu}$ in
the Born-Infeld action would generate a mass for one linear combination of
$U(1)$'s via the Higgs mechanism.  However, the field $C_{\mu\nu}$ also
has a zero mode, seemingly lifting a second $U(1)$.  The actual story must
be more complicated, with the bulk Chern-Simons term playing a role,
because from the point of view of a single brane this would be
simultaneous electric and magnetic Higgsing, an impossibility.

\section{Extensions}

\subsection{Unequal Masses}

Now we consider the general \none\ case, three masses not necessarily
equal.  Examination of the classical $F$-term equations~\eref{fterm}
suggest use of the coordinates
\begin{eqnarray}
z^1 \* \sqrt{m_2 m_3}\, \chi \cos\theta \
,\nonumber\\
z^2 \* \sqrt{m_1 m_3}\, \chi \sin\theta
\cos\phi\ , \nonumber\\
z^3 \* \sqrt{m_1 m_2}\, \chi
\sin\theta \sin \phi \ . \label{ellips}
\end{eqnarray}
We will study the potential with the Ansatz that $\chi$ is constant and
also that $F_{\theta\phi} = \frac{1}{2}n \sin\theta$.  We
insert this into the potential~\eref{cdact}.  Noting that
\begin{eqnarray}
\det G_\perp \* 4\tilde m  |m_1 m_2 m_3 \chi^4| Z \sin^2\theta\ ,
\nonumber\\
\tilde m &\equiv& |m_1| \cos^2 \theta + |m_2| \sin^2 \theta \cos^2 \phi
+ |m_3| \sin^2 \theta \cos^2 \phi  \ , \nonumber\\
T_{mnp} x^m dx^n\wedge dx^p \* |m_1 m_2 m_3| (|m_1| + |m_2| + |m_3|) \chi
\bar \chi^2\ , \nonumber\\
T_{i\bar \jmath \bar k}\OL{T}_{\bar l j k}  z^i \bar
z^{\bar l} \* 2 \tilde m  |m_1 m_2 m_3 \chi^2|\ ,
\end{eqnarray}
the potential becomes
\begin{equation}
-\frac{S}{V} = \frac{4|m_1 m_2 m_3|}{ \pi g n
(2\pi\alpha')^4 }\frac{|m_1| + |m_2| + |m_3|}{3}  | M \chi^2 - 2\pi
\alpha' in\chi/2\sqrt{2} |^2\ .
\end{equation}
This has a supersymmetric minimum at $\chi = 2\pi
\alpha' in/2\sqrt{2}M$.  The two-sphere is now an ellipsoid with axes
\begin{equation}
\frac{\pi \alpha' n}{|M|} \sqrt{|m_2 m_3|}\ ,\quad
\frac{\pi \alpha' n}{|M|} \sqrt{|m_1 m_3|}\ ,\quad
\frac{\pi \alpha' n}{|M|} \sqrt{|m_1 m_2|}\ .
\end{equation}

 When $m_3\to 0$ with $m_1=m_2$ fixed, we obtain \ntwo\ Yang-Mills;
here the ellipsoid degenerates into a line of length $m_1$.  This is
very easy to understand, classically, for the Higgs vacuum.  \ntwo\
Yang-Mills with a massive hypermultiplet has a moduli space, which
classically is just given by the positions of D3-branes (suitably
modified so that they can only move in two dimensions --- this remains
to be understood in our present context.)  On this space is a single
point where there are $N-1$ massless {\it electrically} charged
particles, from the hypermultiplet.  At this point, the $N$ D3-branes
are arranged in a line.  From the classical equations, it is easy to
see that an \ntwo\ breaking mass parameter for the vector multiplet
causes this line to become a noncommutative ellipsoid analogous to the
sphere of \cite{ktsphere}.  In addition, the electromagnetic dual of
this transition corresponds to a well-known fact, both in field theory
\cite{nsewone,klytaps,mdss} and in MQCD
\cite{ewMQCD,ucbmqcd,hsz,deBOz}, concerning the breaking
of pure \ntwo\ Yang-Mills to \none. The moduli space of \ntwo\ has an
associated Seiberg-Witten auxiliary Riemann surface of genus $g$.
There are $N$ special isolated points on the moduli space where $N-1$
mutually local dyons become massless, and the Riemann surface
completely degenerates.  In this degeneration, the $N$ handles of the
surface join along a singular line segment.  The addition of a mass
parameter breaking \ntwo\ supersymmetry to \none\ engenders a
transition whereby the $N$ handles join to form a single surface of
genus zero: the line segment opens up into a closed curve. In M theory
this is represented by a multigenus M5-brane making a transition to a
genus zero M5-brane.  Here we see signs of a similar phenomenon; the
3-branes which represent the moduli space of the \ntwo\ theory
presumably align along a line segment, then join and expand to form an
NS5-brane ellipsoid.

By contrast, when $m_1=m_2\to 0$ with $m_3$ fixed, the ellipsoid
becomes a disk while its overall size shrinks to zero.  In the field
theory, one expects an infrared fixed point, and indeed the
supergravity should go over smoothly to the kink solution of
\cite{ir3} and at small $r$ to the ten-dimensional space of
\cite{pwtend}. Presumably the $G_{\it 3}$ background in \cite{pwtend} is
related to the linearized one we have obtained in \Eref{susyt},
although we have not checked this.

It would be interesting to understand these
connections in more detail, but we should note that our approximations will
break down in these limits.  We have seen that the 5-brane shell has
a finite thickness, and we need this to be less than the shortest axis of
the ellipsoid for our linearized approach to be consistent.

We note also that most of our results generalize easily to this case.
The superpotential, the domain wall tensions and the condensates, are
all related, as is the dipole moment of the 5-brane, to the volume of
the ellipsoid $\propto m_1m_2m_3$.  The flux tubes will presumably
show signs of the extra metastable Regge trajectories seen in
\cite{mdss} by localizing on the ellipsoid, along the lines observed
in MQCD in \cite{hsz}.

\subsection{\nzero}

Let us make a few brief remarks about the nonsupersymmetric case,
$m_4 = m'$ with $m_1 = m_2 = m_3 = m$ kept equal.  The potential is now
\begin{equation}
-\frac{S}{V} = \frac{4}{ \pi g n
(2\pi\alpha')^4 }  \Biggl\{
|M|^2 |z|^4 + \frac{ 2\pi
\alpha' n}{3\sqrt{2}} {\rm Im}\Bigl[ (3 m z \bar z^2 + m' z^3) \OL M
\Bigr] +
\frac{ (2\pi
\alpha' n)^2 }{8}O(z^2)
\Biggr\}\ .
\end{equation}
The quadratic term depends on the boundary conditions, as discussed at the
end of section~IV.A.  Its general form is given by
\begin{equation}
O(z^2) = \frac{1}{3} |z|^2 \sum_{i=1}^4 |m_i|^2  + (L=2)\ .
\label{nzquad}
\end{equation}
In the absence of supersymmetry we must make a particular choice of
the $L=2$ harmonic $\mu_{mn}$, defined in \Eref{ambig}, which
represents a traceless combination of masses for the scalar bilinears.
This harmonic reduces the masses of some of the scalars, and if too
large it can cause the gauge symmetry to break.  However, it
represents an adjustable parameter in the Lagrangian, so we are
completely free to choose it in such a way that it preserves a stable
vacuum, assuming such a choice exists.  Further, if we maintain
$SO(3)$ invariance, the choices are greatly reduced.

Since $\mu_{mn}=0$ in the $SO(3)$-invariant supersymmetric case, and
since the vacua in the supersymmetric case are stable, it is evident
that for small $m'/m$ the continued use of $\mu_{mn}=0$ leads to
stable vacua at nonzero $z$.  Of course, the vacua need no longer be
degenerate.  Whether the single NS5-brane is the preferred vacuum is
less obvious, although it seems likely, since it is an extreme case
among the vacua.  In such a vacuum the spectrum would be altered and
the stable domain walls would be lost, but most of the other features
of the confining vacuum --- flux tubes, baryon vertices, condensates
and instantons --- would be qualitatively unchanged.  The new features
would be the appearance of gluon condensates, such as $\tr F^2$, which
must be zero in the supersymmetric case, and nontrivial dependence on
the phase of $m'/m$.

While the situation is less clear when $m'\sim m$, there are reasons
to expect, on purely physical grounds, that a confining vacuum does in
fact exist.  It seems a worthwhile challenge to seek it in
supergravity.

\subsection{Orbifolds and QCD}

Many authors have studied supersymmetric and nonsupersymmetric
orbifolds of the \nfour\ Yang-Mills theory, its D3-brane
representation and its supergravity dual.  (A list of references may
be found in \cite{magoo}.)  Here we briefly consider two orbifolds of
the results we have obtained above.

In the \nfour\ theory, a $\ZZ_2$ orbifold on four coordinates leaves
an \ntwo\ $SU(N)\times SU(N)$ theory with $({\bf N},\OL{\bf N}) +
(\OL{\bf N},{\bf N})$ hypermultiplets.  We can combine this action
with the mass perturbation in two distinct ways.  The first is a
simultaneous rotation by $\pi$ in the 45- and 78-planes.  This is part
of the $SU(2)$ that acts on the chiral superfields, and so commutes
with the \none\ supersymmetry and leaves it unbroken.  The second is a
rotation by $\pi$ in the 45-plane and $-\pi$ in the 78-plane.  Since
it differs from the first rotation by a $2\pi$ rotation in the
78-plane, we can think of it as the first rotation times $(-1)^F$,
with $F$ being spacetime fermion number.  The first rotation commutes
with the \none\ supersymmetry generator, so the second anticommutes
with it and leaves a nonsupersymmetric theory.  Note that the two
rotations are conjugate to one another, and so in the absence of the
mass term would give equivalent theories.

We are assuming that the $\ZZ_2$ acts on the Chan-Paton factors as
\begin{equation}
\biggl[ \begin{array}{cc} I_N & 0 \\ 0 & -I_N \end{array} \biggr]\ .
\end{equation}
More generally the blocks could be different sizes, leading to an $SU(N)
\times SU(M)$ gauge theory; this gives rise to a twisted state tadpole
and so is more complicated \cite{quiver}.
For the supersymmetric orbifold, the massless fields that survive are
\begin{eqnarray}
A_\mu \* \biggl[ \begin{array}{cc} A_\mu & 0 \\ 0 & A'_\mu \end{array}
\biggr]\ ,\qquad
\Phi_3 = \biggl[ \begin{array}{cc} \Phi & 0 \\ 0 & \Phi' \end{array}
\biggr]\ ,\nonumber\\
\Phi_1 \* \biggl[ \begin{array}{cc} 0 & Q_1 \\ \tilde Q_1 & 0
\end{array}
\biggr]\ ,\qquad
\Phi_2 = \biggl[ \begin{array}{cc} 0 & Q_2 \\ \tilde Q_2 & 0 \end{array}
\biggr]\ . \label{spectrum}
\end{eqnarray}
Thus there are $SU(N) \times SU(N)$ vector multiplets, chiral
multiplets $\Phi$ and $\Phi'$ in the respective adjoints, and
bifundamental chiral multiplets $Q_1$, $Q_2$ in the $({\bf N},\OL{\bf
N})$ and $\tilde Q_1$, $\tilde Q_2$ in the $(\OL{\bf N},{\bf N})$.
The \none\ superpotential is $W\propto \phi (Q_1\tilde Q_2 - Q_2\tilde
Q_1) + \hat\phi(\tilde Q_1 Q_2-\tilde Q_2 Q_1)$.  The $AdS$
description of this theory is $AdS^5 \times S^5/\ZZ_2$.  There is a
fixed plane at $x^4=x^5=x^7=x^8=0$, which in the supergravity is
$AdS_5\times S^1$ where the second factor is a fixed $S^1$ on the
$S^5$.  We may preserve supersymmetry and add the superpotential $m
(\phi^2+\hat\phi^2 + \tilde Q_1 Q_1+\tilde Q_2 Q_2)$.  In the gravity
description, this is simply the perturbation we studied earlier.  The
low energy theory consists of two separate \none\ Yang-Mills theories
with no massless matter.  There are two gluino bilinears operators,
one even and one odd under the $\ZZ_2$.

The second rotation differs by $(-1)^F$, so the action on the bosons
is the same: they are the same as for the theory~\eref{spectrum}.  The
action on the fermions is opposite, so the fermionic partners are all
absent, while fermions appear in all the blocks with 0's.  Thus the
low-energy field theory in this case is nonsupersymmetric $SU(N)\times
SU(N)$ gauge theory with a Dirac fermion $\Psi$ in the $(\OL{\bf N},{\bf
N})$.  This is a $\ZZ_2$ orbifold of \none\ Yang-Mills
\cite{schmaltz}. It has no fermion bilinear odd under the $\ZZ_2$.
Note that for $N=3$ it is QCD with three massless quarks and with the
vector $SU(3)$ of the flavor group gauged.  This gauging removes all
but a $\ZZ_3$ axial symmetry, or more generally a $\ZZ_N$.

In both cases, the renormalization group flow is from an
$SU(N)\times SU(N)$ \ntwo\ supersymmetric theory in the ultraviolet to
a gauge theory with massless fermions but no massless scalars.  In the
limit $m\to\infty$, with $N$ and $\Lambda^3 \equiv
m^3\exp(-8\pi^2/g^2_{{\rm YM}}N)$ fixed, the supersymmetric orbifold
has $N^2$ vacua; these confining vacua, in which the two gluino
bilinears have separate condensates, should survive to the finite $m$
case.  The presence of the massive matter in the bifundamental
representation assures, however, that there is only one type of $\ZZ_N$
flux tube, and only one type of baryon vertex.  Only the condensates
can distinguish vacua separated by $N$ consecutive domain walls.  By
contrast, the nonsupersymmetric orbifold has only one fermion
bilinear, with a consequently unique expectation value.  As we
mentioned, there is an accidental $\ZZ_N$ axial symmetry in the limit
$m\to\infty$, $N,\Lambda$ fixed {\it if $\Psi$ is held
massless} --- more on this below.  In this limit we expect the $\ZZ_N$
to be broken by a fermion bilinear (note this is consistent with the
dynamics of physical QCD) so we expect $N$ confining vacua.  Again
there is only one $\ZZ_N$ string and one baryon vertex.  (In both
theories we expect physical baryon states; however they are rather
similar in the two cases, differing in mass only slightly.)

On the string side of the duality, all of our brane configurations are
invariant under $\ZZ_2$ and so survive in the orbifold theories.  We
do not see a mechanism for new phases to arise geometrically, so the
extra vacua in the supersymmetric case are presumably associated with
expectation values of fields in the twisted sector.  Certainly the
difference of the gluino condensates (and corresponding scalar
operators) is in the twisted sector, while
the sum is discussed in section~V.F.  The two supergravity orbifolds
differ in the behavior of fermionic fields under the reflection, so
that the spherical harmonics, and spectra, will be different in the
two cases.  The physics involving strings and baryons is essentially
the same as before, but it would be interesting to understand how the
domain walls are different.

The nonperturbative condensate in the unorbifolded theory
survives, as a gluino bilinear expectation value, to small $gN$, where
it can break the $\ZZ_N$ nonanomalous $R$-symmetry of \none\ Yang-Mills.
Unfortunately we cannot make the same claim for the condensate in the
nonsupersymmetric orbifold. Only if $\Psi$ is massless for
$m\to\infty$, $N$ and $\Lambda$ fixed, does the theory have a discrete
axial symmetry, which can be broken by a $\bar\Psi\Psi$ condensate.
However, while a gluino is always massless by supersymmetry, no such
symmetry protects the fermion $\Psi$.  We must therefore assume that
$\Psi$ can obtain a mass in perturbation theory through nonplanar
graphs.  Fine tuning is required to obtain the axial symmetry and the
massless fermion at small $gN$, and thus any connection between chiral
symmetry breaking in QCD and the large-$gN$ condensate is
tenuous.

As an aside, we emphasize the physical interest of these questions.
Nonsupersymmetric $SU(N)\times SU(N)$ with a single massless
bifundamental fermion, treated as a function of $N$ and the ratio of
gauge couplings, is a very interesting theory worthy of further
attention.  First, if the gauge couplings are very different, the
physics between the two strong coupling scales approximates physical
QCD.  Second, the theory exhibits both confinement and chiral symmetry
breaking, with the breaking of a discrete axial symmetry and
interesting domain walls.  As such, it closely resembles \none\
supersymmetric Yang-Mills theory.  Third, in contrast to \none\
Yang-Mills, the low-energy theory has only vector-like fermions and
can be investigated with relative ease on the lattice.  To our
knowledge it has not been previously studied.  Of course, lattice
studies of weakly  broken \none\ Yang-Mills are not impossible \cite{montvay};
and one may hope, by comparing the $SU(N)\times SU(N)$ theory with a
{ massive} bilinear fermion to broken \none\ Yang-Mills, to study
the extent to which nonperturbative physics survives orbifolding in
the large $N$ limit.  In short, this theory is physically
interesting, tractable on the lattice, qualitatively related to
supersymmetric Yang-Mills theory even for small $N$, and perhaps
quantitatively related to it at large $N$.

\section{Discussion and Future Directions}

As with all dualities, our work has implications in both directions
--- for supergravity and string theory, and for gauge theory.

\subsection{Strings, gravity, and singularities}

We have found one more example of a recurrent pattern, the resolution
of a naked singularity by brane physics.  Earlier examples are the
nonconformal D$p$ geometries \cite{IMSY}, the Coulomb branch
singularities \cite{coul1,coul3,coul4}, and the enhancon \cite{jpp}.

It is not clear how earlier work on the \noneplus\ theory is related
to ours, because it was all in the context of five-dimensional
supergravity.  The solutions of ref.~\cite{gppz} might lift to ours,
but this requires that a great deal of physics, the entire brane
configuration, be hidden in the `consistent oxidation' of the
five-dimensional solution.  We note that the solutions \cite{gppz} all
have constant dilaton while our 5-branes will produce a locally
varying dilaton even if not a dilaton monopole moment.  As far as is
known, the oxidation cannot produce such an effect.  In
ref.~\cite{gubs} a general criterion was proposed for identifying
physically acceptable naked singularities.  Again this was expressed
in terms of the five-dimensional theory and so cannot be applied to
our solutions without substantial additional work.

There are a number of obvious loose ends in our work.  We have not
obtained the full supergravity background.  We had the good fortune
that we could obtain the relevant physics by working only to first
order, and partly to second order, in the mass perturbation.  It
remains at least to solve the supergravity equations to second order.
We have observed that the equations of the supergravity have many
simplifications in our situation, but we have not understood their
origin.  It seems extremely likely that these will extend to the full
second order calculation.  It may even be possible, with the guidance
from our approximate solution, to find the exact supergravity
solutions.  A related question is the analysis of the supersymmetry
properties of our solutions, to establish why the supersymmetry is
preserved.  This involves the 5-brane world-volume as well as the bulk
supergravity.

We should remind the reader that we have not fully explained a key
point, namely the fact that there are far more brane configurations
than there are field theory vacua: many brane configurations represent
the same vacuum.  For example, we have not explained why the massive
vacuum with a $(1,q)$ 5-brane is the same as that with a $(1,q+N)$
5-brane; we have merely shown that the question never arises purely
within the supergravity regime.  To understand the transition between
these and other descriptions as $\tau$ is varied remains an
interesting challenge.  A related issue is the minimum of the brane
potential at $z=0$.  This is outside the range of validity of our
approximation, but would naively correspond to an unexpanded brane and
so a singular spacetime.  Since there exist expanded brane
representations for all ground states, we presume that a correct
interpretation involves transmutation into a different brane
configuration.  A complete accounting of these transitions is greatly
to be desired.

Another related fact is that many brane configurations (such as $p$
coincident D5-branes, $p\ll\sqrt N$) represent multiple vacua, which
are only split from one another by strong coupling $SU(p)$ physics
that we cannot see in supergravity because $p$ is small.  This physics
will have to be understood separately, and perhaps is only described
by field theory.

Perturbed $AdS$ spacetime is relevant to the generation of hierarchies
in Randall-Sundrum compactification~\cite{RS,verl}.  In the latter
paper it was argued that a large number of D3-branes localized in a
compact space will generate an effective $AdS$ throat.  That work was
in the context of an \nfour\ compactification and so the throat was
stable: no relevant perturbation exists.  The situation becomes more
interesting in more realistic cases.  If the D3-branes are localized
on a space that leaves \none\ unbroken, and the compactification is
generic, the $G_{\it 3}$ mode that we have considered will have a
nonzero boundary value, proportional to $m$, which is of order order
one in four-dimensional Planck units.  Since this is a relevant
perturbation it becomes nonlinear essentially immediately, and the
throat disappears.  If for some reason the perturbation is anomalously
small, then of course, as in any supersymmetric field theory, the
ratio $m/m_{\rm Pl}$ is quantum mechanically stable and the throat
will be larger.  Our work shows that the incipient throat is capped by
an expanded brane.  It is also possible to avoid the instability
using, for example, a discrete symmetry $G_{\it 3} \to -G_{\it
3}$.\footnote{This has also been noted by O. Aharony.}

It would be most interesting to study marginally relevant
interactions.  These may not exist in the \nfour\ theory, but are
present in various elaborations.  On general grounds, not requiring
supersymmetry, such perturbations would naturally become nonlinear
only well down the $AdS$ throat, where again an expanded brane would
presumably form. This behavior would bear some similarity to the
hierarchy mechanism of ref.~\cite{GW}.  Examples of such interactions
were considered in \cite{klebnek,klebtseyt}, and there is strong
reason to expect an expanded brane there as well.

Our work bears directly on recent proposals regarding the vanishing of
the cosmological constant~\cite{stanford}, which involve a naked
singularity in the compact dimensions.  There have been many
criticisms, published and unpublished, of the assumptions made in
ref.~\cite{stanford} regarding the properties of the singularity, but
we can add to these the specific example of what string theory does in
our case.  In terms of our notation from section~III.B, the authors of
refs.~\cite{stanford} assume that the parameter $b$ can take arbitrary
values.  That is, they integrate in the coordinate $r$, and assume
that any singularity encountered gives an acceptable spacetime.
Further, they must assume that $b$ is a fixed parameter, rather than a
dynamical quantity. As we see from our discussion, this is
inconsistent with the requirements of $AdS$/CFT duality, and it is not
what happens in our case: $b$ takes discrete values, depending on the
particular vacuum, and can make dynamical transitions from one value
to another.

Thus the singularity in the end is replaced by an ordinary physical
object, like a hydrogen atom.  For hydrogen, too, one can integrate
the wave equation inward in $r$ for any energy, obtaining a generally
singular solution.  The experimental discreteness of
the spectrum indicates that nature abhors a generic singularity.

It is notable that our system has a large number of ground
states,\footnote {N. Arkani-Hamed, S. Dimopoulos, J. Feng, S. Gubser,
N. Kaloper, E. Silverstein and F. Wilczek have emphasized the
importance of this point.}  of order $e^{\sqrt{N}}$.  In the
supersymmetric case these are all degenerate, but once supersymmetry
is weakly broken they will form a closely spaced, near-continuum of
discrete states.  If a singularity is of this type, the system may
have vacua of exceedingly small cosmological constant, and the
mechanism of \cite{brownteit} may be realizable.  In order for such a
mechanism to solve the problem, these states must be sufficiently
metastable, and there must be a dynamical mechanism to select a state
with a small net cosmological constant (the same problems are in any
event present in the continuum case); for further
discussion of the latter issue see~\cite{raph}.

Finally, we should note that the Myers dielectric effect will arise in
many other situations, such as perturbations of other conformal and
nonconformal theories.  As one example, consider the perturbation of
the BFSS matrix theory given by the dimensional reduction of the
\noneplus\ mass term.  This has been used as a means of analyzing the
structure of the matrix theory bound state~\cite{matpert}.  Now we see
that this deformation has a physical interpretation in its own right:
it is the matrix theory for M theory with a nontrivial boundary
condition on the 4-form field strength.  In this background the
graviton will blow up into a finite-sized M2-brane sphere.  This same
mechanism, the Myers expansion of a highly boosted graviton in a
background field strength, has recently led to a remarkable
explanation of the stringy exclusion principle \cite{SEP}.

\subsection{Gauge theory}

The brane solution gives a beautiful representation of a confining
gauge theory and many of the associated phenomena.  There are many
artifacts of the massive \nfour\ matter, so this is still far from
QCD, but we emphasize that it is a confining four-dimensional gauge
theory in its own right.  We should perhaps note that we use the
Maldacena duality freely in the entire range $1/N < g < N$.
Discussion often focuses on large $N$ with fixed $gN$, but this is
only one part of the interesting range.

Without too much additional work, it should be possible to understand
the breaking of \nfour\ to \ntwo\ Yang-Mills.  Much of the moduli
space, including the part corresponding to the repulson singularity
studied in \cite{jpp}, should be visible in supergravity.  We might
hope that, as in \cite{ewelliptic}, the Seiberg-Witten Riemann surface
will appear in a natural way.  However, where the surface degenerates
(at points with light charged particles) we expect there will be
physics lying outside the supergravity regime.  Since the transition
from \ntwo\ to \none\ is described simply in MQCD, we may hope that
this breakdown will be described simply using semiclassical branes,
giving a picture of the physics studied in \cite{mdss}.  We also
expect a breakdown at Argyres-Douglas fixed points \cite{AD}, where
the physics, presumably a supergravity kink connecting one $AdS$ region
to another, may not be described in a linear approximation.

Once the supersymmetric gravity background is completely understood,
it should also be possible to fully explore the nonsupersymmetric
case.  Since even the confining vacuum is a small perturbation of an
\nfour\ Coulomb branch configuration, we do not believe that small
breaking of supersymmetry is likely to have a large impact on the
theory.  New condensates and new $\theta$ dependence, and a loss of
the degeneracy between vacua, should certainly be seen, but otherwise
we see no reason why the nonsupersymmetric confining vacuum must
appear much different from the supersymmetric one.  However, while
this will be true if the supersymmetry breaking scale $m'$ is small
compared to $m$, it will surely be false if $m'\gg m$.  The
quantitative question of where the transition lies, as a function of
$gN$, remains to be explored.  Even assuming, however, that $m\sim m'$
is within the supergravity regime, this does not make the study of
pure nonsupersymmetric Yang-Mills any easier, since it is a long way
from $gN\gg1$ to the required regime.

Assuming the nonsupersymmetric case can be studied to a degree, an
obvious next step in studying the gravity--gauge theory correspondence
is to add massless charged fermions. It would be most interesting to
see chiral symmetry breaking and to determine if and when the
properties of pions lie within the supergravity regime.  Adding
supersymmetric charged matter is unfortunately not of great use, since
$SU(N)$ with $N_f\ll N$ massless chiral multiplets in the fundamental
plus antifundamental representation has no stable vacuum, while if
$N_f\sim N$ there is as yet no dual string description.  The low
flavor case might be studied by taking type IIB on an orientifold,
which gives an \ntwo\ $Sp(N)$ gauge theory with a hypermultiplet
${\cal A}$ in the antisymmetric tensor representation and four ${\cal
F}_i$ in the fundamental representation \cite{QandA,AFM}.  As in
\noneplus, the ultraviolet theory is finite, and could be perturbed by
supersymmetric masses $m$ for the adjoint chiral multiplet and ${\cal
A}$, and $\mu_i$ for the ${\cal F}_i$.  When $\mu_i\sim m$, the
physics probably resembles \noneplus, but the theory will exit the
supergravity regime as any one of the $\mu_i$ go to zero.  Breaking
the supersymmetry, leaving only the fermions in ${\cal F}_i$ and the
gauge bosons massless, would be a challenge, but might be tractable.

In this paper we have not pursued the obvious and important
connections of our work with noncommutative field theory in two
additional dimensions.  Similar relations are present already in MQCD.
We are especially intrigued that the confining flux tube of large $N$
Yang-Mills is a nearly-BPS instanton string of a sphere-compactified
six-dimensional noncommutative gauge theory (supersymmetric or not.)
Note that the baryon vertex is also an interesting object in this
theory. The massive vacua with multiple coincident 5-branes, and their
solitons, may be of considerable interest for the study of nonabelian
noncommutative field theory.  This may also be true for
lower-dimensional cases, where the classical vacua are still described
by spherical D(p+2) branes.

It is important to note that many of our results bear some resemblance
to those seen in MQCD, which is an unusual compactification of the
$(2,0)$ theory on an M5-brane.  Both the breaking of \nfour\ to \ntwo\
Yang-Mills \cite{ewelliptic} and the breaking of \ntwo\ to \none\
\cite{ewMQCD} have been considered, although the full \noneplus\ model
has not been constructed in MQCD.  It is interesting to consider some
of the similarities and differences with the supergravity dual of
\noneplus.  In our picture, the vacuum is represented by an NS5/D3
hybrid; in MQCD the vacuum is given by a multisheeted M5-brane, which,
when the radius $R_{10}$ of the M-theory circle is small, is an NS5/D4
hybrid.  Second, the flux tubes in \noneplus\ are F1 strings bound to
the NS5/D3 hybrid; in MQCD they are M2-branes bound to the M5 brane.
The baryon vertex in MQCD is an M2 brane which extends off of but has
a boundary on the M5 brane, analogous to our D3 brane whose boundary
is the NS5-brane two-sphere.  Domain walls are in both approaches are
5-branes mediating transitions between different 5-brane vacua.  There
are a number of important differences that this list understates; but
the most interesting difference, perhaps, is the instantons at large
$N$.  As we have seen, the fractional instantons noted in
\cite{brodie} are resummed at large $gN$ into strings wrapping the
NS5-brane two-sphere.  All of these connections hint at the usual
duality between M theory on a torus and type IIB string theory, but
the connection between the two pictures is not as simple as this,
given the absence of a torus in both constructions.  It would be
interesting to make these connections precise.

One of the most interesting aspects of the \noneplus\ theory, and any
similar gravity dual of \nfour\ broken to \ntwo, is that it can be
used to great effect to understand more deeply the connection of field
theory with gravity and string theory.  Our work and that of
\cite{doreykumar,ADK} points in this direction.  The holomorphic
properties of these field theories can be completely understood using
field theory and/or M theory, as in \cite{rdew,ewelliptic,doreya}, for
all values of $g$ and all values of $gN$. Consequently, one can
distinguish clearly those regimes of parameter space in which the
theory is well-described by electric variables, by magnetic variables,
or by IIB string variables respectively.  The various universal and
quasiuniversal physical properties of the theory are given different
descriptions in each regime (see for example our discussion of the
instanton expansion of the superpotential, section~VI.G,) and there
are surely many things to be learned by studying how one regime
converts to another.  In particular, the possibility of {\it
quantitatively} matching one regime to another deserves attention.
It is probable that this will be required if one is to study QCD,
which lies mostly outside the supergravity regime.  If such matching
is only possible for holomorphic and BPS quantities, as so far seems
likely, it poses yet another obstacle in the quest for a description
of the strong interactions.  However, even if the goal of doing
calculations in large-$N$ QCD, using string variables, is shown to be
unrealizable, the fact that field theories exhibit behavior which
differs greatly from the physics we have observed so far in nature is
of potentially great importance, in that it may open new avenues for
solving old problems.

\acknowledgements{}

We would like to thank O. Aharony, N. Arkani-Hamed, I. Bena,
M. Berkooz, K.  Dasgupta, S. Dimopoulos, E. Gimon, S. Gubser,
A. Hashimoto, S. Hellerman, N.  Itzhaki, D. Kabat, N. Kaloper,
R. Myers, N. Nekrasov, K. Pilch, L. Randall, S.  Sethi,
E. Silverstein, W. Taylor, N. Warner, E. Witten, and many participants
of the ITP Program on Supersymmetric Gauge Theories for helpful
discussion of various aspects of this work.  M.J.S. is especially
grateful to D. Kabat for discussions and assistance at early stages of
this work. We also especially thank O. Aharony for pointing out important
deficiencies in our discussion of the superpotential and the condensates.
The work of J.P. was supported by National Science
Foundation grants PHY94-07194 and PHY97-22022. The work of M.J.S. was
supported by National Science Foundation grant NSF PHY95-13835 and by
the W.M. Keck Foundation.


\begin{references}
% LATEX(US)
\bibitem{tHooftlargeN}
G.~'t Hooft,
%``A Planar Diagram Theory For Strong Interactions,''
Nucl.\ Phys.\  {\bf B72}, 461 (1974).
\bibitem{maldacon}
J.~Maldacena,
%``The large N limit of superconformal field theories and supergravity,''
Adv.\ Theor.\ Math.\ Phys.\  {\bf 2}, 231 (1998)
[hep-th/9711200].
\bibitem{gppz}
L.~Girardello, M.~Petrini, M.~Porrati and A.~Zaffaroni,
%``The supergravity dual of N = 1 super Yang-Mills theory,''
hep-th/9909047.
\bibitem{cvew}
C.~Vafa and E.~Witten,
%``A Strong coupling test of S duality,''
Nucl.\ Phys.\  {\bf B431}, 3 (1994)
[hep-th/9408074].
\bibitem{rdew}
R.~Donagi and E.~Witten,
%``Supersymmetric Yang-Mills Theory And Integrable Systems,''
Nucl.\ Phys.\  {\bf B460}, 299 (1996)
[hep-th/9510101].
\bibitem{mjsnotes}
M.~J.~Strassler,
%``Messages for QCD from the superworld,''
Prog.\ Theor.\ Phys.\ Suppl.\  {\bf 131}, 439 (1998)
[hep-lat/9803009].
\bibitem{doreya}
N.~Dorey,
%``An elliptic superpotential for softly broken N = 4 supersymmetric
%Yang-Mills
%theory,''
JHEP {\bf 9907}, 021 (1999)
[hep-th/9906011].
\bibitem{doreykumar}
N.~Dorey and S.~P.~Kumar,
%``Softly-broken N = 4 supersymmetry in the large-N limit,''
JHEP {\bf 0002}, 006 (2000)
[hep-th/0001103].
\bibitem{lattice}
M.~J.~Strassler,
%``QCD, supersymmetric QCD, lattice QCD and string theory: Synthesis on  the
%horizon?,''
Nucl.\ Phys.\ Proc.\ Suppl.\  {\bf 73}, 120 (1999)
[hep-lat/9810059].
\bibitem{highTa}
E.~Witten,
%``Anti-de Sitter space, thermal phase transition, and confinement in  gauge
%theories,''
Adv.\ Theor.\ Math.\ Phys.\  {\bf 2}, 505 (1998)
[hep-th/9803131]
\bibitem{highTb}S.~Rey, S.~Theisen and J.~Yee,
%``Wilson-Polyakov loop at finite temperature in large N gauge theory and
%anti-de Sitter supergravity,''
Nucl.\ Phys.\  {\bf B527}, 171 (1998)
[hep-th/9803135].
\bibitem{highTc}
A.~Brandhuber, N.~Itzhaki, J.~Sonnenschein and S.~Yankielowicz,
%``Wilson loops, confinement, and phase transitions in large N gauge  theories
%from supergravity,''
JHEP {\bf 9806}, 001 (1998)
[hep-th/9803263].
\bibitem{grossooguri}
D.~J.~Gross and H.~Ooguri,
%``Aspects of large N gauge theory dynamics as seen by string theory,''
Phys.\ Rev.\  {\bf D58}, 106002 (1998)
[hep-th/9805129].
\bibitem{myers}
R.~C.~Myers,
%``Dielectric-branes,''
JHEP {\bf 9912}, 022 (1999)
[hep-th/9910053].
\bibitem{ktsphere}
D.~Kabat and W.~I.~Taylor,
%``Linearized supergravity from matrix theory,''
Phys.\ Lett.\  {\bf B426}, 297 (1998)
[hep-th/9712185].
\bibitem{bfss}
T.~Banks, W.~Fischler, S.~H.~Shenker and L.~Susskind,
%``M theory as a matrix model: A conjecture,''
Phys.\ Rev.\  {\bf D55}, 5112 (1997)
[hep-th/9610043].
\bibitem{coul1}
P.~Kraus, F.~Larsen and S.~P.~Trivedi,
%``The Coulomb branch of gauge theory from rotating branes,''
JHEP {\bf 9903}, 003 (1999)
[hep-th/9811120].
\bibitem{coul2}
I.~R.~Klebanov and E.~Witten,
%``AdS/CFT correspondence and symmetry breaking,''
Nucl.\ Phys.\  {\bf B556}, 89 (1999)
[hep-th/9905104].
\bibitem{coul3}
D.~Z.~Freedman, S.~S.~Gubser, K.~Pilch and N.~P.~Warner,
%``Continuous distributions of D3-branes and gauged supergravity,''
hep-th/9906194.
\bibitem{coul4}
A.~Brandhuber and K.~Sfetsos,
%``Wilson loops from multicentre and rotating branes, mass gaps and phase
%structure in gauge theories,''
hep-th/9906201.
\bibitem{jpp}
C.~V.~Johnson, A.~W.~Peet and J.~Polchinski,
%``Gauge theory and the excision of repulson singularities,''
hep-th/9911161;\\
L.~Jarv and C.~V.~Johnson,
%``Orientifolds, M-theory, and the ABCD's of the enhancon,''
hep-th/0002244.
\bibitem{ewelliptic}
E.~Witten,
%``Solutions of four-dimensional field theories via M-theory,''
Nucl.\ Phys.\  {\bf B500}, 3 (1997)
[hep-th/9703166].
\bibitem{RS}
L.~Randall and R.~Sundrum,
%``An alternative to compactification,''
Phys.\ Rev.\ Lett.\  {\bf 83}, 4690 (1999)
[hep-th/9906064].
\bibitem{stanford}
N.~Arkani-Hamed, S.~Dimopoulos, N.~Kaloper and R.~Sundrum,
%``A small cosmological constant from a large extra dimension,''
hep-th/0001197;\\
S.~Kachru, M.~Schulz and E.~Silverstein,
%``Self-tuning flat domain walls in 5d gravity and string theory,''
hep-th/0001206.
\bibitem{twoadj}
A.~Khavaev, K.~Pilch and N.~P.~Warner,
%``New vacua of gauged N = 8 supergravity in five dimensions,''
hep-th/9812035.
\bibitem{karch}
A.~Karch, D.~Lust and A.~Miemiec,
%``New N = 1 superconformal field theories and their supergravity
%description,''
Phys.\ Lett.\  {\bf B454}, 265 (1999)
[hep-th/9901041].
\bibitem{ALIS}
P.~C.~Argyres, K.~Intriligator, R.~G.~Leigh and M.~J.~Strassler,
%``On inherited duality in N = 1 d = 4 supersymmetric gauge theories,''
hep-th/9910250.
\bibitem{tHclass}
G.~'t Hooft,
%``On The Phase Transition Towards Permanent Quark Confinement,''
Nucl.\ Phys.\  {\bf B138}, 1 (1978);
%``A Property Of Electric And Magnetic Flux In Nonabelian Gauge Theories,''
Nucl.\ Phys.\  {\bf B153}, 141 (1979);
%``Topology Of The Gauge Condition And New Confinement Phases In Nonabelian
%Gauge Theories,''
Nucl.\ Phys.\  {\bf B190}, 455 (1981).
\bibitem{ewtheta}
E.~Witten,
%``Dyons Of Charge E Theta / 2 Pi,''
Phys.\ Lett.\  {\bf B86}, 283 (1979).
\bibitem{om}
C.~Montonen and D.~Olive,
%``Magnetic Monopoles As Gauge Particles?,''
Phys.\ Lett.\  {\bf B72}, 117 (1977).
\bibitem{dbranes}
J.~Polchinski,
%``Dirichlet-Branes and Ramond-Ramond Charges,''
Phys.\ Rev.\ Lett.\  {\bf 75}, 4724 (1995)
[hep-th/9510017];
%``TASI lectures on D-branes,''
hep-th/9611050.
\bibitem{witads}
E.~Witten,
%``Anti-de Sitter space and holography,''
Adv.\ Theor.\ Math.\ Phys.\  {\bf 2}, 253 (1998)
[hep-th/9802150].
\bibitem{GPK}
S.~S.~Gubser, I.~R.~Klebanov and A.~M.~Polyakov,
%``Gauge theory correlators from non-critical string theory,''
Phys.\ Lett.\  {\bf B428}, 105 (1998)
[hep-th/9802109].
\bibitem{bala}
V.~Balasubramanian, P.~Kraus and A.~Lawrence,
%``Bulk vs. boundary dynamics in anti-de Sitter spacetime,''
Phys.\ Rev.\  {\bf D59}, 046003 (1999)
[hep-th/9805171]; 
T.~Banks, M.~R.~Douglas, G.~T.~Horowitz and E.~Martinec,
%``AdS dynamics from conformal field theory,''
hep-th/9808016.
\bibitem{ir1}L.~Girardello, M.~Petrini, M.~Porrati and A.~Zaffaroni,
%``Novel local CFT and exact results on perturbations of N = 4 super
%Yang-Mills from AdS dynamics,'' JHEP {\bf 9812}, 022 (1998)
[hep-th/9810126].
\bibitem{ir2}
J.~Distler and F.~Zamora,
%``Non-supersymmetric conformal field theories from stable anti-de Sitter
%spaces,'' Adv.\ Theor.\ Math.\ Phys.\  {\bf 2}, 1405 (1999)
[hep-th/9810206].
\bibitem{ir3}
D.~Z.~Freedman, S.~S.~Gubser, K.~Pilch and N.~P.~Warner,
%``Renormalization group flows from holography supersymmetry and a  c-theorem,''
hep-th/9904017.
\bibitem{gppz1}
L.~Girardello, M.~Petrini, M.~Porrati and A.~Zaffaroni,
%``Confinement and condensates without fine tuning in supergravity duals
%of gauge theories,'' JHEP {\bf 9905}, 026 (1999)
[hep-th/9903026].
\bibitem{gubs}
S.~S.~Gubser,
%``Curvature singularities: The good, the bad, and the naked,''
hep-th/0002160.
\bibitem{garyrob}
G.~T.~Horowitz and R.~Myers,
%``The value of singularities,''
Gen.\ Rel.\ Grav.\  {\bf 27}, 915 (1995)
[gr-qc/9503062].
\bibitem{priv1}
O. de Wolfe, D. Freedman, S. Gubser, K. Pilch, N. Warner, E.
Witten, private communications.
\bibitem{iibact}
E.~Bergshoeff, H.~J.~Boonstra and T.~Ortin,
%``S duality and dyonic p-brane solutions in type II string theory,''
Phys.\ Rev.\  {\bf D53}, 7206 (1996)
[hep-th/9508091].
\bibitem{schwarz}
J.~H.~Schwarz,
%``Covariant Field Equations Of Chiral N=2 D = 10 Supergravity,''
Nucl.\ Phys.\  {\bf B226}, 269 (1983).
\bibitem{KRv}
H.~J.~Kim, L.~J.~Romans and P.~van Nieuwenhuizen,
%``The Mass Spectrum Of Chiral N=2 D = 10 Supergravity On S**5,''
Phys.\ Rev.\  {\bf D32}, 389 (1985).
\bibitem{gunmar}
M.~Gunaydin and N.~Marcus,
%``The Spectrum Of The S**5 Compactification Of The Chiral N=2, D = 10
%Supergravity And The Unitary Supermultiplets Of U(2, 2/4),''
Class.\ Quant.\ Grav.\  {\bf 2}, L11 (1985).
\bibitem{lido}
R.~G.~Leigh,
%``Dirac-Born-Infeld Action From Dirichlet Sigma Model,''
Mod.\ Phys.\ Lett.\  {\bf A4}, 2767 (1989);\\
M.~Li,
%``Boundary States of D-Branes and Dy-Strings,''
Nucl.\ Phys.\  {\bf B460}, 351 (1996)
[hep-th/9510161];\\
M.~R.~Douglas,
%``Branes within branes,''
hep-th/9512077.
\bibitem{bergd}
M.~Cederwall, A.~von Gussich, B.~E.~Nilsson, P.~Sundell and A.~Westerberg,
%``The Dirichlet super-p-branes in ten-dimensional type IIA and IIB
%supergravity,''
Nucl.\ Phys.\  {\bf B490}, 179 (1997)
[hep-th/9611159];\\
E.~Bergshoeff and P.~K.~Townsend,
%``Super D-branes,''
Nucl.\ Phys.\  {\bf B490}, 145 (1997)
[hep-th/9611173].
\bibitem{smat}
J.~Polchinski,
%``S-matrices from AdS spacetime,''
hep-th/9901076;\\
L.~Susskind,
%``Holography in the flat space limit,''
hep-th/9901079.
\bibitem{AOS-J}
M.~Alishahiha, Y.~Oz and M.~M.~Sheikh-Jabbari,
%``Supergravity and large N noncommutative field theories,''
JHEP {\bf 9911}, 007 (1999)
[hep-th/9909215].
\bibitem{CHS2}
C.~G.~Callan, J.~A.~Harvey and A.~Strominger,
%``World sheet approach to heterotic instantons and solitons,''
Nucl.\ Phys.\  {\bf B359}, 611 (1991).
\bibitem{ewbaryons}
E.~Witten,
%``Baryons and branes in anti de Sitter space,''
JHEP {\bf 9807}, 006 (1998)
[hep-th/9805112].
\bibitem{gktII}
S.~S.~Gubser and I.~R.~Klebanov,
%``Baryons and domain walls in an N = 1 superconformal gauge theory,''
Phys.\ Rev.\  {\bf D58}, 125025 (1998)
[hep-th/9808075].
\bibitem{dibaryons}
S.~Gukov, M.~Rangamani and E.~Witten,
%``Dibaryons, strings, and branes in AdS orbifold models,''
JHEP {\bf 9812}, 025 (1998)
[hep-th/9811048].
\bibitem{reyyee}
S.~Rey and J.~Yee,
%``Macroscopic strings as heavy quarks in large N gauge theory and  anti-de
%Sitter supergravity,''
hep-th/9803001.
\bibitem{maldastring}
J.~Maldacena,
%``Wilson loops in large N field theories,''
Phys.\ Rev.\ Lett.\  {\bf 80}, 4859 (1998)
[hep-th/9803002].
\bibitem{throat}
We thank O. Aharony, M. Berkooz and S. Sethi for their
insights. 
\bibitem{ncinst}
N.~Nekrasov and A.~Schwarz,
%``Instantons on noncommutative R**4 and (2,0) superconformal six
%dimensional theory,''
Commun.\ Math.\ Phys.\  {\bf 198}, 689 (1998)
[hep-th/9802068];\\
B.~Pioline and A.~Schwarz,
%``Morita equivalence and T-duality (or B versus Theta),''
JHEP {\bf 9908}, 021 (1999)
[hep-th/9908019].
\bibitem{IMSY}
N.~Itzhaki, J.~M.~Maldacena, J.~Sonnenschein and S.~Yankielowicz,
%``Supergravity and the large N limit of theories with sixteen  supercharges,''
Phys.\ Rev.\  {\bf D58}, 046004 (1998)
[hep-th/9802042].
\bibitem{corr}
G.~T.~Horowitz and J.~Polchinski,
%``A correspondence principle for black holes and strings,''
Phys.\ Rev.\  {\bf D55}, 6189 (1997)
[hep-th/9612146].
\bibitem{ahew}
A.~Hanany and E.~Witten,
%``Type IIB superstrings, BPS monopoles, and three-dimensional gauge
%dynamics,''
Nucl.\ Phys.\  {\bf B492}, 152 (1997)
[hep-th/9611230].
\bibitem{dvalishif}
G.~Dvali and M.~Shifman,
Nucl.\ Phys.\  {\bf B504}, 127 (1997)
[hep-th/9611213];
%``Domain walls in strongly coupled theories,''
Phys.\ Lett.\  {\bf B396}, 64 (1997)
[hep-th/9612128].
\bibitem{ewMQCD}
E.~Witten,
%``Branes and the dynamics of {QCD},''
Nucl.\ Phys.\  {\bf B507}, 658 (1997)
[hep-th/9706109].
\bibitem{Sen}
A.~Sen,
%``SO(32) spinors of type I and other solitons on brane-antibrane pair,''
JHEP {\bf 9809}, 023 (1998)
[hep-th/9808141].
\bibitem{dvakakgab}
G.~Dvali, G.~Gabadadze and Z.~Kakushadze,
%``BPS domain walls in large N supersymmetric {QCD},''
Nucl.\ Phys.\  {\bf B562}, 158 (1999)
[hep-th/9901032].
\bibitem{montvay}
R.~Kirchner, I.~Montvay, J.~Westphalen, S.~Luckmann and K.~Spanderen  [DESY-
                  Munster Collaboration],
%``Evidence for discrete chiral symmetry breaking in N = 1 supersymmetric
%Yang-Mills theory,''
Phys.\ Lett.\  {\bf B446}, 209 (1999)
[hep-lat/9810062];\\
A.~Feo, R.~Kirchner, I.~Montvay and A.~Vladikas  [DESY-Munster
                  Collaboration],
%``Low-energy features of SU(2) Yang-Mills theory with light gluinos,''
hep-lat/9909071.
\bibitem{ADK}
O.~Aharony, N.~Dorey and S.~P.~Kumar, to appear.
\bibitem{brodie}
J.~H.~Brodie,
%``Fractional branes, confinement, and dynamically generated  superpotentials,''
Nucl.\ Phys.\  {\bf B532}, 137 (1998)
[hep-th/9803140].
\bibitem{narayanan}
R.~G.~Edwards, U.~M.~Heller and R.~Narayanan,
%``Evidence for fractional topological charge in SU(2) pure Yang-Mills
%theory,''
Phys.\ Lett.\  {\bf B438}, 96 (1998)
[hep-lat/9806011].
\bibitem{amandajoe}
A.~W.~Peet and J.~Polchinski,
%``UV/IR relations in AdS dynamics,''
Phys.\ Rev.\  {\bf D59}, 065011 (1999)
[hep-th/9809022].
\bibitem{CHS}
C.~G.~Callan, J.~A.~Harvey and A.~Strominger,
%``Worldbrane actions for string solitons,''
Nucl.\ Phys.\  {\bf B367}, 60 (1991).
\bibitem{nsewone}
N.~Seiberg and E.~Witten,
%``Electric - magnetic duality, monopole condensation, and confinement in N=2
%supersymmetric Yang-Mills theory,''
Nucl.\ Phys.\  {\bf B426}, 19 (1994)
[hep-th/9407087].
\bibitem{klytaps}
A.~Klemm, W.~Lerche, S.~Yankielowicz and S.~Theisen,
%``Simple singularities and N=2 supersymmetric Yang-Mills theory,''
Phys.\ Lett.\  {\bf B344}, 169 (1995)
[hep-th/9411048];
P.~C.~Argyres and A.~E.~Faraggi,
%``The vacuum structure and spectrum of N=2 supersymmetric SU(n) gauge theory,''
Phys.\ Rev.\ Lett.\  {\bf 74}, 3931 (1995)
[hep-th/9411057].
\bibitem{mdss}
M.~R.~Douglas and S.~H.~Shenker,
%``Dynamics of SU(N) supersymmetric gauge theory,''
Nucl.\ Phys.\  {\bf B447}, 271 (1995)
[hep-th/9503163].
\bibitem{ucbmqcd}
K.~Hori, H.~Ooguri and Y.~Oz,
%``Strong coupling dynamics of four-dimensional N = 1 gauge theories from  M
%theory fivebrane,''
Adv.\ Theor.\ Math.\ Phys.\  {\bf 1}, 1 (1998)
[hep-th/9706082].
\bibitem{hsz}
A.~Hanany, M.~J.~Strassler and A.~Zaffaroni,
%``Confinement and strings in M{QCD},''
Nucl.\ Phys.\  {\bf B513}, 87 (1998)
[hep-th/9707244].
\bibitem{deBOz}
J.~de Boer and Y.~Oz,
%``Monopole condensation and confining phase of N = 1 gauge theories via
%M-theory fivebrane,''
Nucl.\ Phys.\  {\bf B511}, 155 (1998)
[hep-th/9708044].
\bibitem{pwtend}
K.~Pilch and N.~P.~Warner,
%``A new supersymmetric compactification of chiral IIB supergravity,''
hep-th/0002192.
\bibitem{magoo}
O.~Aharony, S.~S.~Gubser, J.~Maldacena, H.~Ooguri and Y.~Oz,
%``Large N field theories, string theory and gravity,''
Phys.\ Rept.\  {\bf 323}, 183 (2000)
[hep-th/9905111].
\bibitem{quiver}
M.~R.~Douglas and G.~Moore,
%``D-branes, Quivers, and ALE Instantons,''
hep-th/9603167.
\bibitem{schmaltz}
M.~Schmaltz,
%``Duality of non-supersymmetric large N gauge theories,''
Phys.\ Rev.\  {\bf D59}, 105018 (1999)
[hep-th/9805218].
\bibitem{verl}
H.~Verlinde,
%``Holography and compactification,''
hep-th/9906182.
\bibitem{GW}
W.~D.~Goldberger and M.~B.~Wise,
%``Modulus stabilization with bulk fields,''
Phys.\ Rev.\ Lett.\  {\bf 83}, 4922 (1999)
[hep-ph/9907447].
\bibitem{klebnek}
I.~R.~Klebanov and N.~A.~Nekrasov,
%``Gravity duals of fractional branes and logarithmic RG flow,''
hep-th/9911096.
\bibitem{klebtseyt}
I.~R.~Klebanov and A.~A.~Tseytlin,
%``Gravity duals of supersymmetric SU(N) x SU(N+M) gauge theories,''
hep-th/0002159.
\bibitem{brownteit}
J.~D.~Brown and C.~Teitelboim,
%``Dynamical Neutralization Of The Cosmological Constant,''
Phys.\ Lett.\  {\bf B195}, 177 (1987).
\bibitem{raph}
R. Bousso and J. Polchinski, hep-th/0003xyz.
\bibitem{matpert}
M.~Porrati and A.~Rozenberg,
%``Bound states at threshold in supersymmetric quantum mechanics,''
Nucl.\ Phys.\  {\bf B515}, 184 (1998)
[hep-th/9708119];\\
G.~Moore, N.~Nekrasov and S.~Shatashvili,
%``D-particle bound states and generalized instantons,''
Commun.\ Math.\ Phys.\  {\bf 209}, 77 (2000)
[hep-th/9803265];\\
N.~A.~Nekrasov,
%``On the size of a graviton,''
hep-th/9909213.
\bibitem{SEP}
J.~McGreevy, L.~Susskind and N.~Toumbas,
%``Invasion of the Giant Gravitons from Anti-de Sitter Space,''
hep-th/0003075.
\bibitem{AD}
P.~C.~Argyres and M.~R.~Douglas,
%``New phenomena in SU(3) supersymmetric gauge theory,''
Nucl.\ Phys.\  {\bf B448}, 93 (1995)
[hep-th/9505062].
\bibitem{QandA}
O.~Aharony, J.~Sonnenschein, S.~Yankielowicz and S.~Theisen,
%``Field theory questions for string theory answers,''
Nucl.\ Phys.\  {\bf B493}, 177 (1997)
[hep-th/9611222].
\bibitem{AFM}
O.~Aharony, A.~Fayyazuddin and J.~Maldacena,
%``The large N limit of N = 2,1 field theories from three-branes in  F-theory,''
JHEP {\bf 9807}, 013 (1998)
[hep-th/9806159].
\end{references}
\end{document}